\documentclass{amsart}

\usepackage{amssymb}

\usepackage{graphicx}

\usepackage[cmtip,all]{xy}

\usepackage{amsrefs}

\usepackage{pgfplots}
\pgfplotsset{compat=1.14}
\usetikzlibrary{arrows}
\usetikzlibrary{calc}

\usepackage[ruled,linesnumbered]{algorithm2e}

\usepackage{hyperref}

\usepackage[capitalize]{cleveref}

\usepackage{stmaryrd}

\usepackage{mathrsfs} 

\usepackage{mathtools}

\usepackage{ragged2e}

\usepackage{enumitem}

\usepackage{caption}

\usepackage{booktabs}

\usepackage{multirow}

\newcommand{\N}{\mathbb{N}}
\newcommand{\Z}{\mathbb{Z}}

\newcommand{\R}{\mathbb{R}}

\newcommand{\F}{\mathbb{F}}

\newcommand{\m}[1]{\mathcal{#1}}
\newcommand{\mA}{\mathcal{A}}
\newcommand{\mB}{\mathcal{B}}
\newcommand{\mC}{\mathcal{C}}

\newcommand{\mL}{\mathcal{L}}
\newcommand{\mM}{\mathcal{M}}
\newcommand{\mO}{\mathcal{O}}
\newcommand{\bdelta}{{\mbox{\boldmath$\delta$}}}
\newcommand{\tbdelta}{{\mbox{\boldmath{\tiny{$\delta$}}}}}

\newcommand{\Int}[2]{\left\llbracket #1~;~#2\right\rrbracket}

\DeclareMathOperator{\Hom}{Hom}
\DeclareMathOperator{\End}{End}
\DeclareMathOperator{\Aut}{Aut}
\DeclareMathOperator{\Diag}{Diag}

\DeclareMathOperator{\id}{id}
\DeclareMathOperator{\Pic}{Pic}
\DeclareMathOperator{\Sp}{Sp}
\DeclareMathOperator{\argmin}{argmin}
\DeclareMathOperator{\Char}{char}

\copyrightinfo{2024}{Pierrick Dartois}

\theoremstyle{definition}
\newtheorem{Definition}{Definition}

\theoremstyle{plain}
\newtheorem{Proposition}[Definition]{Proposition}
\newtheorem{Lemma}[Definition]{Lemma}

\newtheorem{Theorem}[Definition]{Theorem}

\theoremstyle{definition}
\newtheorem{Remark}[Definition]{Remark}

\newtheorem{Notation}[Definition]{Notation}

\begin{document}

\title[Fast computation of 2-isogenies in dimension 4]{Fast computation of 2-isogenies in dimension 4\\ and cryptographic applications}


\author{Pierrick Dartois}
\address{Univ. Bordeaux, CNRS, INRIA, Bordeaux INP, IMB, UMR 5251, F-33400, Talence, France \and INRIA, IMB, UMR 5251, F-33400, Talence, France}
\curraddr{}
\email{pierrick dot dartois at u-bordeaux dot fr}
\thanks{}



\dedicatory{}

\begin{abstract}
Dimension 4 isogenies have first been introduced in cryptography for the cryptanalysis of Supersingular Isogeny Diffie-Hellman (SIDH) 
and have been used constructively in several schemes, including SQIsignHD, a derivative of SQIsign isogeny based signature scheme. 
Unlike in dimensions 2 and 3, we can no longer rely on the Jacobian model and its derivatives to compute isogenies. In dimension 4 (and higher), we can only use theta-models. Previous works by Romain Cosset, David Lubicz and Damien Robert 
have focused on the computation of $\ell$-isogenies in theta-models of level $n$ coprime to $\ell$ (which requires to use $n^g$ coordinates in dimension $g$). For cryptographic applications, we need to compute chains of $2$-isogenies, requiring to use $\geq 3^g$ coordinates in dimension $g$ with state of the art algorithms.  

In this paper, we present algorithms to compute chains of $2$-isogenies between abelian varieties of dimension $g\geq 1$ with theta coordinates of level $n=2$, generalizing a previous work by Pierrick Dartois, Luciano Maino, Giacomo Pope and Damien Robert in dimension $g=2$. We propose an implementation of these algorithms in dimension $g=4$ to compute endomorphisms of elliptic curve products derived from Kani's lemma with applications to SQIsignHD and SIDH cryptanalysis. We are now able to run a complete key recovery attack on SIDH when the endomorphism ring of the starting curve is unknown within a few seconds on a laptop for all NIST SIKE parameters.
\end{abstract}

\maketitle

\section*{Acknowledgements}
This research has been supported by the Agence Nationale de la Recherche under grant ANR-19-CE48-0008 (CIAO) and the France 2030 program under grant ANR-22-PETQ-0008 (PQ-TLS).


\section{Introduction}

Higher dimensional isogenies have become a popular and widely used tool in cryptography since they were introduced to attack the Supersingular Isogeny Diffie-Hellman (SIDH) key exchange \cite{CastryckDecruSIDH,Maino_et_al_SIDH,RobSIDH}. The attacks against SIDH used a result due to Kani \cite[proof of Theorem 2.3]{Kani1997}, abusively called Kani's lemma, to "embed" a (presumably secret) isogeny $\varphi$ between elliptic curves into an isogeny $F$ between elliptic products of dimension $2, 4$ or $8$. This higher dimensional isogeny $F$ could be computed in polynomial time given images of $\varphi$ on some torsion points, that were given in SIDH. The ability to evaluate $F$ could be used to evaluate $\varphi$ everywhere, leading to a full key recovery.

With this new method involving higher dimensional isogenies, we are now able to evaluate an isogeny everywhere given its images on torsion points. This allows in particular to evaluate non-smooth degree isogenies efficiently \cite{RobPolyLog}, which could not be done with previous state of the art techniques. After its cryptanalytic use against SIDH, Kani's lemma has been leveraged for several constructive applications in cryptography: among others, the public-key encryption schemes FESTA \cite{FESTA} and its improvement QFESTA \cite{QFESTA}, SQIsignHD \cite{SQISignHD_autocite} followed by \cite{SQIsign2D-West,SQIsign2D-East,SQIPrime,IdToIso2D} improving the digital signature NIST candidate SQIsign \cite{SQIsign}, SCALLOP-HD \cite{SCALLOP-HD} improving the SCALLOP group action \cite{SCALLOP}, a key encapsulation mechanism IS-CUBE \cite{IS-CUBE} and a verifiable random function DeuringVRF \cite{Leroux_VRFHD}. From an algorithmic point of view, Kani's lemma offers a way to make the Deuring correspondence between ideals of an elliptic curve endomorphism ring and isogenies defined on this elliptic curve effective for ideals of non-smooth norms \cite{Clapotis}. This effective way to translate ideals into isogenies has been leveraged in SQIsignHD and follow-up works \cite{SQISignHD_autocite,SQIsign2D-West,SQIsign2D-East,SQIPrime}, DeuringVRF \cite{Leroux_VRFHD} and also in the group action setting \cite{SCALLOP-HD} where an orientation of elliptic curves is fixed.

\subsection{Previous works}

In the works of W. Castryck, T. Decru, L. Maino, C. Martindale, L. Panny, G. Pope and B. Wesolowski against SIDH \cite{CastryckDecruSIDH,Maino_et_al_SIDH}, dimension $2$ isogenies were used which constrained the attack. When the endomorphism ring of the starting curve is unknown, a subexponential search for parameter tweaks and a subexponential auxiliary $1$-dimensional isogeny computation were necessary. In~\cite[§ 2, 4]{RobSIDH}, D. Robert proved that using dimension $4$ or $8$ isogenies would relax these constraints and make the attack polynomial. Unlike $2$-dimensional attacks\footnote{See \url{https://github.com/Breaking-SIDH/direct-attack} and \url{https://github.com/GiacomoPope/Castryck-Decru-SageMath}}, Robert's $4$-dimensional attack has not been implemented yet.

Finding fast higher dimensional isogeny formulas was not crucial to cryptanalytic applications but has become relevant to constructive applications. Most applications use $2$-dimensional isogenies and some of them use $4$-dimensional isogenies \cite{SQISignHD_autocite,Leroux_VRFHD}. In dimension 2, state of the art techniques using Richelot isogenies \cite{Smith_thesis,Richelot2009} became the bottleneck of these cryptographic protocols, making them impractical. Recently, efficient formulas using theta coordinates have been introduced to compute $2$-dimensional $2$ and $3$-isogenies \cite{Theta_dim2,Theta_33}.

While still useful in practice, especially for SQIsignHD \cite{SQISignHD_autocite}, $4$-dimensional isogenies have been the focus of very little attention. Algorithmic efforts have been made to compute $\ell$-isogenies between abelian varieties of any dimension in the Theta model \cite{DRisogenies,RobLub2015,RobLub2022}. The state of the art technique \cite{RobLub2022} enables to compute $\ell$-isogenies between abelian varieties of dimension $g$ with theta coordinates of level $n$ coprime to $\ell$ in time $O((n\ell)^g)$. In particular, we can compute $2$-isogenies using $n^g$ theta coordinates of level $n$ with $n\geq 3$. Generalizing this method to level $n=\ell=2$ is essential to reduce complexity. This has been done and implemented in dimension~2~\cite{Theta_dim2}, leading to very fast computations. The algorithmic approach of~\cite{Theta_dim2} was based on formulas valid in any dimension introduced in a note by Damien Robert~\cite[Chapter 7]{Robert_note_theta}. However, no proof was provided for these formulas. 

\subsection{Our contribution}

In this paper, we generalize the algorithms of \cite{Theta_dim2} to compute $2$-isogenies in any dimension $g$ with level $2$ theta coordinates. We provide proofs for the formulas introduced in \cite[Chapter 7]{Robert_note_theta}. We also introduce and prove formulas to change theta coordinates required for isogeny computations. Finally, we provide an implementation of $4$-dimensional $2$-isogenies with applications to SQIsignHD\footnote{This implementation can be found here \url{https://github.com/Pierrick-Dartois/Theta_dim4}.} and SIDH torsion point attacks with a random starting curve \cite{RobSIDH} along with implementation details that were missing in the original SQIsignHD paper \cite[Appendix~F]{SQISignHD_autocite}.

\cref{sec: Preliminaries} recalls foundational notions on polarized abelian varieties and theta functions, following the approach of Mumford \cite{Mumford66}. \cref{sec: change of basis} introduces a formula to compute new theta coordinates associated to a change of symmetric theta structure. In \cref{sec: main section}, we provide formulas and an algorithmic approach to compute chains of $2$-isogenies in any dimension $g$. In \cref{sec: application to SQIsignHD}, we apply this approach to the computation of $4$-dimensional $2^e$-isogenies between elliptic products derived from Kani's lemma, e.g. in the context of SQIsignHD and SIDH torsion attacks. Our SIDH attacks run in a few seconds on a laptop for all NIST SIKE parameters. \cref{sec: basis change with full torsion} and \cref{sec: appendix half torsion} give implementation details, focusing on how to compute the necessary changes of theta coordinates in this context. Finally, \cref{sec: optimal strategies} explains how optimal divide an conquer strategies can be adapted to compute chains of $4$-dimensional isogenies.

\section{Preliminaries on the Theta model}\label{sec: Preliminaries}

\subsection{Polarized abelian varieties}

In the following, $k$ is an algebraically closed field of characteristic $\Char(k)\neq 2$ and $A$ is an abelian variety of dimension $g$ defined over $k$. Abelian varieties are projective connected group varieties. By rigidity, their group structure is always abelian \cite[Corollary 2.4]{Milne1986}. In particular, elliptic curves are abelian varieties of dimension 1.

An \emph{isogeny} between abelian varieties is a group variety homomorphism (homomorphism of varieties respecting the group law) which is surjective and has finite kernel. In particular, if $f: A\longrightarrow B$ is an isogeny, then $A$ and $B$ have the same dimension. As in dimension 1, in higher dimension, we can compute $f$ with the knowledge of $\ker(f)$. This will be the goal of this paper.

A \emph{line bundle} $\mL$ over $A$ is a locally free sheaf of $\mO_A$-modules of rank $1$. This means that $\mL$ is locally isomorphic to the sheaf of regular functions $\mO_A$ of $A$. Isomorphism classes of line bundles on $A$ form a group denoted by $\Pic(A)$ \cite[Chapter 7, p.~178]{GW}. Line bundles can be seen as divisors, since there is a group isomorphism between $\Pic(A)$ and equivalence classes of divisors modulo principal divisors \cite[Proposition II.6.15]{Hartshorne}. We denote by $\Pic^0(A)$ the subgroup of line bundles invariant by translation:
\[\Pic^0(A)=\{[\mL]\mid \forall a\in A(k),\quad t_a^*\mL\simeq\mL\}.\]
$\Pic^0(A)$ identifies with the group of $k$-rational points $\widehat{A}(k)$ of the \emph{dual abelian variety} $\widehat{A}$. If $\mL$ is a line bundle on $A$ generated by global sections $s_1, \cdots, s_n\in\Gamma(A,\mL)$ \emph{i.e.} when $s_{1,x},\cdots, s_{n,x}$ generate the stalk $\mL_x$ for all $x\in A$, these global sections define a morphism of $k$-varieties $A\longrightarrow \mathbb{P}_k^{n-1}$ \cite[Theorem II.7.1]{Hartshorne}. 
We say that $\mL$ is \emph{very ample} when it induces such a map $A\longrightarrow \mathbb{P}_k^{n-1}$ which is a closed immersion and that $\mL$ is \emph{ample} when one of its powers is very ample. Abelian varieties are projective \cite[Theorem 7.1]{Milne1986} so (very) ample line bundles always exist on $A$.  

Given a line bundle $\mL$ on $A$, one defines a group homomorphism $\varphi_\mL: A(k)\longrightarrow \Pic(A), a\longmapsto [t_a^*\mL\otimes\mL^{-1}]$. As a consequence of the theorem of the square \cite[Theorem 6.7]{Milne1986}, this homomorphism maps into $\Pic^0(A)$ \cite[Proposition 10.1]{Milne1986}. When $\mL$ is ample, $\varphi_\mL$ has finite kernel. In that case, $\varphi_\mL$ defines an isogeny $A\longrightarrow \widehat{A}$, which is called a \emph{polarization} of $A$. We say that $(A,\varphi_\mL)$ or $(A, \mL)$ is a \emph{polarized abelian variety}. Intuitively, a polarization on $A$ might be seen as a way to orient $A$. We can derive models from them (\emph{eg}. the Theta model). We say that $\mL$ or $\varphi_\mL$ is \emph{separable} when $\Char(k)\not|\deg(\varphi_\mL)$. \textbf{Throughout this paper, we shall assume that line bunldes are generated by global sections, ample and separable.} 

We say that a polarization $\varphi_\mL$ is \emph{principal} and that $(A,\mL)$ is a \emph{principally polarised abelian variety (PPAV)} when $\varphi_\mL$ is an isomorphism $A\overset{\sim}{\longrightarrow}\widehat{A}$. All elliptic curves are principally polarized but this is not the case of all abelian varieties. If $(A,\mL_0)$ and $(B,\mM_0)$ are PPAVs and $f: A\longrightarrow B$ is an isogeny and $d\in\N^*$, we say that $f$ is $d$-\emph{isogeny} when $\widehat{f}\circ\varphi_{\mM_0}\circ f=[d]\circ \varphi_{\mL_0}$ \emph{i.e.} $\widetilde{f}\circ f=[d]$, where $\widetilde{f}:=\varphi_{\mL_0}^{-1}\circ\widehat{f}\circ\varphi_{\mM_0}$. By abuse, we shall call $\widetilde{f}$ the dual of $f$ (instead of $\widehat{f}$). If $\mL$ and $\mM$ are polarisations on $A$ and $B$ respectively, we say that $f$ is a \emph{polarised isogeny} from $(A,\mL)$ to $(B,\mM)$ and denote $f: (A,\mL)\longrightarrow (B,\mM)$ when $\widehat{f}\circ\varphi_{\mM}\circ f=\varphi_{\mL}$. If $\mL=\mL_0^d$ and $\mM=\mM_0$, then $f$ is a $d$-isogeny if and only if it is a polarised isogeny $(A,\mL)\longrightarrow (B,\mM)$. Indeed, we have $\varphi_{\mL}=\varphi_{\mL_0^d}=[d]\circ\varphi_{\mL_0}$ by the theorem of the square \cite[Theorem 6.7]{Milne1986}.

\subsection{The theta group}

Let $\mL$ be an ample and separable line bundle on $A$ and $K(\mL):=\{x\in A(k)\mid t_x^*\mL\simeq\mL\}$ be the kernel of $\varphi_\mL$. The \emph{theta group} of $\mL$, denoted by $G(\mL)$ is made of pairs $(x,\phi_x)$, where $x\in K(\mL)$ and $\phi_x$ is an isomorphism $\mL\overset{\sim}{\longrightarrow} t_x^*\mL$ (which always exists by the definition of $K(\mL)$). $G(\mL)$ is indeed a group for the group law given by $(x,\phi_x)\cdot(y,\phi_y):=(x+y,t_x^*\phi_y\circ\phi_x)$ for all $(x,\phi_x),(y,\phi_y)\in G(\mL)$, where $t_x^*\phi_y\circ\phi_x$ is the map:
\[\mL\overset{\phi_x}{\longrightarrow} t_x^*\mL\overset{t_x^*\phi_y}{\longrightarrow} t_x^*(t_y^*\mL)=t_{x+y}^*\mL,\]
$t_x$ being the translation by $x$.

There is an exact sequence:
\[1\longrightarrow k^*\longrightarrow G(\mL)\longrightarrow K(\mL)\longrightarrow 0,\]
where the first arrow is $\lambda\longmapsto (0,\lambda\id_\mL)$ and the last arrow is the \emph{forgetful map} $\rho_\mL: (x,\phi_x)\longmapsto x$. 

$G(\mL)$ is not abelian. To measure the commutativity defect of two elements, we introduce the \emph{commutator pairing} $e_\mL: K(\mL)\times K(\mL)\longrightarrow k^*$, given for all $x, y\in K(\mL)$ by $e_\mL(x,y):=\widetilde{x}\cdot\widetilde{y}\cdot\widetilde{x}^{-1}\cdot\widetilde{y}^{-1}$, where $\widetilde{x}$ and $\widetilde{y}$ are respectively lifts of $x$ and $y$ in $G(\mL)$. By the above exact sequence, this quantity $e_\mL(x,y)$ defines a scalar in $k^*$ and does not depend on the lifts $\widetilde{x}$ and $\widetilde{y}$ of $x$ and $y$. The commutator pairing is bilinear, skew-symmetric and non-degenerate \cite[Theorem~1]{Mumford66}.

An isogeny between polarised abelian varieties $f: (A,\mL)\longrightarrow (B,\mM)$ satisfies $f^*\mM\simeq\mL$. Then, it is easy to check that $K(\mL)$ contains $K:=\ker(f)$. In addition, $K$ can be lifted in $G(\mL)$ as follows. Given an isomorphism $\alpha: f^*\m{M} \overset{\sim}{\longrightarrow} \mL$, we define $\widetilde{K}:=\{(x,t_x^*\alpha\circ\alpha^{-1})\mid x\in K\}\subset G(\mL)$. The forgetful map induces an isomorphism $\widetilde{K}\overset{\sim}{\longrightarrow} K$. We say that $\widetilde{K}$ is a \emph{level subgroup} lying above $K$. In particular, $\widetilde{K}$ is abelian so $e_\mL$ is trivial on $K$. In this case, we say that $K$ is \emph{isotropic}. Conversely, it can be proved that a subgroup $K\subseteq K(\mL)$ admits a level subgroup lying above it if and only if $K$ is isotropic. There is a one to one correspondence between level subgroups $\widetilde{K}$ in $G(\mL)$ and pairs $(f,\alpha)$ as above \cite[Proposition~1, p.~291]{Mumford66}.

\subsection{Theta structures}\label{sec: theta structures}

The commutator pairing being non-degenerate, it can be proved that $K(\mL)$ admits a \emph{symplectic decomposition}, namely there exists maximal isotropic subgroups $K_1(\mL)$ and $K_2(\mL)$ such that $K(\mL)=K_1(\mL)\oplus K_2(\mL)$ and such that $e_\mL$ induces an isomorphism $K_2(\mL)\cong\widehat{K}_1(\mL):=\Hom(K_1(\mL),k^*)$. By the finite abelian groups structure theorem, there exists a unique tuple of integers 
$\bdelta=(d_1,\cdots, d_r)$ such that $d_1|\cdots|d_r$ and $K_1(\mL)\simeq Z_1(\bdelta)$ and $K_2(\mL)\simeq Z_2(\bdelta)$, where:
\[Z_1(\bdelta):=\prod_{i=1}^r\Z/d_i\Z \quad \mbox{and} \quad Z_2(\bdelta):=\widehat{Z}_1(\bdelta)=\Hom(Z_1(\bdelta),k^*).\]
We say that $\mL$ is \emph{of type $\bdelta$}. It follows that $K(\mL)\subseteq A[d_r]$, with $A[d_r]\simeq (\Z/d_r\Z)^{2g}$ by \cite[Remarks 8.4 and 8.5]{Milne1986} and by separability of $\mL$. Hence, we can assume that $r=g$ without loss of generality. 

Let $Z(\bdelta):=Z_1(\bdelta)\times Z_2(\bdelta)$. As $K(\mL)$, $Z(\bdelta)$ can be equipped with a non-degenerate skew-symmetric pairing $e_{\tbdelta}: Z(\bdelta)\times Z(\bdelta)\longrightarrow k^*$, given by:
\begin{equation}\forall (\textbf{i},\chi),(\textbf{i}',\chi')\in Z(\bdelta),\quad e_{\tbdelta}((\textbf{i},\chi),(\textbf{i}',\chi'))=\chi'(\textbf{i})\chi(\textbf{i}')^{-1}.\label{eq: def e delta}
\end{equation}
$K(\mL)$ is not only isomorphic to $Z(\bdelta)$. Actually, there exists a \emph{symplectic isomorphism} $\sigma: K(\mL)\overset{\sim}{\longrightarrow}Z(\bdelta)$ mapping $K_i(\mL)$ to $Z_i(\bdelta)$ and such that $e_\tbdelta(\sigma(x),\sigma(y))=e_\mL(x,y)$ for all $x, y\in K(\mL)$. 

We define the \emph{Heisenberg group} as $\m{H}(\bdelta):=k^*\times Z(\bdelta)$, with the following non abelian group law:
\[(\alpha,\textbf{i},\chi)\cdot (\beta,\textbf{i}',\chi'):=(\alpha\beta\chi'(\textbf{i}),\textbf{i}+\textbf{i}',\chi\chi'),\] 
for all $(\alpha, \textbf{i},\chi), (\beta,\textbf{i}',\chi')\in k^*\times Z_1(\bdelta)\times Z_2(\bdelta)$. As for $G(\mL)$, there is an exact sequence:
\[1\longrightarrow k^*\longrightarrow \m{H}(\bdelta)\longrightarrow Z(\bdelta)\longrightarrow 0.\]
If $\mL$ is of type $\bdelta$, a \emph{theta structure} is an isomorphism $\Theta_\mL: \m{H}(\bdelta)\overset{\sim}{\longrightarrow}G(\mL)$ inducing an isomorphism of exact sequences:
\[\xymatrix{
1 \ar[r] & k^* \ar[r] \ar@{=}[d] & \m{H}(\bdelta) \ar[r] \ar[d]^{\Theta_\mL} & Z(\bdelta) \ar[r] \ar[d]^{\overline{\Theta}_\mL} & 0 \\
1 \ar[r] & k^* \ar[r] & G(\mL) \ar[r] & K(\mL) \ar[r] & 0
}\] 
Such theta structures always exist and are in bijection with triples $(\overline{\Theta}_\mL,s_1,s_2)$, where $\overline{\Theta}_\mL$ is a symplectic isomorphism $Z(\bdelta)\overset{\sim}{\longrightarrow} K(\mL)$ and $s_i$ are sections $K_i(\mL)\overset{\sim}{\longrightarrow}\widetilde{K}_i(\mL)$, the $\widetilde{K}_i(\mL)\subset G(\mL)$ being level subgroups lying above $K_i(\mL)$ for $i\in\{1,2\}$. Note that the $K_i(\mL)$ are fully determined by $\overline{\Theta}_\mL$ via the formula $K_i(\mL)=\overline{\Theta}_\mL(Z_i(\bdelta))$. In the following, we denote $K_i(\overline{\Theta}_\mL)$ or $K_i(\Theta_\mL)$ instead of $K_i(\mL)$ and $\widetilde{K}_i(\Theta_\mL)$ instead of $\widetilde{K}_i(\mL)$ to stress this dependence.

\subsection{Theta functions}\label{sec: theta functions}

The Heisenberg group $\m{H}(\bdelta)$ acts on the space $V(\bdelta)$ of functions $Z_1(\bdelta)\longrightarrow k$ as follows:
\begin{equation}\forall f\in V(\bdelta), (\alpha,\textbf{i},\chi)\in\m{H}(\bdelta), \quad (\alpha,\textbf{i},\chi)\cdot f: \textbf{j} \longmapsto \alpha \chi(\textbf{j})^{-1}f(\textbf{j}-\textbf{i}).\label{eq: theta group action}
\end{equation}
This action defines the only irreducible representation of $\m{H}(\bdelta)$ on which $k^*$ acts naturally \cite[Proposition 3, p. 295]{Mumford66}. Similarly, the theta group $G(\mL)$ acts on the ring of global sections $\Gamma(A,\mL)$ as follows:
\[\forall s\in\Gamma(A,\mL), (x,\phi_x)\in G(\mL), \quad (x,\phi_x)\cdot s=t_{-x}^*(\phi_x(s)).\]
When $\mL$ is of type $\bdelta$, this representation is irreducible \cite[Theorem 2, p. 297]{Mumford66} and naturally isomorphic to $V(\bdelta)$. Hence, there is an isomorphism $\beta: V(\bdelta)\overset{\sim}{\longrightarrow} \Gamma(A,\mL)$ respecting the group actions of $\m{H}(\bdelta)$ and $G(\mL)$, namely such that:
\begin{equation}
\forall v\in V(\bdelta), h\in \m{H}(\bdelta), \quad \beta(h\cdot v)=\Theta_{\mL}(h)\cdot\beta(v).\label{eq: beta}
\end{equation}
As a consequence of Schur's lemma \cite[Lemma XVIII.5.9]{Lang_Alg}, $\beta$ is unique up to scalar multiplication. Consider the basis of $V(\bdelta)$ given by Kronecker functions $(\delta_\textbf{i})_{\textbf{i}\in Z_1(\tbdelta)}$ and the basis of $\Gamma(A, \mL)$ given by $\theta_\textbf{i}^\mL:=\beta(\delta_\textbf{i})$ for all $\textbf{i}\in Z_1(\bdelta)$. We call this basis $(\theta_\textbf{i}^\mL)_{\textbf{i}\in Z_1(\tbdelta)}$ the basis of \emph{theta functions} associated to $\Theta_\mL$ or the basis of \emph{$\Theta_\mL$-coordinates}. It is defined up to multiplication by a scalar in $k^*$. 

When $\mL$ is generated by global sections, theta functions define a map $A\longrightarrow \mathbb{P}_k^{d-1}$ \cite[Theorem II.7.1]{Hartshorne},
where $d:=\prod_{i=1}^g d_i$ and $\bdelta:=(d_1,\cdots, d_g)$, so they give a way to represent the polarised abelian variety $(A,\mL)$ in the projective space. When $d_g\geq 3$, $\mL$ is generated by global sections and the induced map $A\longrightarrow \mathbb{P}_k^{d-1}$ is a closed immersion \cite[p.~163]{Mumford70}. When $2|\bdelta$, $\mL$ is also generated by global sections \cite[p.~60]{Mumford70} but the theta functions only define a closed immersion of the Kummer variety $A/\pm\lhook\joinrel\longrightarrow \mathbb{P}_k^{d-1}$ under the assumption that $(A,\mL)$ is not a product of polarised abelian varieties and that $\mL$ is \emph{totally symmetric}, as we shall see in \cref{sec: symmetric theta structures} \cite[Theorem~4.8.1]{ComplexAV}.

When we work with a polarization $\mL$ of type $\bdelta$, we obtain $d=\prod_{i=1}^g d_i$ theta coordinates to represent points. In practice, to minimize computational complexity, we assume $\bdelta=\underline{2}:=(2,\cdots,2)$ to obtain $2^g$ theta coordinates only. In that case, we say that $\Theta_\mL$ is a theta structure of \emph{level 2}. We work on the Kummer variety so points are represented up to sign. 

Theta functions are convenient because we can easily compute the action of the theta group on these functions. The theta group action can be leveraged to derive arithmetic formulas (including differential addition, doubling, isogeny computation, change of coordinates).

\subsection{The theta null point}

As we have seen previously, we may see theta functions as projective coordinates. The \emph{theta null point} is the projective point $(\theta_\textbf{i}^\mL(0))_{\textbf{i}\in Z_1(\tbdelta)}$. In \cite[p.~51]{DRphd}, Robert proves that the theta null point determines the sections $s_i: K_i(\Theta_\mL)\overset{\sim}{\longrightarrow} \widetilde{K}_i(\Theta_\mL)$ induced by the theta structure $\Theta_\mL$\footnote{Provided it is not identically zero, which is always the case in practice.}. When $4|\bdelta$, the theta null point even determines the symplectic isomorphism $\overline{\Theta}_\mL: Z(\bdelta)\overset{\sim}{\longrightarrow} K(\mL)$, so the whole theta structure. When $2|\bdelta$, $\overline{\Theta}_\mL$ is only determined up to signs.

In practice, this means that if $x=\overline{\Theta}_\mL(\textbf{j},\chi)\in K(\mL)$, then we have the equalities (as affine points):
\begin{equation}
(\theta_\textbf{i}^\mL(x))_{\textbf{i}\in Z_1(\tbdelta)}=(\Theta_\mL(1,\textbf{j},\chi)\cdot\theta_\textbf{i}^\mL(0))_{\textbf{i}}=(\chi(\textbf{i}+\textbf{j})^{-1}\cdot\theta_{\textbf{i}+\textbf{j}}^\mL(0))_{\textbf{i}},\label{eq: action translate null}
\end{equation} 
where the first equality is proved in \cite[p.~51]{DRphd} and the last follows from \cref{eq: theta group action}. More generally, the theta group action can be used to translate any point $y\in A(k)$ by a point of $K(\mL)$:
\begin{equation}
(\theta_\textbf{i}^\mL(x+y))_{\textbf{i}\in Z_1(\tbdelta)}=(\Theta_\mL(1,\textbf{j},\chi)\cdot\theta_\textbf{i}^\mL(y))_{\textbf{i}}=(\chi(\textbf{i}+\textbf{j})^{-1}\cdot\theta_{\textbf{i}+\textbf{j}}^\mL(y))_{\textbf{i}}.\label{eq: action translate}
\end{equation}

\subsection{The product theta structure}\label{sec: product}

Let $(A_1,\mL_1), \cdots, (A_r,\mL_r)$ be polarised abelian varieties, $A:=\prod_{i=1}^r A_i$ and $\mL:=\bigotimes_{i=1}^r \pi_i^*\mL_i$, where $\pi_i:A\longrightarrow A_i$ is the projection for all $i\in\Int{1}{r}$. Then $(A,\mL)$ is an abelian variety equipped with the \emph{product polarization}. We have natural isomorphisms $K(\mL)\cong\bigoplus_{i=1}^r K(\mL_i)$ and 
\[G(\mL)\cong \prod_{i=1}^r G(\mL_i)/\{(\lambda_1,\cdots,\lambda_r)\in k^*\mid \lambda_1\cdots\lambda_r=1\}.\]
Let $\bdelta^{(i)}$ be the type of $\mL_i$ for all $i\in\Int{1}{r}$ and $\bdelta:=\bdelta^{(1)}\vee\cdots\vee\bdelta^{(r)}$ be the concatenation of the $\bdelta^{(i)}$. Then, we also have:
\[\m{H}(\bdelta)\cong \prod_{i=1}^r \m{H}(\bdelta^{(i)})/\{(\lambda_1,\cdots,\lambda_r)\in k^*\mid \lambda_1\cdots\lambda_r=1\}.\]
If $\Theta_{\mL_1}, \cdots, \Theta_{\mL_r}$ are theta structures on $G(\mL_1), \cdots, G(\mL_r)$ respectively, the \emph{product theta structure} $\Theta_\mL:=\prod_{i=1}^r\Theta_{\mL_i}$ is the isomorphism $\m{H}(\bdelta)\overset{\sim}{\longrightarrow} G(\mL)$ induced by $(h_1,\cdots, h_r) \longmapsto (\Theta_{\mL_1}(h_1),\cdots, \Theta_{\mL_r}(h_r))$. This theta structure induces a natural symplectic decomposition of $K(\mL)=K_1(\Theta_\mL)\oplus K_2(\Theta_\mL)$, where $K_i(\Theta_\mL):=\prod_{j=1}^r K_i(\Theta_{\mL_j})$ for $i\in\{1,2\}$.

\begin{Lemma}\cite[p. 70]{DRphd}\label{lemma: product theta structure}
For all $\textbf{i}:=(\textbf{i}_1,\cdots, \textbf{i}_r)\in Z_1(\bdelta^{(1)})\times\cdots\times Z_1(\bdelta^{(r)})$,
\[\theta_\textbf{i}^\mL=\bigotimes_{j=1}^r \pi_j^*\theta_{\textbf{i}_j}^{\mL_j}.\]
\end{Lemma}

\begin{proof}
With the notations of \cref{sec: theta functions}, we have $V(\bdelta)\cong \bigoplus_{j=1}^r V(\bdelta^{(j)})$ and $\Gamma(A,\mL)=\bigotimes_{j=1}^r \pi_j^*\Gamma(A_j,\mL_j)$. For all $j\in\Int{1}{r}$, let $\beta_j: V(\bdelta^{(j)}) \overset{\sim}{\longrightarrow} \Gamma(A_j,\mL_j)$ be an isomorphism satisfying \cref{eq: beta} for $\Theta_{\mL_j}$. Then, the isomorphism $\beta: V(\bdelta)\overset{\sim}{\longrightarrow}\Gamma(A,\mL)$, $v_1\otimes\cdots\otimes v_r\longmapsto \pi_1^*\beta_1(v_1)\otimes\cdots\otimes\pi_r^*\beta_r(v_r)$ also satisfies \cref{eq: beta} for the product theta structure~$\Theta_{\mL}$. The result follows.
\end{proof}

Our goal is to work with a product of elliptic curves. By \cref{lemma: product theta structure}, we just have to multiply theta coordinates of elliptic curves to obtain theta coordinates on the product. However, a question remains: how can we translate elliptic curves Montgomery coordinates into theta coordinates?

\subsection{From Montgomery to theta coordinates in level 2}\label{sec: Montgomery theta}

In \cite[Chapter 7, Appendix A]{Robert_note_theta}, formulas were introduced to convert Montgomery coordinates into level 2 theta coordinates and vice versa. Let $E$ be an elliptic curve in the Montgomery model and $(T'_1,T'_2)$ be a basis of $E[4]$ such that $T'_2:=(-1:1)$. Let us write $T'_1:=(r:s)$. Then, we may define a level 2 theta structure on $E$ with theta null point $(a:b):=(r+s:r-s)$. The conversion map from Montgomery to theta coordinates is then $(x:z)\longmapsto (a(x-z):b(x+z))$.  Conversely, if $(a:b)$ is the theta null point, the conversion map from theta to Montgomery coordinates is $(\theta_0,\theta_1)\longmapsto (a\theta_1+b\theta_0:a\theta_1-b\theta_0)$.

We see here is that a basis of the $4$-torsion can determine a theta structure of level 2. This is a consequence of a general result for symmetric theta structures (\cref{thm: compatible symmetric theta structures}.(ii)).

\subsection{The isogeny theorem}\label{sec: isogeny theorem}

Here we explain how to compute isogenies with theta coordinates. Let $f:(A,\mL)\longrightarrow (B,\mM)$ be an isogeny between polarised abelian varieties. Let $\bdelta$ and $\bdelta_\mM$ be respectively the types of $\mL$ and $\mM$. We want to express the $f^*\theta_\textbf{i}^\mM$ for all $\textbf{i}\in Z_1(\bdelta_\mM)$ (seen as functions $x\longmapsto \theta_\textbf{i}^\mM(f(x))$) in the basis $(\theta_\textbf{j}^\mL)_{\textbf{j}\in Z_1(\tbdelta)}$ (seen as functions $x\longmapsto \theta_\textbf{j}^\mL(x)$).

We begin by choosing \emph{compatible} theta structures on $G(\mL)$ and $G(\mM)$ determining these theta functions. Let $K:=\ker(f)$. Assume that $K\subseteq K(\mL)$ is isotropic, let $\widetilde{K}$ be a level subgroup lying over $K$ and $\mathcal{Z}(\widetilde{K})$ be the centralizer of $\widetilde{K}$ in $G(\mL)$. Then, the isomorphism $\alpha: f^*\mM\overset{\sim}{\longrightarrow}\mL$ associated to $\widetilde{K}$ by \cite[Proposition~1, p.~291]{Mumford66} induces a surjective map $\alpha_f: \mathcal{Z}(\widetilde{K})\relbar\joinrel\twoheadrightarrow G(\mM)$ of kernel $\widetilde{K}$ \cite[Proposition 2, p. 291]{Mumford66}. For $i\in\{1,2\}$, let $\widetilde{K}_i(\Theta_\mL)$ (respectively $\widetilde{K}_i(\Theta_\mM)$) be the level subgroups lying above $K_i(\Theta_\mL)$ (respectively $K_i(\Theta_\mM)$) induced by the theta structure (as we saw in \cref{sec: theta structures}). 

\begin{Definition}\label{def: compatibility}
We say that two theta structures $\Theta_\mL$ and $\Theta_\mM$ on $G(\mL)$ and $G(\mM)$ respectively are \emph{compatible} when: \textbf{(i)} $\widetilde{K}=(\widetilde{K}\cap\widetilde{K}_1(\Theta_\mL))\oplus(\widetilde{K}\cap\widetilde{K}_2(\Theta_\mL))$ and \textbf{(ii)} $\alpha_f$ maps $\mathcal{Z}(\widetilde{K})\cap \widetilde{K}_i(\Theta_\mL)$ to $\widetilde{K}_i(\Theta_\mM)$ for $i\in\{1,2\}$.
\end{Definition}

Let $K^\bot:=\{x\in K(\mL)\mid \forall y\in K, \quad e_\mL(x,y)=1\}$ be the \emph{orthogonal of $K$} and let us write $K:=K_1\oplus K_2$ and $K^\bot:=K^{\bot,1}\oplus K^{\bot, 2}$, with $K_i, K^{\bot, i}\subseteq K_i(\Theta_\mL)$ for $i\in\{1,2\}$. Then, if we fix a theta structure $\Theta_\mL$ on $G(\mL)$, there is a one to one correspondence between theta structures $\Theta_\mM$ on $G(\mM)$ compatible with $\Theta_\mL$ and isomorphisms $\sigma: K^{\bot,1}/K_1\overset{\sim}{\longrightarrow} Z_1(\bdelta_\mM)$ \cite[Proposition 3.6.2]{DRphd}. In \cite[Theorem~4, p.~302]{Mumford66}, Mumford proved the following theorem to compute isogenies with theta coordinates, which has been reformulated by Robert \cite[Theorem 3.6.4]{DRphd}. 

\begin{Theorem}\label{thm: isogeny theorem}
Let $\Theta_\mL$ and $\Theta_\mM$ be compatible theta structures on $G(\mL)$ and $G(\mM)$ respectively and let $\sigma: K^{\bot,1}/K_1\overset{\sim}{\longrightarrow} Z_1(\bdelta_\mM)$ be the isomorphism induced by $\Theta_\mM$. Then, there exists $\lambda\in k^*$ such that for all $\textbf{i}\in Z_1(\bdelta_\mM)$,
\begin{equation}f^*\theta_\textbf{i}^{\mM}=\lambda\sum_{\textbf{j}\in\overline{\Theta}_\mL^{-1}(\sigma^{-1}(\{\textbf{i}\}))}\theta_\textbf{j}^\mL.\label{eq: iso theorem}
\end{equation}
\end{Theorem}

When $K\subseteq K_2(\Theta_\mL)$, there is always only one index $\textbf{j}$ in the sum of \cref{eq: iso theorem} and the isogeny is simpler to compute. In this paper, we always work in that case. When $K\not\subseteq K_2(\Theta_\mL)$, we may compute a change of theta coordinates. This is explained in \cref{sec: change of basis}.

\subsection{Symmetric theta structures}\label{sec: symmetric theta structures}

A line bundle $\mL$ on $A$ is \emph{symmetric} when $[-1]^*\mL\simeq\mL$. It is \emph{totally symmetric} when there exists a symmetric line bundle $\mM$ such that $\mL\simeq\mM^2$. By \cite[Proposition~1, p.~305]{Mumford66}, $\mL$ is totally symmetric if and only if it descends to a line bundle $\m{N}$ on the Kummer variety $K_A:=A/\pm$ via the projection $\pi: A\longrightarrow K_A$ \emph{i.e.} such that $\mL\simeq\pi^*\m{N}$.

Let $\mL$ be a symmetric line bundle of type $\bdelta$. In \cite[p. 308]{Mumford66}, Mumford defines $\delta_{-1}$, an automorphism of $G(\mL)$ making the following diagram commute:
\[\xymatrix{
1 \ar[r] & k^* \ar@{=}[d] \ar[r] & G(\mL) \ar[d]^{\delta_{-1}} \ar[r]^{\rho_\mL} & K(\mL) \ar[d]^{[-1]} \ar[r] & 0\\
1 \ar[r] & k^* \ar[r] & G(\mL) \ar[r]^{\rho_\mL} & K(\mL) \ar[r] & 0
}\]
He also defines an Heisenberg group analogue $D_{-1}\in\Aut(\m{H}(\bdelta))$. We say that a theta structure $\Theta_\mL$ on $G(\mL)$ is \emph{symmetric} when $\Theta_\mL\circ D_{-1}=\delta_{-1}\circ\Theta_\mL$. We say that an element $g\in G(\mL)$ is \emph{symmetric} when $\delta_{-1}(g)=g^{-1}$. A theta structure $\Theta_\mL$ is symmetric if and only if its induced level subgroups $\widetilde{K}_i(\Theta_\mL)$ are symmetric \emph{i.e.} made of symmetric elements \cite[Proposition 4.2.9]{DRphd}.

Let $f:A\longrightarrow B$ be an isogeny of kernel $K$. Let $\mL$ be a symmetric line bundle on $A$. Then, there exists a symmetric line bundle $\mM$ on $B$ such that $f^*\mM\simeq\mL$ if and only if there is a symmetric level subgroups $\widetilde{K}$ lying above $K$ \cite[Proposition 4.2.12]{DRphd}. Assuming $\mM$ exists, if $\Theta_\mL$ is a symmetric theta structure on $G(\mL)$, then any theta structure $\Theta_\mM$ on $G(\mM)$ which is compatible with $\Theta_\mL$ in the sense of \cref{def: compatibility}, is automatically symmetric \cite[Remark 4.2.15]{DRphd}. This property is very convenient to compute isogeny chains, because we can obtain a symmetric theta structure on the codomain from a symmetric theta structure on the domain.

Now, we assume that $\mL$ is totally symmetric. We explain compatibility conditions between theta structures on $G(\mL)$ and $G(\mL^2)$. Note that $K(\mL^2)=[2]^{-1}K(\mL)$ \cite[Proposition~4, p.~310]{Mumford66}, so that $K(\mL)\subset K(\mL^2)$. In \cite[pp. 309-310]{Mumford66}, Mumford defines group homomorphisms $\varepsilon_2$ and $\eta_2$ between $G(\mL)$ and $G(\mL^2)$ making the following diagrams commute:
\[\xymatrix{
1 \ar[r] & k^* \ar[d]^{\lambda \mapsto \lambda^2} \ar[r] & G(\mL) \ar[d]^{\varepsilon_2} \ar[r]^{\rho_\mL} & K(\mL) \ar@{^{(}->}[d] \ar[r] & 0\\
1 \ar[r] & k^* \ar[r] & G(\mL^2) \ar[r]^{\rho_{\mL^2}} & K(\mL^2) \ar[r] & 0
}\]
\[\xymatrix{
1 \ar[r] & k^* \ar[d]^{\lambda \mapsto \lambda^{2}} \ar[r] & G(\mL^2) \ar[d]^{\eta_2} \ar[r]^{\rho_{\mL^2}} & K(\mL^2) \ar[d]^{[2]} \ar[r] & 0\\
1 \ar[r] & k^* \ar[r] & G(\mL) \ar[r]^{\rho_\mL} & K(\mL) \ar[r] & 0
}\]
In \cite[p.~316]{Mumford66}, he also defines their Heisenberg analogues $E_2: \m{H}(\bdelta)\longrightarrow \m{H}(2\bdelta)$ and $H_2: \m{H}(2\bdelta)\longrightarrow \m{H}(\bdelta)$. 

\begin{Definition}
We say that theta structures $\Theta_\mL$ and $\Theta_{\mL^2}$ on $G(\mL)$ and $G(\mL^2)$ respectively are \emph{compatible} when $\Theta_{\mL^2}\circ E_2=\varepsilon_2\circ\Theta_\mL$ and $\Theta_{\mL}\circ H_2=\eta_2\circ\Theta_{\mL^2}$. We also say that $(\Theta_\mL,\Theta_{\mL^2})$ is a \emph{pair of symmetric theta structures} (for $(\mL,\mL^2)$).
\end{Definition}

\begin{Theorem}\label{thm: compatible symmetric theta structures}
\begin{enumerate}[label=(\roman*)]

\item[]

\item Every symmetric theta structure $\Theta_{\mL^2}$ on $G(\mL^2)$ induces a unique symmetric theta structure $\Theta_\mL$ on $G(\mL)$ that is compatible with $\Theta_{\mL^2}$.

\item The resulting theta structure $\Theta_\mL$ on $G(\mL)$ only depends on the symplectic isomorphism $\overline{\Theta}_{\mL^2}: Z(2\bdelta)\overset{\sim}{\longrightarrow} K(\mL^2)$.

\item Every symmetric theta structure on $G(\mL)$ is induced by a symmetric theta structure on $G(\mL^2)$, or equivalently, by a symplectic isomorphism $Z(2\bdelta)\overset{\sim}{\longrightarrow} K(\mL^2)$.
\end{enumerate}
\end{Theorem}

\begin{proof}
(i) is \cite[Remark~1, p.~317]{Mumford66}, (ii) is \cite[Remark~3, p.~319]{Mumford66} and (iii) is \cite[Remark~4, p.~319]{Mumford66}.
\end{proof}

\subsection{Addition and duplication formulas}\label{sec: duplication formulas}

Let $\mL$ be a totally symmetric line bundle of type $\bdelta$ on $A$. Let $(\Theta_\mL,\Theta_{\mL^2})$ be a pair of symmetric theta structures for $(\mL,\mL^2)$. Then, by \cite[Corollary 4.3.7]{DRphd}, we have for all $x, y\in A(k)$, and all $\textbf{i},\textbf{j}\in Z_1(\bdelta)$,
\begin{equation}
\theta_\textbf{i}^\mL(x+y)\theta_\textbf{j}^\mL(x-y)=\sum_{\left\{\substack{\textbf{u},\textbf{v}\in Z_1(2\tbdelta)\\\textbf{u}+\textbf{v}=2\textbf{i}\\\textbf{u}-\textbf{v}=2\textbf{j}}\right.}\theta_\textbf{u}^{\mL^2}(x)\theta_\textbf{v}^{\mL^2}(y).\label{eq: duplication 1}
\end{equation}
We have an injective map $(\Z/2\Z)^g\lhook\joinrel\longrightarrow Z_1(2\bdelta)$, mapping $\textbf{t}:=(t_1,\cdots,t_g)$ to $\textbf{t}\bdelta=(t_1d_1,\cdots, t_gd_g)$. We can then define the following change of variables: for all $\chi\in \widehat{(\Z/2\Z)^g}$ and $\textbf{i}\in Z_1(2\bdelta)$,
\begin{equation}
U_{\chi,\textbf{i}}^{\mL^2}:=\sum_{\textbf{t}\in (\Z/2\Z)^g} \chi(\textbf{t})\theta_{\textbf{i}+\textbf{t}\tbdelta}^{\mL^2}.\label{eq: dual theta coordinates}
\end{equation}
Using this change of variable, we can rewrite and revert \cref{eq: duplication 1}:

\begin{Theorem}\cite[Theorem 4.4.3]{DRphd}\label{thm: duplication formulas}
Let $x, y\in A(k)$. Then there exists $\lambda_1, \lambda_2\in k^*$ such that for all $\textbf{i}, \textbf{j}\in Z_1(2\bdelta)$ such that $\textbf{i}\equiv \textbf{j} \mod \underline{2}$ and $\chi\in \widehat{(\Z/2\Z)^g}$, we have:
\begin{equation}
\theta_{(\textbf{i}+\textbf{j})/2}^\mL(x+y)\theta_{(\textbf{i}-\textbf{j})/2}^\mL(x-y)=\lambda_1\sum_{\chi\in \widehat{(\Z/2\Z)^g}}U_{\chi,\textbf{i}}^{\mL^2}(x)U_{\chi,\textbf{j}}^{\mL^2}(y)\label{eq: duplication 2}
\end{equation}
\begin{equation}
U_{\chi,\textbf{i}}^{\mL^2}(x)U_{\chi,\textbf{j}}^{\mL^2}(y)=\lambda_2\sum_{\textbf{t}\in (\Z/2\Z)^g}\chi(\textbf{t})\theta_{(\textbf{i}+\textbf{j}+\textbf{t}\tbdelta)/2}^{\mL}(x+y)\theta_{(\textbf{i}-\textbf{j}+\textbf{t}\tbdelta)/2}^{\mL}(x-y).\label{eq: duplication 3}
\end{equation}
\end{Theorem}

These formulas can be used to compute differential addition. Knowing the coordinates $\theta_\textbf{i}^\mL(x), \theta_\textbf{i}^\mL(y)$ and $\theta_\textbf{i}^\mL(x-y)$, we can obtain the $\theta_\textbf{i}^\mL(x+y)$ \cite[Algorithm 4.4.10]{DRphd}. In particular, for doubling ($x=y$), $(\theta_\textbf{i}^\mL(x-y))_{\textbf{i}}=(\theta_\textbf{i}^\mL(0))_{\textbf{i}}$ is the \emph{theta null point}, so we only need to know the $\theta_\textbf{i}^\mL(x)$, provided the theta null point has been precomputed (see \cref{alg: doubling}).

\subsection{Heisenberg group automorphisms}

We denote by $\Aut_{k^*}(\m{H}(\bdelta))$ the group of automorphisms of $\m{H}(\bdelta)$ fixing $k^*$. Every such automorphism $\psi\in\Aut_{k^*}(\m{H}(\bdelta))$ induces a symplectic isomorphism $\overline{\psi}\in\Sp(Z(\bdelta))$, and by \cite[Proposition 3.5.1]{DRphd}, we have an exact sequence:
\[0\longrightarrow Z(\bdelta)\longrightarrow \Aut_{k^*}(\m{H}(\bdelta))\longrightarrow \Sp(Z(\bdelta))\longrightarrow 1.\]
Every $\psi\in\Aut_{k^*}(\m{H}(\bdelta))$, can be written explicitly as 
\begin{equation}
\psi(\alpha,\textbf{i},\chi):=(\alpha\xi(\textbf{i},\chi),\overline{\psi}(\textbf{i},\chi))\label{eq: automorphism}
\end{equation}
for all $(\alpha,\textbf{i},\chi)\in\m{H}(\bdelta)$, where $\overline{\psi}:=(\overline{\psi}_1,\overline{\psi}_2)\in\Sp(Z(\bdelta))$ is the induced symplectic isomorphism and $\xi: Z(\bdelta)\longrightarrow k^*$ is a \emph{semi-character}, satisfying the following property:
\[\xi(\textbf{i}_1+\textbf{i}_2,\chi_1\cdot\chi_2)=\frac{\xi(\textbf{i}_1,\chi_2)\xi(\textbf{i}_2,\chi_2)\overline{\psi}_2(\textbf{i}_2,\chi_2)(\overline{\psi}_1(\textbf{i}_1,\chi_1))}{\chi_2(\textbf{i}_1)},\]
where $\overline{\psi}_2(\textbf{i}_2,\chi_2)$ is a character, for all $(\textbf{i}_1,\chi_1), (\textbf{i}_2,\chi_2)\in Z(\bdelta)$ \cite[Remark 3.5.2]{DRphd}. If $\mL$ has type $\bdelta$, then $\Aut_{k^*}(\m{H}(\bdelta))$ acts faithfully and transitively on the set of theta structures on $G(\mL)$ by right-composition \cite[p. 52]{DRphd}.

Now, what happens if we restrict to symmetric theta structures? An automorphism $\psi\in\Aut_{k^*}(\m{H}(\bdelta))$ is \emph{symmetric} if $\psi\circ D_{-1}=D_{-1}\circ\psi$ (where $D_{-1}$ has been defined in \cref{sec: symmetric theta structures}). We denote by $\Aut_{k^*}^0(\m{H}(\bdelta))$ the subgroup of symmetric automorphisms. Let $\mL$ be a totally symmetric theta structure of type~$\bdelta$. Then, as previously, $\Aut_{k^*}^0(\m{H}(\bdelta))$ acts faithfully and transitively on the set of symmetric theta structures on $G(\mL)$. By \cite[p. 67]{DRphd}, we also have an exact sequence:
\[0\longrightarrow Z(\bdelta)[2]\longrightarrow \Aut_{k^*}^0(\m{H}(\bdelta))\longrightarrow \Sp(Z(\bdelta))\longrightarrow 1.\]

\begin{Lemma}\label{lemma: symmetric semi-character}
Let $\psi\in\Aut_{k^*}(\m{H}(\bdelta))$ and $\overline{\psi}$, $\xi$ as in \cref{eq: automorphism}. Then $\psi$ is symmetric if and only if 
\begin{equation}
\forall(\textbf{i},\chi)\in Z(\bdelta), \quad \xi(\textbf{i},\chi)^2=\chi(\textbf{i})^{-1}\overline{\psi}_2(\textbf{i},\chi)(\overline{\psi}_1(\textbf{i},\chi)). \label{eq: symmetric semi-character}
\end{equation}
\end{Lemma}

\begin{proof}
Let $(\alpha,\textbf{i},\chi)\in\m{H}(\bdelta)$. Then, by the definition of $D_{-1}$ \cite[p.~316]{Mumford66}, we have $D_{-1}(\alpha,\textbf{i},\chi)=(\alpha,-\textbf{i},\chi^{-1})=\alpha^2/\chi(\textbf{i})(\alpha,\textbf{i},\chi)^{-1}$, so that:
\[\psi\circ D_{-1}(\alpha,\textbf{i},\chi)=\psi\left(\frac{\alpha^2}{\chi(\textbf{i})}(\alpha,\textbf{i},\chi)^{-1}\right)=\frac{\alpha^2}{\chi(\textbf{i})}\psi(\alpha,\textbf{i},\chi)^{-1}\]
\[\mbox{and} \qquad D_{-1}\circ\psi(\alpha,\textbf{i},\chi)=\frac{\alpha^2\xi(\textbf{i},\chi)^2}{\overline{\psi}_2(\textbf{i},\chi)(\overline{\psi}_1(\textbf{i},\chi))}\psi(\alpha,\textbf{i},\chi)^{-1}\]
The result follows.
\end{proof} 

\subsection{Action of Heisenberg automorphisms on theta functions}

Let $\mL$ be a line bundle of type $\bdelta$ on $A$, $\Theta_\mL$ be a theta structure on $G(\mL)$, $\psi\in\Aut_{k^*}(\m{H}(\bdelta))$ and $\Theta'_\mL:=\Theta_\mL\circ\psi$. The group actions introduced in \cref{sec: theta functions} give a way to compute the change of coordinates matrix between theta functions $(\theta_\textbf{i})_{\textbf{i}\in Z_1(\tbdelta)}$ and $({\theta'}_\textbf{i})_{\textbf{i}\in Z_1(\tbdelta)}$ associated to $\Theta_\mL$ and $\Theta'_\mL$ respectively. 

By the definition of the theta group action (\cref{eq: theta group action}), we have:
\begin{equation}
\delta_\textbf{i}=(1,\textbf{i},1)\cdot\delta_0 \quad \mbox{and} \quad (1,0,\chi)\cdot\delta_0=\delta_0\label{eq: action Kronecker}
\end{equation}
for all $\textbf{i}\in Z_1(\bdelta)$ and $\chi\in Z_2(\bdelta)$, where the $\delta_\textbf{i}$ are the Kronecker functions. It follows that the action of the level subgroup $\widetilde{K}_2(\Theta'_\mL)$ associated to $\Theta'_\mL$ stabilizes~${\theta'}_0^{\mL}$. Furthermore, $T_\textbf{i}:=\sum_{\textbf{j}\in Z_2(\tbdelta)}\Theta_\mL\circ\psi(1,0,\textbf{j})\cdot\theta_\textbf{i}$ is stable under the action of $\widetilde{K}_2(\Theta'_\mL)$ for all $\textbf{i}\in Z_1(\bdelta)$. But this level subgroup is maximal (since $K_2(\Theta_\mL)$ is maximal isotropic in $K(\mL)$), so \cite[Proposition~3, p.~295]{Mumford66} ensures that the subspace of $V(\bdelta)$ stabilized by $\widetilde{K}_2(\Theta'_\mL)$ has dimension $1$. The following result follows:

\begin{Proposition}\cite[p.~53]{DRphd}\label{prop: theta change of basis}
There exists $\textbf{i}_0\in Z_1(\bdelta)$ and $\lambda\in k^*$ such that:
\[\theta'_0=\lambda\sum_{\chi\in Z_2(\tbdelta)}\Theta_\mL\circ\psi(1,0,\chi)\cdot\theta_{\textbf{i}_0}.\]
\end{Proposition}

Once we found $\theta'_0$, we can obtain $\theta'_\textbf{i}$ for all $\textbf{i}\in Z_1(\bdelta)$, by the formula $\theta'_\textbf{i}=\Theta_\mL\circ\psi(1,\textbf{i},1)\cdot\theta'_0$ derived from \cref{eq: action Kronecker}.

\section{Change of theta coordinates}\label{sec: change of basis}

In this section, we derive explicit change of coordinates formulas from \cref{prop: theta change of basis} in the case of symmetric theta structures. Given a totally symmetric line bundle $\mL$ of type $\bdelta$ on an abelian variety $A$ and a symmetric theta structure $\Theta_\mL$ on $G(\mL)$, we know that $\Theta_\mL$ is completely determined by a symplectic isomorphism $\overline{\Theta}_{\mL^2}:Z(2\bdelta)\overset{\sim}{\longrightarrow} K(\mL^2)$ by \cref{thm: compatible symmetric theta structures}. Our change of coordinates formula (\cref{thm: symplectic change of basis}) only depends on a symplectic isomorphism of $\Sp(Z(2\bdelta))$ \emph{i.e.}\ on a symplectic change of basis in $Z(2\bdelta)$ (or $K(\mL^2)$). This formula was already known to Igusa \cite[Theorem V.2]{igusaTheta} and Cosset \cite[Proposition 3.1.24]{Cosset_thesis} but was proved in the analytic setting of complex theta functions. Our proof only uses the algebraic setting of \cite{Mumford66}. In this setting, we obtain more convenient formulas to implement for isogeny computations over finite fields.

\subsection{Action of Heisenberg automorphisms on pairs of symmetric theta structures}\label{sec: action Heisenberg}

Throughout this section, we fix a totally symmetric line bundle $\mL$ of type $\bdelta:=(d_1, \cdots, d_g)$ on an abelian variety $A$. We study the action of $\Aut_{k^*}(\m{H}(2\bdelta))$ on pairs of symmetric theta structures $(\Theta_\mL,\Theta_{\mL^2})$ for $(\mL,\mL^2)$. 

First, we give more details on the maps $E_2$ and $H_2$ introduced in \cref{sec: symmetric theta structures} and defined in \cite[p.~316]{Mumford66}. We have a natural embedding $Z_1(\bdelta)\lhook\joinrel\longrightarrow Z_1(2\bdelta)$ via the multiplication by $2$ map and conversely, we have a natural surjective map $Z_1(2\bdelta)\relbar\joinrel\twoheadrightarrow Z_1(\bdelta)$ mapping $\textbf{i}:=(i_1 \mod 2d_1, \cdots, i_g \mod 2d_g)$ to $\overline{\textbf{i}}:=\textbf{i}\mod \bdelta = (i_1 \mod d_1, \cdots, i_g \mod d_g)$. Looking at the dual, we also have a natural embedding $Z_2(\bdelta)\lhook\joinrel\longrightarrow Z_2(2\bdelta)$ mapping every $\chi\in Z_2(\bdelta)$ to $2\star\chi:\textbf{i}\in Z_1(2\bdelta)\longmapsto \chi(\overline{\textbf{i}})\in k^*$ and a surjective map $Z_2(2\bdelta)\relbar\joinrel\twoheadrightarrow Z_2(\bdelta)$ mapping every $\chi\in Z_2(2\bdelta)$ to $\overline{\chi}: \textbf{i}\in Z_1(\bdelta)\longmapsto \chi(2\textbf{i})\in k^*$. Then $E_2:\m{H}(\bdelta)\longrightarrow\m{H}(2\bdelta)$ and $H_2: \m{H}(2\bdelta)\longrightarrow \m{H}(\bdelta)$ are given by: 
\[\forall (\alpha, \textbf{i}, \chi)\in \m{H}(\bdelta), \quad E_2(\alpha,\textbf{i}, \chi):=(\alpha^2,2\textbf{i},2\star\chi),\]
\[\forall (\alpha, \textbf{i}, \chi)\in \m{H}(2\bdelta), \quad H_2(\alpha, \textbf{i}, \chi):=(\alpha^2,\overline{\textbf{i}},\overline{\chi}).\]
Let $\overline{E}_2: Z(\bdelta)\longrightarrow Z(2\bdelta)$ and $\overline{H}_2: Z(2\bdelta)\longrightarrow Z(\bdelta)$ be the homomorphisms induced by $E_2$ and $H_2$ respectively. We also define $D_n: \m{H}(\bdelta)\longrightarrow\m{H}(\bdelta)$ by $D_n(\alpha,\textbf{i},\chi):=(\alpha^{n^2},n\textbf{i},\chi^n)$ for all $(\alpha, \textbf{i}, \chi)\in \m{H}(\bdelta)$ and $n\in\Z$.

\begin{Lemma}\label{lemma: properties E H D}\cite[p.~316]{Mumford66}
Assume that $2|\bdelta$. Then:

\begin{enumerate}[label=(\roman*)]
\item $\ker(H_2)=\{h\in\m{H}(2\bdelta)\mid h^2=1\}$.
\item $E_2\circ D_{-1}^{\m{H}(\tbdelta)}=D_{-1}^{\m{H}(2\tbdelta)}\circ E_2$ and $H_2\circ D_{-1}^{\m{H}(2\tbdelta)}=D_{-1}^{\m{H}(\tbdelta)}\circ H_2$ (where the exponents indicate the group of definition).
\item $E_2\circ H_2=D_2^{\m{H}(2\tbdelta)}$ and $H_2\circ E_2=D_2^{\m{H}(\tbdelta)}$.
\item For all $h\in\m{H}(\bdelta)$, $D_n(h)=h^{n(n+1)/2}D_{-1}(h)^{n(n-1)/2}$.
\end{enumerate}
\end{Lemma}

\begin{Proposition}\label{prop: compatible change of basis}
Assume that $2|\bdelta$. Then:

\begin{enumerate}[label=(\roman*)]
\item For all $\psi\in\Aut^0_{k^*}(\m{H}(2\bdelta))$, there exists a unique $\psi'\in\Aut^0_{k^*}(\m{H}(\bdelta))$ such that $\psi'\circ H_2=H_2\circ \psi$ and $\psi\circ E_2=E_2\circ\psi'$.

\item Let $\overline{\psi}, \overline{\psi}', \xi, \xi'$ be respectively the symplectic automorphisms and semi-characters associated to $\psi$ and $\psi'$ as in \cref{eq: automorphism}. Then, we have $\overline{\psi}'\circ\overline{H}_2=\overline{H}_2\circ\overline{\psi}$, $\overline{E}_2\circ\overline{\psi}'=\overline{\psi}\circ\overline{E}_2$ and:
\[\forall (\textbf{i},\chi)\in Z(2\bdelta), \quad \xi'(\overline{\textbf{i}},\overline{\chi})=\chi(\textbf{i})^{-1}\overline{\psi}_2(\textbf{i},\chi)(\overline{\psi}_1(\textbf{i},\chi)).\]

\item Let $(\Theta_\mL, \Theta_{\mL^2})$ and $(\Theta'_\mL, \Theta'_{\mL^2})$ be two pairs of symmetric theta structures for $(\mL,\mL^2)$. Then, there exists $\psi\in\Aut^0_{k^*}(\m{H}(2\bdelta))$ such that $\Theta'_{\mL^2}=\Theta_{\mL^2}\circ\psi$ and $\Theta'_{\mL}=\Theta_{\mL}\circ\psi'$, where $\psi'\in\Aut^0_{k^*}(\m{H}(\bdelta))$ is induced by $\psi$. 
\end{enumerate}
\end{Proposition}

\begin{proof}
\textbf{(i)} Since $2|\bdelta$, \cref{lemma: properties E H D}.(i) ensures that $\ker(H_2)=\m{H}(2\bdelta)[2]$, so that $\ker(H_2\circ \psi)=\m{H}(2\bdelta)[2]=\ker(H_2)$ as well since $\psi$ is an automorphism. Hence, $H_2\circ \psi$ factors through $H_2$ and this defines an automorphism $\psi': \m{H}(\bdelta)\overset{\sim}{\longrightarrow} \m{H}(\bdelta)$ such that $\psi'\circ H_2=H_2\circ\psi$. This automorphism $\psi'$ is trivial on $k^*$ because $\psi$ is and $H_2$ acts as $\lambda\longmapsto\lambda^2$. Furthermore, $\psi'$ is symmetric by \cref{lemma: properties E H D}.(ii) and since $H_2$ is surjective, so $\psi'\in\Aut^0_{k^*}(\m{H}(\bdelta))$. The uniqueness is a consequence of the surjectivity of $H_2$.

We now prove that $\psi\circ E_2=E_2\circ\psi'$. By surjectivity of $H_2$, it suffices to prove that $\psi\circ E_2\circ H_2=E_2\circ\psi'\circ H_2$ \emph{i.e.} that $\psi\circ D_2=D_2\circ\psi$ since $\psi'\circ H_2=H_2\circ \psi$ and $E_2\circ H_2=D_2$ by \cref{lemma: properties E H D}.(iii). Let $h\in \m{H}(2\bdelta)$. Then, $D_2(h)=h^3 D_{-1}(h)$ by \cref{lemma: properties E H D}.(iv) and since $\psi$ is symmetric:
\[\psi\circ D_2(h)=\psi(h^3D_{-1}(h))=\psi(h)^3\psi\circ D_{-1}(h)=\psi(h)^3 D_{-1}\circ\psi(h)=D_2\circ\psi(h)\]
The result follows.

\textbf{(ii)} Let $(\alpha,\textbf{i},\chi)\in\m{H}(2\bdelta)$. The equalities $\overline{\psi}'\circ\overline{H}_2=\overline{H}_2\circ\overline{\psi}$ and $\overline{E}_2\circ\overline{\psi}'=\overline{\psi}\circ\overline{E}_2$ immediately follow from $\psi'\circ H_2=H_2\circ \psi$ and $\psi\circ E_2=E_2\circ\psi'$. By \cref{lemma: symmetric semi-character}:
\[\xi'(\overline{\textbf{i}},\overline{\chi})=\xi(\textbf{i},\chi)^2=\chi(\textbf{i})^{-1}\overline{\psi}_2(\textbf{i},\chi)(\overline{\psi}_1(\textbf{i},\chi)).\]
The result follows.

\textbf{(iii)} Let $\psi:=\Theta_{\mL^2}^{-1}\circ\Theta'_{\mL^2}$. Then, $\Theta_{\mL^2}$ and $\Theta'_{\mL^2}$ being symmetric, we have:
\[\psi\circ D_{-1}=\Theta_{\mL^2}^{-1}\circ\Theta'_{\mL^2}\circ D_{-1}=\Theta_{\mL^2}^{-1}\circ\gamma_{-1}\circ\Theta'_{\mL^2}=D_{-1}\circ\Theta_{\mL^2}^{-1}\circ\Theta'_{\mL^2}=D_{-1}\circ\psi,\]
so $\psi\in\Aut^0_{k^*}(\m{H}(2\bdelta))$. In addition, by compatibility of the pairs $(\Theta_\mL, \Theta_{\mL^2})$ and $(\Theta'_\mL, \Theta'_{\mL^2})$:
\[\Theta'_{\mL}\circ H_2=\eta_2\circ \Theta'_{\mL^2}=\eta_2\circ \Theta_{\mL^2}\circ \psi=\Theta_{\mL}\circ H_2\circ\psi=\Theta_{\mL}\circ \psi'\circ H_2,\]
so that $\Theta'_{\mL}=\Theta_{\mL}\circ \psi'$ since $H_2$ is surjective. This completes the proof.
\end{proof}

\subsection{Change of symmetric theta structures}\label{subsec: change of basis}

Let $\bdelta:=(d_1,\cdots, d_g)$ with $d_1|\cdots|d_g$ and $\zeta\in k^*$ be a $d_g$-th primitive root of unity. Let us fix a canonical symplectic basis of $Z(\bdelta)$ as follows. For $l\in\Int{1}{g}$, let $\textbf{e}_l$ be the vector of $Z_1(\bdelta)$ with $1$ at index $l$ and $0$ everywhere else. For all $l\in\Int{1}{g}$, let $\chi_l\in Z_2(\bdelta)$ be the character such that $\chi_l(\textbf{e}_m)=\zeta^{d_g/d_l\delta_{l,m}}$ for all $l\in\Int{1}{g}$. Then $Z_1(\bdelta)$ can be canonically identified with $Z_2(\bdelta)$ via the map $\textbf{i}\in Z_1(\bdelta)\longmapsto \chi^\textbf{i}:=\prod_{l=1}^g \chi_l^{i_l}$. We then have
\begin{equation}
\forall \textbf{i},\textbf{j}\in Z_1(\bdelta), \quad \chi^\textbf{i}(\textbf{j})=\zeta^{\langle \textbf{i}|\textbf{j}\rangle} \quad \mbox{with} \quad \langle \textbf{i}|\textbf{j}\rangle:=\sum_{l=1}^g \frac{d_g}{d_l}i_l j_l.
\label{eq: def chi index and scalar prod}
\end{equation}
Such a basis is called a \emph{$\zeta$-canonical symplectic basis}.

\begin{Lemma}\label{lemma: symplectic matrix}
Let $\sigma: Z(\bdelta)\overset{\sim}{\longrightarrow} Z(\bdelta)$ be an automorphism of $Z(\bdelta)$ and $M$ be its matrix in the $\zeta$-canonical symplectic basis $(\textbf{e}_1,\cdots, \textbf{e}_g, \chi_1, \cdots, \chi_g)$. Then $\sigma$ is symplectic if and only if
\[\,^t M\cdot J_\Delta\cdot M\equiv J_\Delta \mod d_g, \qquad \mbox{where} \qquad J_\Delta:=\left(\begin{array}{cc}
0 & \Delta\\
-\Delta & 0
\end{array}\right)\]
and $\Delta:=\Diag(d_g/d_1,\cdots, d_g/d_{g-1},1)$. 

If we write
\[M:=\left(\begin{array}{cc} 
A & C\\
B & D
\end{array}\right),\]
this is equivalent to $\,^t B\Delta A\equiv \,^t A\Delta B$, $\,^t D\Delta C\equiv\,^t C\Delta D$ and $\,^t A\Delta D-\,^t B\Delta C\equiv \Delta$ modulo $d_g$. 
\end{Lemma}

\begin{proof}
Let $l, m\in\Int{1}{g}$. Then, by the definition of $M$, we have $\sigma(\textbf{e}_l,1)=(A\textbf{e}_l,\chi^{B\textbf{e}_l})$ and $\sigma(\textbf{e}_m,1)=(A\textbf{e}_m,\chi^{B\textbf{e}_m})$, where $\chi^{B\textbf{e}_l}$ and $\chi^{B\textbf{e}_m}$ have been defined in \cref{eq: def chi index and scalar prod}. It follows that, on the one hand:
\begin{align*}
e_\tbdelta(\sigma(\textbf{e}_l,1),\sigma(\textbf{e}_m,1))&=e_{\tbdelta}((A\textbf{e}_l,\chi^{B\textbf{e}_l}),(A\textbf{e}_m,\chi^{B\textbf{e}_m}))\\
&=\chi^{B\textbf{e}_j}(A\textbf{e}_l)\chi^{-B\textbf{e}_l}(A\textbf{e}_m) \quad (\mbox{by \cref{eq: def e delta}}) \\
&=\zeta^{\langle A\textbf{e}_l|B\textbf{e}_m\rangle -\langle B\textbf{e}_l|A\textbf{e}_m\rangle} \quad (\mbox{by \cref{eq: def chi index and scalar prod}})\\
&=\zeta^{\,^t \textbf{e}_l (\,^t A\Delta B -\,^t B\Delta A)\textbf{e}_m}.
\end{align*}
On the other hand, $e_\tbdelta((\textbf{e}_l,1),(\textbf{e}_m,1))=1$ by \cref{eq: def e delta}. We similarly obtain:
\begin{align*}
e_\tbdelta(\sigma(0,\chi_l),\sigma(0,\chi_m))&=e_{\tbdelta}((C\textbf{e}_l,\chi^{D\textbf{e}_l}),(C\textbf{e}_m,\chi^{D\textbf{e}_m}))=\chi^{D\textbf{e}_m}(C\textbf{e}_l)\chi^{-D\textbf{e}_l}(C\textbf{e}_m)\\
&=\zeta^{\langle C\textbf{e}_l|D\textbf{e}_m\rangle -\langle D\textbf{e}_l|C\textbf{e}_m\rangle}=\zeta^{\,^t \textbf{e}_l (\,^t C\Delta D -\,^t D\Delta C)\textbf{e}_m}
\end{align*}
and $e_\tbdelta((0,\chi_l),(0,\chi_m))=1$. Finally,
\begin{align*}
e_\tbdelta(\sigma(\textbf{e}_l,1),\sigma(0,\chi_m))&=e_{\tbdelta}((A\textbf{e}_l,\chi^{B\textbf{e}_l}),(C\textbf{e}_m,\chi^{D\textbf{e}_m}))=\chi^{D\textbf{e}_m}(A\textbf{e}_l)\chi^{-B\textbf{e}_l}(C\textbf{e}_m)\\
&=\zeta^{\langle A\textbf{e}_l|D\textbf{e}_m\rangle -\langle B\textbf{e}_l|C\textbf{e}_m\rangle}=\zeta^{\,^t \textbf{e}_l (\,^t A\Delta D-\,^t B\Delta C)\textbf{e}_m}
\end{align*}
and $e_\tbdelta((\textbf{e}_l,1),(0,\chi_m))=\zeta^{d_g/d_l\delta_{l, m}}$. Hence, $\sigma$ is symplectic if and only if
\begin{align*}
& \forall x,y\in Z(\bdelta), \quad e_{\tbdelta}(x,y)=e_{\tbdelta}(\sigma(x),\sigma(y))\\
\Longleftrightarrow & \ \forall l, m\in \Int{1}{g}, \quad \left\{\begin{array}{rcl} e_\tbdelta((\textbf{e}_l,1),(\textbf{e}_m,1)) & = & e_\tbdelta(\sigma(\textbf{e}_l,1),\sigma(\textbf{e}_m,1)) \\ 
e_\tbdelta((0,\chi_l),(0,\chi_m)) & = & e_\tbdelta(\sigma(0,\chi_l),\sigma(0,\chi_m)) \\
e_\tbdelta((\textbf{e}_l,1),(0,\chi_m)) & = & e_\tbdelta(\sigma(\textbf{e}_l,1),\sigma(0,\chi_m))
\end{array}\right. \\
\Longleftrightarrow & \ \forall l, m\in \Int{1}{g}, \quad \left\{\begin{array}{rcl} \,^t \textbf{e}_l (\,^t A\Delta B -\,^t B\Delta A)\textbf{e}_m & \equiv & 0 \mod d_g\\ 
\,^t \textbf{e}_l (\,^t C\Delta D -\,^t D\Delta C)\textbf{e}_m & \equiv & 0 \mod d_g \\
\,^t \textbf{e}_l (\,^t A\Delta D-\,^t B\Delta C)\textbf{e}_m & \equiv & \frac{d_g}{d_l}\delta_{l, m} \mod d_g 
\end{array}\right.\\
\Longleftrightarrow & \left\{\begin{array}{rcl} \,^t A\Delta B -\,^t B\Delta A & \equiv & 0 \mod d_g \\ 
\,^t C\Delta D -\,^t D\Delta C & \equiv & 0 \mod d_g  \\
\,^t A\Delta D-\,^t B\Delta C & \equiv & \Delta \mod d_g
\end{array}\right.
\end{align*}
The result follows.
\end{proof}

We now finally prove an explicit change of coordinates formula between symmetric theta structures $\Theta_\mL$ and $\Theta'_\mL$. This formula provides a \emph{change of coordinates matrix} from $\Theta_\mL$ to $\Theta'_\mL$. In the rest of the paper, \textbf{we shall only consider symmetric theta structures} (and simply mention them as theta structures) in order to be able to use the following theorem. 

\begin{Theorem}[Change of theta coordinates]\label{thm: symplectic change of basis}
Let $\Theta_{\mL^2}$ be a symmetric theta structure on $G(\mL^2)$ and $\Theta_\mL$ be the induced compatible theta structure on $G(\mL)$. Let $\psi\in\Aut^0_{k^*}(\m{H}(2\bdelta))$ and $\psi'\in\Aut^0_{k^*}(\m{H}(\bdelta))$ be the induced symmetric automorphism (following \cref{prop: compatible change of basis}.(i)). Let $\zeta$ be a primitive $2d_g$-th root of unitiy and 
\[M:=\left(\begin{array}{cc} 
A & C\\
B & D
\end{array}\right)\]
be the matrix of $\overline{\psi}\in\Sp(Z(2\bdelta))$ in the $\zeta$-canonical symplectic basis. Let $(\theta_\textbf{i}^{\mL})_{\textbf{i}\in Z_1(\tbdelta)}$ and $({\theta'}_\textbf{i}^{\mL})_{\textbf{i}\in Z_1(\tbdelta)}$ be respectively the $\Theta_\mL$ and $\Theta'_\mL$-coordinates (where $\Theta'_{\mL}:=\Theta_\mL\circ\psi'$). Then, there exists $\textbf{i}_0\in Z_1(\bdelta)$ and $\lambda\in k^*$ such that for all $\textbf{i}\in Z_1(\bdelta)$,
\[{\theta'}_\textbf{i}^{\mL}=\lambda\sum_{\textbf{j}\in Z_1(\tbdelta)}\zeta^{\langle \textbf{i}|\textbf{j}\rangle -\langle A\textbf{i}+C\textbf{j}+2\textbf{i}_0|B\textbf{i}+D\textbf{j}\rangle}\theta_{A\textbf{i}+C\textbf{j}+\textbf{i}_0}^\mL.\]
We can choose any value of $\textbf{i}_0\in Z_1(\bdelta)$ such that
\[\sum_{\textbf{j}\in Z_1(\tbdelta)}\zeta^{-\langle C\textbf{j}+2\textbf{i}_0|D\textbf{j}\rangle}\theta_{\textbf{i}_0+C\textbf{j}}^\mL\neq 0.\]
\end{Theorem}

\begin{proof}
By \cref{prop: theta change of basis}, there exists $\textbf{i}_0\in Z_1(\bdelta)$ and $\lambda\in k^*$ such that ${\theta'}_0^{\mL}=\lambda T_{\textbf{i}_0}$, where
\begin{equation}
T_{\textbf{i}_0}:=\sum_{\textbf{j}\in Z_2(\tbdelta)}\Theta_{\mL}\circ\psi'(1,0,\textbf{j})\cdot \theta_{\textbf{i}_0}^{\mL}
\label{eq: Ti0}
\end{equation}
is non-zero. 

Recall the notations introduced in the beginning of \cref{sec: action Heisenberg,subsec: change of basis}. As explained before, we can identify $Z_1(\bdelta)$ with $Z_2(\bdelta)$ via the map $\textbf{j}\longmapsto\chi^{\textbf{j}}$, where $\chi^\textbf{j}(\textbf{i}):=\zeta^{2\langle \textbf{i}|\textbf{j}\rangle}$ for all $\textbf{i}, \textbf{j}\in Z_1(\bdelta)$ ($\zeta^2$ being a primitive $d_g$-th root of unity). Similarly, we identify $Z_1(2\bdelta)$ with $Z_2(2\bdelta)$ via the map $\textbf{j}\longmapsto\widetilde{\chi}^{\textbf{j}}$, where $\widetilde{\chi}^\textbf{j}(\textbf{i})=\zeta^{\langle \textbf{i}|\textbf{j}\rangle}$ for all $\textbf{i}, \textbf{j}\in Z_1(2\bdelta)$. Now, by \cref{eq: automorphism} and \cref{prop: compatible change of basis}.(ii), we can express $\psi'$ as follows. For all $\textbf{i}, \textbf{j}\in Z_1(\bdelta)$, we have:
\begin{equation}
\psi'(1,\textbf{i},\chi^\textbf{j})=\left(\widetilde{\chi}^{\textbf{v}}(\textbf{u})^{-1}\overline{\psi}_2(\textbf{u},\widetilde{\chi}^{\textbf{v}})(\overline{\psi}_1(\textbf{u},\widetilde{\chi}^{\textbf{v}})),\overline{\psi}'(\textbf{i},\chi^{\textbf{j}})\right),\label{eq: exp psi'}
\end{equation}
with $\textbf{u}, \textbf{v}\in Z_1(2\bdelta)$ such that $\overline{\textbf{u}}=\textbf{i}$ and $\overline{\textbf{v}}=\textbf{j}$. Then, $M$ being the matrix of $\overline{\psi}$ in the $\zeta$-canonical symplectic basis of $Z(2\bdelta)$, we have:
\begin{equation}
\overline{\psi}_1(\textbf{u},\widetilde{\chi}^{\textbf{v}})=A\textbf{u}+C\textbf{v} \quad \mbox{and} \quad \overline{\psi}_2(\textbf{u},\widetilde{\chi}^{\textbf{v}})=\widetilde{\chi}^{B\textbf{u}+D\textbf{v}}.\label{eq: concrete exp psi}
\end{equation}
In addition, by \cref{prop: compatible change of basis}.(ii), we have:
\begin{align*}
\overline{\psi}'(\textbf{i},\chi^{\textbf{j}})&=\overline{\psi}'(\overline{\textbf{u}},\overline{\widetilde{\chi}^{\textbf{v}}})=\overline{\psi}'\circ\overline{H}_2(\textbf{u},\widetilde{\chi}^{\textbf{v}})=\overline{H}_2\circ\overline{\psi}(\textbf{u},\widetilde{\chi}^{\textbf{v}}) \\
&=\overline{H}_2(A\textbf{u}+C\textbf{v},\widetilde{\chi}^{B\textbf{u}+D\textbf{v}}) \quad (\mbox{by \cref{eq: concrete exp psi}})\\
&=(\overline{A\textbf{u}+C\textbf{v}},\overline{\widetilde{\chi}^{B\textbf{u}+D\textbf{v}}})=(A\textbf{i}+C\textbf{j},\chi^{B\textbf{i}+D\textbf{j}})
\end{align*}
From this and \cref{eq: exp psi'}, it follows that for all $(\textbf{i},\textbf{j})\in Z_1(\bdelta)$,
\begin{align}\psi'(1,\textbf{i},\chi^\textbf{j})&=(\zeta^{-\langle \textbf{u}|\textbf{v}\rangle}\widetilde{\chi}^{B\textbf{u}+D\textbf{v}}(A\textbf{u}+C\textbf{v}),A\textbf{i}+C\textbf{j},\chi^{B\textbf{i}+D\textbf{j}}) \nonumber \\
&=(\zeta^{-\langle \textbf{u}|\textbf{v}\rangle+\langle B\textbf{u}+D\textbf{v}|A\textbf{u}+C\textbf{v}\rangle},A\textbf{i}+C\textbf{j},\chi^{B\textbf{i}+D\textbf{j}})  \nonumber \\
&=(\zeta^{-\langle \textbf{i}|\textbf{j}\rangle+\langle B\textbf{i}+D\textbf{j}|A\textbf{i}+C\textbf{j}\rangle},A\textbf{i}+C\textbf{j},\chi^{B\textbf{i}+D\textbf{j}}) \label{eq: concrete exp psi'}
\end{align}
For the last equality, using the fact that $2|\bdelta$ since $\mL$ is totally symmetric, we can check that $-\langle \textbf{u}|\textbf{v}\rangle+\langle B\textbf{u}+D\textbf{v}|A\textbf{u}+C\textbf{v}\rangle$ modulo $2d_g$ only depends on the values of $\textbf{u}$ and $\textbf{v}$ modulo $\bdelta$. Consequently, by \cref{eq: Ti0} and the definition of the theta group action (\cref{eq: theta group action}),
\begin{align}T_{\textbf{i}_0}&=\sum_{\textbf{j}\in Z_2(\tbdelta)}\Theta_{\mL}(\zeta^{\langle D\textbf{j}|C\textbf{j}\rangle},C\textbf{j},\chi^{D\textbf{j}})\cdot \theta_{\textbf{i}_0}^{\mL} \nonumber\\
&=\sum_{\textbf{j}\in Z_1(\tbdelta)}\zeta^{\langle D\textbf{j}|C\textbf{j}\rangle}\chi^{D\textbf{j}}(C\textbf{j}+\textbf{i}_0)^{-1}\theta_{C\textbf{j}+\textbf{i}_0}^{\mL} \nonumber \\
&=\sum_{\textbf{j}\in Z_1(\tbdelta)}\zeta^{\langle D\textbf{j}|C\textbf{j}\rangle}\zeta^{-2\langle C\textbf{j}+\textbf{i}_0|D\textbf{j}\rangle}\theta_{C\textbf{j}+\textbf{i}_0}^{\mL} \nonumber\\
&=\sum_{\textbf{j}\in Z_1(\tbdelta)}\zeta^{-\langle C\textbf{j}+2\textbf{i}_0|D\textbf{j}\rangle}\theta_{C\textbf{j}+\textbf{i}_0}^{\mL} \label{eq: concrete Ti0}
\end{align}

And, if $T_{\textbf{i}_0}\neq 0$, then by \cref{eq: action Kronecker} we have for all $\textbf{i}\in Z_1(\bdelta)$,
{\allowdisplaybreaks
\begin{align*}
{\theta'}_\textbf{i}^{\mL}&=\Theta'_{\mL}(1,\textbf{i},1)\cdot {\theta'}_0^{\mL}=\lambda \Theta_{\mL}\circ\psi'(1,\textbf{i},1)\cdot T_{\textbf{i}_0}\\
&=\lambda\sum_{\textbf{j}\in Z_1(\tbdelta)}\zeta^{-\langle C\textbf{j}+2\textbf{i}_0|D\textbf{j}\rangle}\Theta_{\mL}\circ\psi'(1,\textbf{i},1)\cdot\theta_{C\textbf{j}+\textbf{i}_0}^{\mL} \quad (\mbox{by~\cref{eq: concrete Ti0}})\\
&=\lambda\sum_{\textbf{j}\in Z_1(\tbdelta)}\zeta^{-\langle C\textbf{j}+2\textbf{i}_0|D\textbf{j}\rangle}\Theta_{\mL}(\zeta^{\langle B\textbf{i}|A\textbf{i}\rangle},A\textbf{i},\chi^{B\textbf{i}})\cdot\theta_{C\textbf{j}+\textbf{i}_0}^{\mL} \quad (\mbox{by~\cref{eq: concrete exp psi'}})\\
&=\lambda\sum_{\textbf{j}\in Z_1(\tbdelta)}\zeta^{-\langle C\textbf{j}+2\textbf{i}_0|D\textbf{j}\rangle}\zeta^{\langle B\textbf{i}|A\textbf{i}\rangle}\chi^{B\textbf{i}}(A\textbf{i}+C\textbf{j}+\textbf{i}_0)^{-1}\theta_{A\textbf{i}+C\textbf{j}+\textbf{i}_0}^{\mL}\quad (\mbox{by~\cref{eq: theta group action}})\\
&=\lambda\sum_{\textbf{j}\in Z_1(\tbdelta)}\zeta^{-\langle C\textbf{j}+2\textbf{i}_0|D\textbf{j}\rangle}\zeta^{\langle B\textbf{i}|A\textbf{i}\rangle}\zeta^{-2\langle B\textbf{i}|A\textbf{i}+C\textbf{j}+\textbf{i}_0\rangle}\theta_{A\textbf{i}+C\textbf{j}+\textbf{i}_0}^{\mL}\\
&=\lambda\sum_{\textbf{j}\in Z_1(\tbdelta)}\zeta^{-\langle C\textbf{j}+2\textbf{i}_0|D\textbf{j}\rangle}\zeta^{-\langle B\textbf{i}|A\textbf{i}+2C\textbf{j}+2\textbf{i}_0\rangle}\theta_{A\textbf{i}+C\textbf{j}+\textbf{i}_0}^{\mL}\\
&=\lambda\sum_{\textbf{j}\in Z_1(\tbdelta)}\zeta^{-\langle C\textbf{j}+2\textbf{i}_0|B\textbf{i}+D\textbf{j}\rangle-\langle B\textbf{i}|A\textbf{i}+C\textbf{j}\rangle}\theta_{A\textbf{i}+C\textbf{j}+\textbf{i}_0}^{\mL}\\
&=\lambda\sum_{\textbf{j}\in Z_1(\tbdelta)}\zeta^{-\langle A\textbf{i}+C\textbf{j}+2\textbf{i}_0|B\textbf{i}+D\textbf{j}\rangle+\langle A\textbf{i}|B\textbf{i}+D\textbf{j}\rangle-\langle B\textbf{i}|A\textbf{i}+C\textbf{j}\rangle}\theta_{A\textbf{i}+C\textbf{j}+\textbf{i}_0}^{\mL}\\
&=\lambda\sum_{\textbf{j}\in Z_1(\tbdelta)}\zeta^{-\langle A\textbf{i}+C\textbf{j}+2\textbf{i}_0|B\textbf{i}+D\textbf{j}\rangle+\langle A\textbf{i}|D\textbf{j}\rangle-\langle B\textbf{i}|C\textbf{j}\rangle}\theta_{A\textbf{i}+C\textbf{j}+\textbf{i}_0}^{\mL}\\
&=\lambda\sum_{\textbf{j}\in Z_1(\tbdelta)}\zeta^{-\langle A\textbf{i}+C\textbf{j}+2\textbf{i}_0|B\textbf{i}+D\textbf{j}\rangle+\,^t \textbf{i}(\,^tA\Delta D-\,^tB\Delta C)\textbf{j}}\theta_{A\textbf{i}+C\textbf{j}+\textbf{i}_0}^{\mL}\\
&=\lambda\sum_{\textbf{j}\in Z_1(\tbdelta)}\zeta^{-\langle A\textbf{i}+C\textbf{j}+2\textbf{i}_0|B\textbf{i}+D\textbf{j}\rangle+\,^t \textbf{i}\Delta \textbf{j}}\theta_{A\textbf{i}+C\textbf{j}+\textbf{i}_0}^{\mL} \quad (\mbox{by \cref{lemma: symplectic matrix}})\\
&=\lambda\sum_{\textbf{j}\in Z_1(\tbdelta)}\zeta^{-\langle A\textbf{i}+C\textbf{j}+2\textbf{i}_0|B\textbf{i}+D\textbf{j}\rangle+\langle \textbf{i}|\textbf{j}\rangle}\theta_{A\textbf{i}+C\textbf{j}+\textbf{i}_0}^{\mL}
\end{align*}}
This completes the proof.
\end{proof}

\section{An algorithm to compute $2^e$-isogenies with level 2 theta coordinates}\label{sec: main section}

In this section, we explain how to compute a $2^e$-isogeny $F:\mA\longrightarrow\mB$ between principally polarised abelian varieties (PPAV) of any dimension $g$ with the Theta model of level $2$. As in dimension 1, we compute $F$ as a chain of $2$-isogenies. In \cref{sec: algo overview}, we give an overview of this isogeny chain computation given an isotropic subgroup $K''\subset\mA[2^{e+2}]$ such that $\ker(F)=[4]K''$. In \cref{sec: generic 2-isogeny}, we give a generic algorithm to compute a $2$-isogeny in the Theta model of level $2$ and we adapt this algorithm in \cref{sec: gluing} to the case of \emph{gluing isogenies}, namely isogenies defined over products of abelian varieties (\emph{eg.} elliptic products, as in SQIsignHD) whose codomain is not isomorphic to a product of abelian varieties. Finally, we explain how to compute the dual of a $2$-isogeny in \cref{sec: dual isogenies}.

\subsection{Algorithmic overview of a $2^e$-isogeny computation}\label{sec: algo overview}

Let $F:\mA\longrightarrow\mB$ be a $2^e$-isogeny between PPAV with kernel $K\subset \mA[2^e]$ of rank $g$. This means that $K$ is a free $(\Z/2^e\Z)$-module of rank $g$. Then, we can decompose $F$ as a chain of $2$-isogenies $F:=f_e\circ \cdots\circ f_1$, with $\ker(f_i)=[2^{e-i}]f_{i-1}\circ\cdots\circ f_1(K)$ for all $i\in\Int{1}{e}$ \cite[Lemma 51]{SQISignHD_autocite}. 

In the following, we say that we \emph{compute} $F$ when we compute all the $2$-isogenies $f_i$ which form a \emph{chain representation} of $F$. As we shall see in \cref{sec: generic 2-isogeny}, we can easily evaluate $f_i$ once we know the dual theta null point of its codomain (Algorithm~\ref{alg: isogeny eval}). Hence, we shall \emph{represent} the $f_i$ with this data: \emph{computing} $f_i$ means computing the dual theta null point of its codomain. In addition to a list of dual theta null points (representing the $f_i$), the representation of $F$ may include relevant change of theta coordinates matrices of theta coordinates to ensure we can apply $f_1$ and compose the $f_i$ in compatible theta coordinates. This way, once we have computed (a chain representation of) $F$, we can evaluate it easily on points.  

To compute the $2$-isogeny $f_i: A_i\longrightarrow A_{i+1}$, one needs to know an isotropic subgroup $K''_i\subset A_i[8]$ such that $[4]K''_i=\ker(f_i)$. Otherwise, we would have to compute square roots, and guess signs which is more costly (see \cref{rem: sqrt codomain}). To simplify the isogeny theorem formula (see \cref{thm: isogeny theorem} and \cref{lemma: f change of level}), we also need to make sure that the level $2$ theta structure $\Theta_{\mL_i}$ on $(A_i,\mL_i)$ satisfies $K_2(\Theta_{\mL_i})=\ker(f_i)$. 

To ensure these conditions for all $i\in\Int{1}{e}$, we proceed as follows. We assume that we know an isotropic subgroup $K''\subset\mA[2^{e+2}]$ such that $K=[4]K''$. To compute $f_1$, we first make sure the level $2$ theta structure $ \Theta_{\mL_1}$ on $(\mA,\mL_1)$ satisfies $K_2(\Theta_{\mL_1})=[2^{e+1}]K''$ (where, for instance, $\mL_1=\mL_0^2$ and $\mL_0$ is the principal polarization). Consider a totally symmetric level $2$ theta structure $\Theta_{\mL_1}$ on $\mA$ that is naturally given to us (\emph{eg.} the product theta structure on a product of elliptic curves). By \cref{thm: compatible symmetric theta structures}.(ii), $\Theta_{\mL_1}$ is completely determined by a symplectic basis $\mathscr{B}$ of $\mA[4]$. We compute a symplectic matrix $M\in\Sp_{2g}(\Z/4\Z)$ mapping $\mathscr{B}$ to a symplectic basis $\mathscr{C}:=(S'_1,\cdots,S'_g,T'_1,\cdots, T'_g)$ of $\mA[4]$ such that $(T'_1,\cdots, T'_g)$ is a basis of $[2^e]K''$. Let $\Theta'_{\mL_1}$ be the theta structure induced by the action of $M$ on $\Theta_{\mL_1}$. By \cref{thm: symplectic change of basis}, we know how to compute the new theta coordinates associated to $\Theta'_{\mL_1}$ and we now have $K_2(\Theta'_{\mL_1})=[2^{e+1}]K''$ by construction. We can then compute $f_1$ with the algorithms of \cref{sec: generic 2-isogeny,sec: gluing}. 

By \cite[Theorem 56]{SQISignHD_autocite}, $([2]f_1(S'_1), \cdots, [2]f_1(S'_g), f_1(T'_1), \cdots, f_1(T'_g))$ is a symplectic basis of $A_2[2]$, so we easily obtain a level $2$ theta structure on $(A_2,\mL_2,\Theta_{\mL_2})$ such that $K_2(\Theta_{\mL_2})=\ker(f_2)=\langle f_1(T'_1), \cdots, f_1(T'_g)\rangle$ (see \cref{sec: generic 2-isogeny}). Hence, once we have evaluated $[2^{e-2}]f_1(K'')$, we can compute $f_2$ without computing a change of theta coordinates. The same applies for all $i\geq 2$: we can compute $f_i$ given $K''_i:=[2^{e-i}]f_{i-1}\circ\cdots\circ f_1(K'')$, without computing a change of theta coordinates. 

As in dimension 1, the $K''_i$ are the leaves of a computation tree whose nodes are basis of rank $g$ isotropic subgroups (with $K''$ as root node), left edges are doublings and right edges are evaluations by the $f_i$ (see \cref{fig: computation tree}). Evaluating this tree can be done in quasi-linear time $O(e\log(e))$ with optimal strategies depending on the relative cost of evaluation and doublings. We refer \cref{sec: optimal strategies} for a detailed explanation on how these strategies are obtained and used to compute $2^e$-isogenies (see Algorithm~\ref{alg: apply strategy} in particular).




\subsection{Generic algorithm to compute a $2$-isogeny in dimension $g$}\label{sec: generic 2-isogeny}

Let $(A,\mL_0)$ and $(B,\mM_0)$ be two PPAVs and $\mL:=\mL_0^2$ and $\mM:=\mM_0^2$. Both $\mL$ and $\mM$ are totally symmetric line bundles of level $2$. Let $f: A\longrightarrow B$ be a $2$-isogeny with respect to the principal polarizations $\varphi_{\mL_0}$ and $\varphi_{\mM_0}$. Then $f$ is also an isogeny of polarised abelian varieties with respect to $\varphi_{\mL^2}$ and $\varphi_{\mM}$. 

Let $K:=\ker(f)$. In the following, we assume that $K=K_2(\Theta_\mL)$. Let $(T_1, \cdots, T_g)$ be a basis of $K$ and $\mathscr{B}'':=(S''_1,\cdots, S''_g,T''_1,\cdots, T''_g)$ be a symplectic basis of $K(\mL^4)=A[8]$ with respect to an $8$-th root of unity $\zeta_8\in k^*$ such that $[4]T''_l=T_l$ for all $l\in\Int{1}{g}$. Then $\mathscr{B}''$ induces a symmetric theta structure $\Theta_{\mL^2}$ on $G(\mL^2)$ by \cref{thm: compatible symmetric theta structures}. In addition, $\mathscr{C}:=([2]f(S''_1), \cdots, [2]f(S''_g), f(T''_1), \cdots, f(T''_g))$ is a $\zeta_8^2$-symplectic basis of $B[4]$ which induces a theta structure $\Theta_\mM$ on $G(\mM)$ which is compatible with $\Theta_{\mL^2}$ by \cite[Theorem 56]{SQISignHD_autocite}. $\Theta_\mM$ is also symmetric by \cite[Remark 4.2.15]{DRphd}. 

\begin{Lemma}\label{lemma: f change of level}
With this choice of theta structure $\Theta_\mM$, we have for all $\textbf{i}\in (\Z/2\Z)^g$,
\[f^*\theta_\textbf{i}^{\mM}=\theta_{2\textbf{i}}^{\mL^2}.\]
\end{Lemma}

\begin{proof}
Recall the notations from \cref{sec: isogeny theorem}: let $K^\bot$ be the orthogonal of $K$ in $K(\mL^2)=[2]^{-1}K(\mL)$ (for $e_{\mL^2}$) and consider the decompositions $K:=K_1\oplus K_2$ and $K^\bot:=K^{\bot,1}\oplus K^{\bot,2}$ induced by the decomposition $K(\mL^2)=K_1(\Theta_{\mL^2})\oplus K_2(\Theta_{\mL^2})$. Then $K^\bot=[2]K_1(\Theta_{\mL^2})\oplus K_2(\Theta_{\mL^2})=K_1(\Theta_\mL)\oplus K_2(\Theta_{\mL^2})$, so that $K^{\bot,1}=K_1(\Theta_\mL)$ and $K_1=\{0\}$. Using the notations from \cref{thm: isogeny theorem} (applied to $\Theta_{\mL^2}$ and $\Theta_{\mM}$), we get that $\Theta_\mM$ is determined by an isomorphism $\sigma: K_1(\Theta_\mL)\overset{\sim}{\longrightarrow} (\Z/2\Z)^g$. By the definition of $\Theta_\mM$ with respect to $\Theta_{\mL^2}$ and $\sigma$ \cite[Theorem 4, p. 302]{Mumford66}, we get that:
\[\forall \textbf{i}\in (\Z/2\Z)^g, \quad \overline{\Theta}_\mM(\textbf{i},1)=f(\sigma^{-1}(\textbf{i})).\]
Hence, with our choice of theta structure $\Theta_\mM$, we get that for all $l\in\Int{1}{g}$, $\sigma([4]S''_l)=\textbf{e}_l$, where $\textbf{e}_l\in (\Z/2\Z)^g$ is the vector with component $1$ at index $l$ and $0$ elsewhere. Furthermore, for all $l\in\Int{1}{g}$, $\overline{\Theta}_{\mL^2}^{-1}([2]S''_l)=\textbf{e}_l\in (\Z/4\Z)^g$ so $\overline{\Theta}_{\mL^2}^{-1}\circ\sigma^{-1}$ is the map $\textbf{i}\in (\Z/2\Z)^g\longmapsto 2\textbf{i}\in (\Z/4\Z)^g$. The result follows.
\end{proof}

By the above lemma, to compute the $\theta_{\textbf{i}}^{\mM}(f(x))$, we have to compute the $\theta_{2\textbf{i}}^{\mL^2}(x)$ knowing the $\theta_{\textbf{i}}^{\mL}(x)$. We may use the duplication formulas introduced in \cref{sec: duplication formulas} for that. 

\begin{Notation} We introduce two operators on $k^{(\Z/2\Z)^g}$:
\begin{itemize}
\item the \emph{Hadamard} operator $H: (x_\textbf{i})_{\textbf{i}\in (\Z/2\Z)^g}\longmapsto \left(\sum_{\textbf{i}\in (\Z/2\Z)^g}(-1)^{\langle \textbf{i}| \textbf{j}\rangle}x_\textbf{i}\right)_{\textbf{j}\in (\Z/2\Z)^g}$;
\item the \emph{squaring} operator $S: (x_\textbf{i})_{\textbf{i}\in (\Z/2\Z)^g}\longmapsto (x_\textbf{i}^2)_{\textbf{i}\in (\Z/2\Z)^g}$.
\end{itemize}
We also denote by $\star$ the component-wise product on $k^{(\Z/2\Z)^g}$. Note that $H\circ H=2^g\id$, so $H$ is an involution up to a projective factor.

For all $l\in\Int{1}{g}$, we denote by $\chi_l\in \widehat{(\Z/2\Z)^g}$ the character $\textbf{i}\longmapsto (-1)^{i_l}$. The $\chi_l$ form a basis of $\widehat{(\Z/2\Z)^g}$ since every character $\chi\in \widehat{(\Z/2\Z)^g}$ can be written as $\chi=\chi^\textbf{j}$ for a unique $\textbf{j}\in (\Z/2\Z)^g$, where $\chi^\textbf{j}:=\prod_{l=1}^g\chi_l^{j_l}: \textbf{i}\longmapsto (-1)^{\langle \textbf{i}|\textbf{j} \rangle}$. Using this canonical isomorphism between $(\Z/2\Z)^g$ and $\widehat{(\Z/2\Z)^g}$, we can index dual theta points by characters. In particular, we may write $(U_{\chi, 0}^{\mM}(x))_{\chi}=H((\theta_\textbf{i}^{\mM}(x))_\textbf{i})$. To lighten the notations, we shall also drop the $0$ index and denote $U_{\chi, 0}^{\mM}(x)$ as $U_{\chi}^{\mM}(x)$. In the following, we shall use both notations $(U_{\chi}^{\mM}(x))_{\chi}$ and $H((\theta_\textbf{i}^{\mM}(x))_\textbf{i})$ and switch between them, depending on context.
\end{Notation}

\begin{Proposition}\cite[Equation (3)]{Theta_dim2}\label{prop: isogeny eval}
Let $x\in A(k)$ and $(\theta_\textbf{i}^\mM(0_B))_{\textbf{i}\in (\Z/2\Z)^g}$ be the theta null point of $(B,\mM,\Theta_\mM)$. Then, we have (up to a projective constant):
\begin{equation}
H((\theta_\textbf{i}^{\mM}(f(x)))_\textbf{i})\star H((\theta_\textbf{i}^{\mM}(0_B))_\textbf{i})=H\circ S((\theta_\textbf{i}^\mL(x))_\textbf{i}).\label{eq: isogeny eval}
\end{equation}
\end{Proposition} 

\begin{proof}
\cref{eq: duplication 3} ensures that (up to a projective constant), we have for all $\chi\in \widehat{(\Z/2\Z)^g}$,
\[U_{\chi}^{\mL^2}(x)U_{\chi}^{\mL^2}(0_A)=\sum_{\textbf{t}\in (\Z/2\Z)^g}\chi(\textbf{t})\theta_\textbf{t}^{\mL}(x)^2,\]
so that $(U_{\chi}^{\mL^2}(x))_{\chi}\star (U_{\chi}^{\mL^2}(0_A))_{\chi}=H\circ S((\theta_\textbf{i}^\mL(x))_\textbf{i})$. Furthermore, \cref{eq: dual theta coordinates} and \cref{lemma: f change of level}, ensure that $(U_{\chi}^{\mL^2}(x))_{\chi}=H((\theta_{2\textbf{i}}^{\mL^2}(x))_\textbf{i})=H((\theta_\textbf{i}^{\mM}(f(x))_\textbf{i})$, up to a projective constant. Finally, we also have 
\[(U_{\chi}^{\mL^2}(0_A))_{\chi}=H((\theta_\textbf{i}^{\mM}(f(0_A))_\textbf{i})=H((\theta_\textbf{i}^{\mM}(0_B))_\textbf{i}).\] 
The result follows.
\end{proof}

Using \cref{eq: isogeny eval}, we easily obtain an algorithm to evaluate $f$ when the codomain theta null point (or its dual) is known. This is a simple generalization of \cite[Algorithm 6]{Theta_dim2}. This algorithm only works when the dual theta constants of $B$ do not vanish. We treat the vanishing case in the next section.

\begin{algorithm}
\SetAlgoLined
\KwData{\justifying A theta point $(\theta_\textbf{i}^\mL(x))_{\textbf{i}}$ of $A$ and the dual theta null point $(U_\chi^\mM(0_B))_\chi$ of $B$ with non-vanishing coordinates.}
\KwResult{\justifying $(\theta_\textbf{i}^\mL(f(x)))_{\textbf{i}}$.}
\justifying Precompute $C_\chi\longleftarrow 1/U_\chi^\mM$ for all $\chi\in \widehat{(\Z/2\Z)^g}$\;
Compute $(Z_\chi)_\chi\longleftarrow H\circ S((\theta_\textbf{i}^\mL(x))_\textbf{i})$\;
Compute $(Y_\chi)_\chi\longleftarrow (C_\chi\cdot Z_\chi)_\chi$\;
\KwRet $H((Y_\chi)_{\chi})$\;
\caption{Generic isogeny evaluation algorithm.}\label{alg: isogeny eval}
\end{algorithm}

We now explain how to compute the dual theta null point of $B$. Let $\m{B}:=(S'_1,\cdots, S'_g, T'_1, \cdots, T'_g)$ be a symplectic basis of $K(\mL^2)$ adapted to the decomposition $K(\mL^2)=K_1(\Theta_{\mL^2})\oplus K_2(\Theta_{\mL^2})$. Let $\zeta_4\in k$ such that $\zeta_4^2=-1$ and $e_4(S'_l,T'_m)=\zeta_4^{\delta_{l,m}}$ for all $l,m\in\Int{1}{g}$. Then, $([2]T'_1,\cdots, [2]T'_g)$ is a basis of $[2]K_2(\Theta_{\mL^2})=K_2(\Theta_\mL)=K$. In addition, $\m{B}$ determines a symplectic isomorphism $\overline{\Theta}_{\mL^2}: Z(\underline{4})\overset{\sim}{\longrightarrow} K(\mL^2)$ mapping the $\zeta_4$-canonical symplectic basis of $Z(\underline{4})$ (as defined in \cref{subsec: change of basis}) to $\m{B}$. By \cref{thm: compatible symmetric theta structures}.(ii), $\overline{\Theta}_{\mL^2}$ determines the symmetric theta structure $\Theta_\mL$ on $G(\mL)$. Via this isomorphism $\overline{\Theta}_{\mL^2}$, for all $l\in\Int{1}{g}$, the character $\chi'_l: \textbf{j}\longmapsto \zeta_4^{j_l}$ corresponds to $T'_l$, so the character $\chi_l={\chi'_l}^2: \textbf{j}\longmapsto (-1)^{j_l}$ corresponds to $T_l:=[2]T'_l$.

\begin{Lemma}\label{lemma: computing theta null point}
Let $T''_l$ such that $[2]T''_l=T'_l$ for all $l\in\Int{1}{g}$. Then for all $l\in\Int{1}{g}$ and $\chi\in \widehat{(\Z/2\Z)^g}$,
\begin{equation}
U_{\chi\chi_l}^{\mM}(0_B)\cdot H\circ S((\theta_\textbf{i}^{\mL}(T''_l))_\textbf{i})_\chi=U_{\chi}^{\mM}(0_B)\cdot H\circ S((\theta_\textbf{i}^{\mL}(T''_l))_\textbf{i})_{\chi\chi_l}.\label{eq: computing theta null point}
\end{equation}
\end{Lemma}

\begin{proof}
Let $l\in\Int{1}{g}$ and $\chi\in \widehat{(\Z/2\Z)^g}$. Then, by \cref{eq: isogeny eval}, we get that 
\begin{equation}
U_{\chi}^{\mM}(0_B)\cdot U_{\chi}^{\mM}(f(T''_l))=H\circ S((\theta_\textbf{i}^\mL(T''_l))_\textbf{i})_\chi.\label{eq: computing theta null point 2}
\end{equation}
Since $[4]T''_l=T_l\in K$, $f(T''_l)$ has order $4$ so $f(T''_l)\equiv f(T''_l)+[2]f(T''_l)$ in the Kummer variety $A/\pm$ and $\theta_\textbf{i}^{\mM}(f(T''_l))=\theta_\textbf{i}^{\mM}(f(T''_l)+[2]f(T''_l))$ for all $\textbf{i}\in (\Z/2\Z)^g$. In addition, $\Theta_{\mL^2}$ and $\Theta_\mM$ are compatible so $[2]f(T''_l)\in K_2(\Theta_\mM)$ since $[2]T''_l=T'_l\in K_2(\Theta_{\mL^2})$. Assuming we have made the canonical choice of theta structure $\Theta_\mM$, the symplectic isomorphism $\overline{\Theta}_\mM: Z(\underline{2})\overset{\sim}{\longrightarrow} K(\mM)$ maps $\chi_l$ to $[2]f(T''_l)$ \cite[Theorem 56]{SQISignHD_autocite}. Hence, \cref{eq: action translate} ensures that:
\[\forall \textbf{i}\in (\Z/2\Z)^g, \quad \theta_\textbf{i}^{\mM}(f(T''_l))=\theta_\textbf{i}^{\mM}(f(T''_l)+[2]f(T''_l))=\chi_l(\textbf{i})^{-1}\theta_{\textbf{i}}^{\mM}(f(T''_l)),\]
so that,
\begin{align*}U_{\chi}^{\mM}(f(T''_l))&=\sum_{\textbf{t}\in (\Z/2\Z)^g}\chi(\textbf{t})\theta_\textbf{t}^{\mM}(f(T''_l))=\sum_{\textbf{t}\in (\Z/2\Z)^g}\chi(\textbf{t})\chi_l(\textbf{t})^{-1}\theta_{\textbf{t}}^{\mM}(f(T''_l))\\
&=U_{\chi\chi_l^{-1}}^{\mM}(f(T''_l))=U_{\chi\chi_l}^{\mM}(f(T''_l)),
\end{align*}
since $\chi_l^{-1}=\chi_l$. Combining this with \cref{eq: computing theta null point 2}, we finally obtain \cref{eq: computing theta null point}.
\end{proof}

Given a basis of $8$-torsion points $(T''_1,\cdots, T''_g)$ lying above the basis of $K$, as in \cref{lemma: computing theta null point}, we can compute the dual theta null point $(U_{\chi}^{\mM}(0_B))_\chi$ of $B$ with \cref{eq: computing theta null point}. First, we select $\chi^0\in \widehat{(\Z/2\Z)^g}$ and $l\in\Int{1}{g}$ such that $H\circ S((\theta_\textbf{i}^{\mL}(T''_l))_\textbf{i})_{\chi^0}\neq 0$ (so that $U_{\chi^0}^{\mM}(0_B)\neq 0$ by \cref{eq: computing theta null point 2}), and then we compute $U_{\chi^0\chi_l}^{\mM}(0_B)/U_{\chi^0}^{\mM}(0_B)=H\circ S((\theta_\textbf{i}^{\mL}(T''_l))_\textbf{i})_{\chi^0\chi_l}/H\circ S((\theta_\textbf{i}^{\mL}(T''_l))_\textbf{i})_{\chi^0}$. Then, taking $\chi:=\chi^0\chi_l$, we find $l'\in\Int{1}{g}$ such that $H\circ S((\theta_\textbf{i}^{\mL}(T''_{l'}))_\textbf{i})_{\chi}\neq 0$ and obtain $U_{\chi\chi_{l'}}^{\mM}(0_B)/U_{\chi}^{\mM}(0_B)$. Multiplying this by the previous quotient, we can obtain $U_{\chi\chi_{l'}}^{\mM}(0_B)/U_{\chi^0}^{\mM}(0_B)$.  We repeat the same procedure until we have covered all indices in $\chi\in \widehat{(\Z/2\Z)^g}$, so we are finally able to compute $(U_{\chi}^{\mM}(0_B)/U_{\chi^0}^{\mM}(0_B))_{\chi}$.

To perform this computation, we fill in a computation tree whose nodes are characters of $\widehat{(\Z/2\Z)^g}$ related to each other by multiplication by a $\chi_l$ for $l\in\Int{1}{g}$. Each edge between a parent $\chi$ and a child $\chi\chi_l$ stores the value $U_{\chi\chi_l}^{\mM}(0_B)/U_{\chi}^{\mM}(0_B)$. The tree filling algorithm is summarized in Algorithm~\ref{alg: tree filling}. The full algorithm computing the codomain dual theta null point is Algorithm~\ref{alg: dual theta null point}, generalizing \cite[Algorithm~5]{Theta_dim2}.

\begin{algorithm}
\SetAlgoLined
\KwData{\justifying Theta coordinates $\theta_\textbf{i}^\mL$ of $8$-torsion points $T''_1,\cdots, T''_g$ such that $K=\langle [4]T''_1,\cdots, [4]T''_g\rangle$.}
\KwResult{\justifying Full computation tree $\m{T}$ as described above.}
Initialize the computation tree $\m{T}\longleftarrow \emptyset$\;
\While{$\m{T}$ does not cover $\widehat{(\Z/2\Z)^g}$}{
Select $\chi^0\in \widehat{(\Z/2\Z)^g}$ and initialize $\m{T}$ at root $\chi^0$\;
\While{all terminal nodes of $\m{T}$ are not not marked as leaves}{
\For{every terminal node $\chi$ of $\m{T}$ not marked as a leaf}{
$leaf\longleftarrow$ True\;
\For{$l=1$ \KwTo $g$}{
\If{$\chi\chi_l\not\in\m{T}$ and $H\circ S((\theta_\textbf{i}^{\mL}(T''_l))_\textbf{i})_{\chi\chi_l}\neq 0$}{
$E(\chi,\chi\chi_l)\longleftarrow H\circ S((\theta_\textbf{i}^{\mL}(T''_l))_\textbf{i})_{\chi\chi_l}/H\circ S((\theta_\textbf{i}^{\mL}(T''_l))_\textbf{i})_{\chi}$\;
Add $\chi\chi_l$ as the child of $\chi$ in $\m{T}$ and store $E(\chi,\chi\chi_l)$ on the edge\;
$leaf \longleftarrow$ False\;
}
}
\If{$leaf$}{
Mark $\chi$ as a leaf\;
}
}
}
}
\KwRet $\m{T}$\;
\caption{Tree filling algorithm for the codomain dual theta null point computation.}\label{alg: tree filling}
\end{algorithm}

\begin{Remark}
Algorithm \ref{alg: dual theta null point} does not always terminate when some theta constants $U_{\chi}^{\mM}(0_B)$ vanish. This can happen when we compute a gluing isogeny (see \cref{sec: gluing}). In this case, we may need more than $g$ points to fill in the tree. This is not a problem because \cref{eq: computing theta null point} generalizes to sums of $T''_l$ and products of~$\chi_l$.
\end{Remark}

\begin{algorithm}
\SetAlgoLined
\KwData{\justifying A computation tree $\m{T}$ as above and a root value $u$.}
\KwResult{\justifying $(U_{\chi}^{\mM}(0_B))_{\chi\in \m{T}}$.}

Let $\chi^0$ be the root of $\m{T}$. Set $U_{\chi^0}^\mM(0_B)\longleftarrow u$\;
\eIf{$\m{T}=\{\chi^0\}$}{
\KwRet $u$\;
}{
\For{every child $\chi$ of $\chi^0$}{
Let $U_{\chi}^{\mM}(0_B)\longleftarrow E(\chi^0,\chi)\cdot u$\;
Recurse on subtree $\m{T}_\chi$ with root $\chi$ and root value $U_{\chi}^{\mM}(0_B)$\;
}
\KwRet all the computed values $(U_{\chi}^{\mM}(0_B))_{\chi\in \m{T}}$\;
}

\caption{Tree evaluation recursive algorithm.}\label{alg: tree evaluation}
\end{algorithm}

\begin{algorithm}
\SetAlgoLined
\KwData{\justifying Theta coordinates $\theta_\textbf{i}^\mL$ of $8$-torsion points $T''_1,\cdots, T''_g$ such that $K=\langle [4]T''_1,\cdots, [4]T''_g\rangle$.}
\KwResult{\justifying $(U_{\chi}^{\mM}(0_B))_{\chi\in \widehat{(\Z/2\Z)^g}}$.}

Call Algorithm \ref{alg: tree filling} to get a computation tree $\m{T}$\;
Call Algorithm \ref{alg: tree evaluation} on $\m{T}$ with root value $U_{\chi^0}^{\mM}(0_B)=1$ to compute $(U_{\chi}^{\mM}(0_B))_{\chi\in \widehat{(\Z/2\Z)^g}}$\;
\KwRet $(U_{\chi}^{\mM}(0_B))_{\chi\in \widehat{(\Z/2\Z)^g}}$\;

\caption{Codomain dual theta null point computation algorithm.}\label{alg: dual theta null point}
\end{algorithm}

\begin{Remark}[Codomain computation without $8$-torsion points]\label{rem: sqrt codomain}
When the $8$-torsion points $T''_1, \cdots, T''_g$ are not known but only a $2$-torsion basis of the kernel is known, we can still compute the codomain of $f$. Using \cref{eq: isogeny eval}, we get that $S\circ H((\theta_\textbf{i}^{\mM}(0_B))_\textbf{i})=H\circ S((\theta_\textbf{i}^\mL(0_A))_\textbf{i})$, so we can compute the codomain theta null point $(\theta_\textbf{i}^{\mM}(0_B))_\textbf{i}$ with $2^g-1$ square root computations and choices of sings (projectively). This method is not only more costly but also not sufficient to determine $(\theta_\textbf{i}^{\mM}(0_B))_\textbf{i}$ in general because all sign choices may not be valid. 

As Robert did in \cite[Chapter 7, Appendix B.2]{Robert_note_theta}, we can prove that we can make at least $g(g+1)/2$ arbitrary sign choices among $2^g-1$. Indeed, we may act on a symplectic basis of $B[4]$ inducing the theta structure $\Theta_{\mM}$ via the symplectic matrix: 
\[M:=\left(\begin{array}{cc}I_g & 0\\ \beta & I_g\end{array}\right)\in\Sp_{2g}(\Z/4\Z),\]
which fixes $K_2(\mM^2)$. By \cref{thm: symplectic change of basis}, the new resulting theta coordinates are ${\theta'}_\textbf{i}^{\mM}=\zeta^{-\langle \textbf{i}|\beta\textbf{i}\rangle}\theta_\textbf{i}^{\mM}$ for all $\textbf{i}\in(\Z/2\Z)^g$ (up to a projective factor), where $\zeta^2=-1$. Since $M$ is symplectic, we have $\,^t \beta=\beta$ by \cref{lemma: symplectic matrix} so we have $g(g+1)/2$ values to choose. We have:
\[\forall \textbf{i}\in (\Z/2\Z)^g, \quad \langle \textbf{i}|\beta\textbf{i}\rangle=\sum_{l=1}^g i_l^2\beta_{l,l}+2\sum_{1\leq l<m\leq g}i_l i_m\beta_{l,m},\]
so the $\langle \textbf{i}|\beta\textbf{i}\rangle$ are determined by the $\beta_{l,l}$ and the $\beta_{l,m} \mod 2$. For all $l\in\Int{1}{g}$ and $m\in\Int{l+1}{g}$, we may fix $\beta_{l,l}\in\{0,2\}$ and $\beta_{l,m}\in\{0,1\}$ and obtain ${\theta'}_{\textbf{e}_l}^{\mM}=(-1)^{-\beta_{l,l}/2}\theta_{\textbf{e}_l}^{\mM}$, ${\theta'}_{\textbf{e}_l+\textbf{e}_m}^{\mM}=(-1)^{-(\beta_{l,l}+\beta_{m,m}+2\beta_{l,m})/2}\theta_{\textbf{e}_l+\textbf{e}_m}^{\mM}$ and ${\theta'}_\textbf{i}^{\mM}=\pm\theta_\textbf{i}^{\mM}$ for all $\textbf{i}\in(\Z/2\Z)^g$. This amounts to choosing $g(g+2)/2$ signs among $2^g-1$ and fixing the others.

In dimension $g=2$, it was already remarked in \cite[§ 4.2]{Theta_dim2} that all $g(g+1)/2=3=2^g-1$ arbitrary sign choices are valid. This is no longer true in dimension $g>2$. In dimension $g=3$, only $6$ among $7$ sign choices determine the last one with an explicit formula \cite{Kunz_et_al_2024}. In dimension $g\geq 4$, we have no such explicit formulas so the theta null point is harder to guess. 

\end{Remark}

\subsection{Gluing isogenies}\label{sec: gluing}

In the previous section, we have seen how to compute $2$-isogenies when none of the dual theta coordinates $(U_{\chi}^\mM(0_B))_{\chi}$ vanish. In this section, we treat the vanishing case which is frequent when we want to compute a \emph{gluing isogeny}, namely isogenies defined over a product of abelian varieties whose codomain is not isomorphic to a product of abelian varieties $f: A_1\times A_2\longrightarrow B$. There is a heuristic explanation of this phenomenon. The level 2 theta coordinates that we use represent points on the Kummer abelian variety, so up to a sign ambiguity. On a product of Kummer varieties $A_1/\pm\times A_2/\pm$ we have two sign ambiguities and on $B/\pm$, only one so we need additional information to evaluate $f$ and remove one sign ambiguity. This additional information will be provided by point translates.

Assuming we have already computed the dual theta constants $U_{\chi}^\mM(0_B)$ (which may vanish), we immediately see that Algorithm~\ref{alg: isogeny eval} may no longer be used (to avoid divisions by zero). However, in order to evaluate $x\in A(k)$, we may still use \cref{eq: isogeny eval} with translates of $x$. Let $T'_1, \cdots, T'_g$ be points such that $\langle [2]T'_1,\cdots, [2]T'_g\rangle= K$ as in the previous section. Then, for all $l\in\Int{1}{g}$ and $\chi\in \widehat{(\Z/2\Z)^g}$, we have:
\begin{equation} U_{\chi}^{\mM}(f(x+T'_l))\cdot U_{\chi}^{\mM}(0_B)=H\circ S((\theta_\textbf{i}^\mL(x+T'_l))_\textbf{i})_\chi.\label{eq: translate isogeny eval int}
\end{equation}
As we have seen in the proof of \cref{lemma: computing theta null point}, the theta structure $\overline{\Theta}_\mM$ maps $f(T'_l)$ to $\chi_l$ so there exists a projective constant $\lambda_l\in k^*$ such that $U_{\chi}^{\mM}(f(x+T'_l))= \lambda_l U_{\chi\chi_l}^{\mM}(f(x))$ for all $\chi\in \widehat{(\Z/2\Z)^g}$. It follows by \cref{eq: isogeny eval} and \cref{eq: translate isogeny eval int} that for all $l\in\Int{1}{g}$ and $\chi\in \widehat{(\Z/2\Z)^g}$, we have:
\begin{equation} \lambda_l\cdot H\circ S((\theta_\textbf{i}^\mL(x))_\textbf{i})_{\chi\chi_l}\cdot U_{\chi}^{\mM}(0_B)=H\circ S((\theta_\textbf{i}^\mL(x+T'_l))_\textbf{i})_\chi\cdot U_{\chi\chi_l}^{\mM}(0_B)\label{eq: compute lambda_i}, \end{equation}
so we can compute the $\lambda_l$ once we know the coordinates of $x$ and the $x+T'_l$ (and $0_B$). Since $\chi_l^2=1$, we also obtain by \cref{eq: isogeny eval} that for all $l\in\Int{1}{g}$, we have:
\begin{equation} \forall \chi\in \widehat{(\Z/2\Z)^g}, \quad \lambda_l\cdot U_{\chi}^{\mM}(f(x))\cdot U_{\chi\chi_l}^{\mM}(0_B)=H\circ S((\theta_\textbf{i}^\mL(x+T'_l))_\textbf{i})_{\chi\chi_l}.\label{eq: translate isogeny eval} \end{equation}
Hence, to compute $U_{\chi}^{\mM}(f(x))$, we may use \cref{eq: isogeny eval} when $U_{\chi}^{\mM}(0_B)\neq 0$ and otherwise, find $l\in\Int{1}{g}$ such that $U_{\chi\chi_l}^{\mM}(0_B)\neq 0$ and use \cref{eq: translate isogeny eval}. We summarize the evaluation procedure in Algorithm \ref{alg: gluing isogeny eval}.

\begin{Remark}\label{rem: additional points evaluation} In practice, we do not need to use all the translates $x+T'_l$ to compute the coordinates of $f(x)$. When $g=2$ and $f: E_1\times E_2\longrightarrow B$ is a gluing of elliptic curves, one point ($T'_1$) is sufficient \cite[Algorithm 8]{Theta_dim2}. When $g=4$ and $f: A_1\times A_2\longrightarrow B$ is a gluing of abelian surfaces, two points ($T'_1$ and $T'_2$) are sufficient.
\end{Remark}

\begin{Remark}\label{rem: sum of points}
Computing the $x+T'_l$ with the Theta model may not be easy since we also need to know the $x-T'_l$ to apply differential addition formulas (\cref{thm: duplication formulas}). This is not an issue when we work on elliptic curve products because we can use standard arithmetic. In practice, we compute chains of $2$-isogenies starting from an elliptic curve product so we can always perform the additions of preimages of the points on this elliptic curve product and then push the result through several $2$-isogenies.
\end{Remark}

\begin{algorithm}
\SetAlgoLined
\KwData{\justifying $4$-torsion points $T'_1,\cdots, T'_g$ such that $K=\langle [2]T'_1,\cdots, [4]T'_g\rangle$, a subset of indices $L\subseteq\Int{1}{g}$, theta points of $A$ $(\theta_\textbf{i}^\mL(x))_{\textbf{i}}$ and $(\theta_\textbf{i}^\mL(x+T'_l))_{\textbf{i}}$ for all $l\in L$ and the dual theta null point $(U_{\chi}^{\mM}(0_B))_\chi$ of $B$.}
\KwResult{\justifying $(\theta_\textbf{i}^\mM(f(x)))_{\textbf{i}}$.}
Precompute $C_\chi \longleftarrow 1/U_{\chi}^{\mM}(0_B)$ for all $\chi\in \widehat{(\Z/2\Z)^g}$ such that $U_{\chi}^{\mM}(0_B)\neq 0$\;
Compute $H\circ S((\theta_\textbf{i}^\mL(x))_\textbf{i})$ and $H\circ S((\theta_\textbf{i}^\mL(x+T'_l))_\textbf{i})$ for all $l\in L$\;
\For{$l\in L$}{
Find $\chi\in \widehat{(\Z/2\Z)^g}$ such that $H\circ S((\theta_\textbf{i}^\mL(x+T'_l))_\textbf{i})_\chi\cdot U_{\chi\chi_l}^{\mM}(0_B)\neq 0$\;
$\lambda_l^{-1}\longleftarrow H\circ S((\theta_\textbf{i}^\mL(x))_\textbf{i})_{\chi\chi_l}\cdot U_{\chi}^{\mM}(0_B)/(H\circ S((\theta_\textbf{i}^\mL(x+T'_l))_\textbf{i})_\chi\cdot U_{\chi\chi_l}^{\mM}(0_B))$\;
}
\For{$\chi\in \widehat{(\Z/2\Z)^g}$}{
\eIf{$U_{\chi}^{\mM}(0_B)\neq 0$}{
$U_{\chi}^{\mM}(f(x))\longleftarrow C_\chi\cdot H\circ S((\theta_\textbf{i}^\mL(x))_\textbf{i})_{\chi}$\;
}{
Find $l\in L$ such that $U_{\chi\chi_l}^{\mM}(0_B)\neq 0$\;
$U_{\chi}^{\mM}(f(x))\longleftarrow \lambda_l^{-1}C_{\chi\chi_l}\cdot H\circ S((\theta_\textbf{i}^\mL(x+T'_l))_\textbf{i})_{\chi\chi_l}$\;
}
}
\KwRet $H((U_{\chi}^{\mM}(f(x)))_\chi)$\;
\caption{Gluing isogeny evaluation algorithm.}\label{alg: gluing isogeny eval}
\end{algorithm}

The evaluation procedure of a gluing isogeny differs significantly from the generic one. However, the codomain theta null point computation is very similar. As in the previous section, let $T''_1, \cdots, T''_g$ be $8$-torsion points such that $T'_l=[2]T''_l$ for all $l\in\Int{1}{g}$ ($K=\langle [4]T''_1, \cdots, [4]T''_g\rangle$). For all multi-index $\textbf{j}\in (\Z/2\Z)^g$, we denote $T''_\textbf{j}:=\sum_{l=1}^g [j_l]T''_l$ and recall that $\chi^\textbf{j}:=\prod_{l=1}^g\chi_l^{j_l}$. Then, \cref{eq: computing theta null point} is still valid for multi-indices: for all $\textbf{j}\in (\Z/2\Z)^g$ and $\chi\in \widehat{(\Z/2\Z)^g}$,
\[U_{\chi\chi^\textbf{j}}^{\mM}(0_B)\cdot H\circ S((\theta_\textbf{i}^{\mL}(T''_\textbf{j}))_\textbf{i})_\chi=U_{\chi}^{\mM}(0_B)\cdot H\circ S((\theta_\textbf{i}^{\mL}(T''_\textbf{j}))_\textbf{i})_{\chi\chi^\textbf{j}}.\]
Using the above equation, we may obtain the dual theta constants $U_{\chi}^{\mM}(0_B)$ from the theta coordinates of (sums of) the $T''_1, \cdots, T''_g$ with tree filling algorithms as in the generic case (see Algorithms \ref{alg: tree filling}, \ref{alg: tree evaluation} and \ref{alg: dual theta null point}).

\begin{Remark}\label{rem: additional kernel points gluing}
For $g=2$, when we glue a product of elliptic curves, only two points $T''_1, T''_2$ (and no sum of points) are needed \cite[Algorithm 7]{Theta_dim2}. For $g=4$, in practice, when we glue two abelian surfaces only one point $T''_1+T''_2$, in addition to $T''_1, \cdots, T''_4$ is needed for the codomain computation. As mentioned in \cref{rem: sum of points}, when we compute a chain of $2$-isogenies starting from an elliptic product, sums of preimages of the $T''_1, \cdots, T''_g$ may be computed on the domain and then pushed through several $2$-isogenies.
\end{Remark}

\subsection{Computing dual isogenies}\label{sec: dual isogenies}

\sloppy
Once we have computed a $2$-isogeny $f: (A,\mL^2)\longrightarrow (B,\mM)$ as in \cref{sec: generic 2-isogeny,sec: gluing}, it is then easy to compute its dual $\widetilde{f}: B \longrightarrow A$ with the data we already have. By the following lemma, we only have to precompute the inverse theta constants $(1/\theta_\textbf{i}^{\mL}(0_A))_\textbf{i}$ to be able to evaluate $\widetilde{f}$. Up to Hadamard transforms, the formulas are similar to those of \cref{sec: generic 2-isogeny}.

\begin{Lemma}\label{lemma: dual isogeny computation}
Let $f: (A,\mL^2)\longrightarrow (B,\mM)$ be a $2$-isogeny. As in \cref{sec: generic 2-isogeny}, let $(\Theta_\mL, \Theta_{\mL^2})$ be a pair of symmetric theta structures for $(\mL,\mL^2)$ such that $\ker(f)=K_2(\Theta_{\mL})$ and $\Theta_\mM$ be a theta structure on $G(\mM)$ compatible with $\Theta_{\mL^2}$ with respect to $f$. Then: 
\begin{enumerate}[label=(\roman*)]
\item $\widetilde{f}$ is a polarised abelian variety $(B,\mM^2) \longrightarrow (A,\mL)$ of kernel $\ker(\widetilde{f})=K_1(\Theta_{\mM})$.
\item We have for all $y\in B(k)$,
\[(\theta_\textbf{i}^{\mL}(\widetilde{f}(y)))_\textbf{i}\star(\theta_\textbf{i}^{\mL}(0_A))_\textbf{i}=H\circ S((U_{\chi}^{\mM}(y))_\chi),\]
up to a projective constant.
\end{enumerate}
\end{Lemma} 

\begin{proof}
\textbf{(i)} Recall that $\mL=\mL_0^2$ and $\mM=\mM_0^2$ where $\varphi_{\mL_0}$ and $\varphi_{\mM_0}$ are principal polarizations. Since $f$ is a $2$-isogeny, we have $f\circ \widetilde{f}=[2]$, where $\widetilde{f}=\varphi_{\mL_0}^{-1}\circ\widehat{f}\circ\varphi_{\mM_0}$. Hence, $\widehat{\widetilde{f}}=\widehat{\varphi_{\mM_0}}\circ f\circ\widehat{\varphi_{\mL_0}^{-1}}$ but $\widehat{\varphi_{\mM_0}}=\varphi_{\mM_0}$ and $\widehat{\varphi_{\mL_0}^{-1}}=\varphi_{\mL_0}^{-1}$, so that $\widehat{\widetilde{f}}=\varphi_{\mM_0}\circ f\circ\varphi_{\mL_0}^{-1}$. It follows that:
\[[2]=f\circ \widetilde{f}=\varphi_{\mM_0}^{-1}\circ\widehat{\widetilde{f}}\circ\varphi_{\mL_0}\circ\widetilde{f},\]
and $\widehat{\widetilde{f}}\circ\varphi_{\mL_0}\circ\widetilde{f}=[2]\varphi_{\mM_0}$. Since $\mL=\mL_0^2$ and $\mM=\mM_0^2$, we have $\varphi_{\mL}=[2]\circ\varphi_{\mL_0}$ and $\varphi_{\mM^2}=[4]\circ\varphi_{\mL_0}$ by the theorem of the square \cite[Theorem 6.7]{Milne1986}. We conclude that $\widehat{\widetilde{f}}\circ\varphi_{\mL}\circ\widetilde{f}=\varphi_{\mM^2}$ so $\widetilde{f}$ is a polarised abelian variety $(A,\mL^2)\longrightarrow (B,\mM)$.

Furthermore, $f(A[2])\subseteq\ker(\widetilde{f})$ since $\widetilde{f}\circ f=[2]$ and $\widetilde{f}$ is separable since $\Char(k)$ is odd so $\#\ker(\widetilde{f})=\deg(\widetilde{f})=\deg(f)=2^g$. We also have $f(A[2])=f(K_1(\Theta_\mL))=K_1(\Theta_\mM)$ by construction and $\# K_1(\Theta_\mM)=2^g$ so the inclusion $f(A[2])\subseteq\ker(\widetilde{f})$ is an equality, which proves (i).

\textbf{(ii)} Consider the symplectic basis $\mathscr{B}''=(S''_1,\cdots, S''_g,T''_1,\cdots, T''_g)$ of $K(\mL^2)=A[8]$ introduced in \cref{sec: generic 2-isogeny} to define $\Theta_{\mL^2}$ and a symplectic basis $\mathscr{B}'''=(S'''_1,\cdots, S'''_g,T'''_1,\cdots, T'''_g)$ of $K(\mL^4)=A[16]$ such that $S''_l=[2]S'''_l$ and $T''_l=[2]T'''_l$ for all $l\in\Int{1}{g}$. By \cref{thm: compatible symmetric theta structures}.(ii), $\mathscr{B}'''$ induces a symmetric theta structure $\Theta_{\mL^4}$ on $G(\mL^4)$. In addition, by \cref{thm: compatible symmetric theta structures}.(i), $\Theta_{\mL^4}$ induces a symmetric theta structure $\Theta'_{\mL^2}$ on $G(\mL^2)$ and by \cref{thm: compatible symmetric theta structures}.(ii), $\Theta'_{\mL^2}$ is determined by $[2]\mathscr{B}'''=\mathscr{B}''$ so $\Theta'_{\mL^2}=\Theta_{\mL^2}$ and $\Theta_{\mL^4}$ is compatible with $\Theta_{\mL^2}$. Then, one can prove exactly as in \cite[Theorem 56]{SQISignHD_autocite}, that $\mathscr{C}'=(f(S''_1),\cdots, f(S''_g),f(T'''_1),\cdots, f(T'''_g))$ is a symplectic basis of $B[8]$ which induces a symmetric theta structure $\Theta_{\mM^2}$ on $G(\mM^2)$ which is compatible with $\Theta_{\mL^4}$ with respect to $f$. Since $\mathscr{C}=[2]\mathscr{C}'$ is the symplectic basis determining $\Theta_{\mM}$, we can conclude by \cref{thm: compatible symmetric theta structures} that $\Theta_{\mM^2}$ is compatible with $\Theta_{\mM}$.

Let $\zeta_{8}:=e_{16}(S''_1,T''_1)$ and $\psi\in\Aut^0_{k^*}(\m{H}(\underline{8}))$ such that $\overline{\psi}$ has matrix 
\[M_{\psi}:=\left(\begin{array}{cc}
0 & -I_g\\
I_g & 0
\end{array}\right)\in\Sp(Z(\underline{8})),\]
in the $\zeta_{8}$-canonical symplectic basis. Let $\psi'\in\Aut^0_{k^*}(\m{H}(\underline{4}))$ be the symmetric Heisenberg automorphism induced by $\psi$ (\cref{prop: compatible change of basis}.(i)). Let $\Theta'_{\mM^2}:=\Theta_{\mM^2}\circ\psi$, $\Theta'_{\mM}:=\Theta_{\mM}\circ\psi'$ and $\Theta'_{\mL}:=\Theta_{\mL}\circ\psi'$. Then, by \cref{thm: symplectic change of basis}, the $\Theta'_{\mM}$-coordinates are the dual of the $\Theta_{\mM}$-coordinates (obtained after a Hadamard transform) and similarly for the $\Theta'_{\mL}$ and $\Theta_{\mL}$-coordinates. Furthermore, $M_{\psi}\cdot\mathscr{C}'=(-f(T'''_1), \cdots, -f(T'''_g),f(S''_1),\cdots, f(S''_g))$ induces $\Theta'_{\mM^2}$ and $M_{\psi}\cdot\mathscr{B}=(-T'_1,\cdots, -T'_g,S'_1,\cdots, S'_g)$ induces $\Theta'_{\mL}$. Hence, \cite[Theorem 56]{SQISignHD_autocite} ensures that $\Theta'_{\mM^2}$ and $\Theta'_{\mL}$ are compatible with respect to $\widetilde{f}$. We also have $\ker(\widetilde{f})=K_1(\Theta_{\mM})=K_2(\Theta'_{\mM})$ by (i). We conclude by \cref{prop: isogeny eval} that for all $y\in B(k)$,
\[H(U_{\chi}^{\mL}(\widetilde{f}(y)))_\chi)\star H((U_{\chi}^{\mL}(0_A))_\chi)=H\circ S((U_{\chi}^{\mM}(y))_\chi),\]
up to a projective constant, $U_{\chi}^{\mL}$ and $U_{\chi}^{\mM}$ being the $\Theta'_{\mL}$ and $\Theta'_{\mM}$-coordinates respectively. Since $H((U_{\chi}^{\mL}(x))_\chi)=(\theta_\textbf{i}^{\mL}(x))_\textbf{i}$ for all $x\in A(k)$, this completes the proof.
\end{proof}

\subsection{Complexity}

In \cref{tab: operation counts}, we give some operation counts for $2$-isogeny computations. Unlike what we have assumed so far, in practice, the base field $k$ that we use is not algebraically closed. All operations take place in the field of definition of torsion points used to compute isogenies (that we also denote by~$k$). We denote by $\textbf{M}$, $\textbf{S}$, $\textbf{I}$ and $\textbf{a}$ the cost (in bit operations) of multiplication, squaring, inversion and addition/subtraction over $k$. In general, inversions are much more costly than multiplications so we compute them by batch. This enables to replace $n\textbf{I}$ by $3(n-1)\textbf{M}+\textbf{I}$ as explained in \cref{sec: batch inversion}. We can even work projectively and remove all inversions\footnote{This optimisation is not implemented in our dimension 4 code. Inversion free algorithms are provided in \cite[Chapter~6]{MyPhD}.}. While additions are much less costly than multiplications, Hadamard transforms require a lot of them, which can impact concrete performance. For that reason, we propose in \cref{sec: recursive Hadamard} a recursive method to compute Hadamard transforms which reduces their cost from $2^{2g}\textbf{a}$ to $g 2^g \textbf{a}$. 

\begin{table}
  \setlength{\tabcolsep}{3pt}
  \centering
  \caption{Cost of algorithms involved in $2$-isogeny computations. Here, $L\subseteq\Int{1}{g}$ is the subset given on entry of \cref{alg: gluing isogeny eval}. We assume $\# L=2$ for $g=4$ (\cref{rem: additional points evaluation}). Inversions in gray can be easily removed by working purely projectively. For the other algorithms, inversion free versions are introduced in \cite[Chapter~6]{MyPhD}.}
  \scriptsize{
\begin{tabular}{ccc|c}
\toprule
 & & & Total cost \\
 \midrule
\multirow{7}{*}{Dim. $g$} & \multirow{2}{*}{Doubling (Alg. \ref{alg: doubling})} & Precomp. & $\textcolor{gray}{\textbf{I}}+3(2^{g+1}-1)\textbf{M}+2^g\textbf{S}+g 2^{g+1}\textbf{a}$\\
 & & Main & $2^{g+1}\textbf{M}+2^{g+1}\textbf{S}+g 2^{g+1}\textbf{a}$\\
  \cmidrule{2-4}
  & \multicolumn{2}{c|}{Generic codomain (Alg. \ref{alg: dual theta null point})} & $\leq \textbf{I}+(2^{g+2}-5)\textbf{M}+g 4^g\textbf{S}+g^2 2^g \textbf{a}$\\
  \cmidrule{2-4}
  & \multirow{2}{*}{Generic eval. (Alg. \ref{alg: isogeny eval})} & Precomp. & $\textcolor{gray}{\textbf{I}}+3(2^{g}-1)\textbf{M}$\\
 & & Main & $2^g\textbf{M}+2^g\textbf{S}+g 2^{g+1}\textbf{a}$\\
 \cmidrule{2-4}
  & \multirow{2}{*}{Gluing eval. (Alg. \ref{alg: gluing isogeny eval})} & Precomp. & $\leq \textcolor{gray}{\textbf{I}}+3(2^{g}-1)\textbf{M}$\\
 & & Main & $\leq \textbf{I}+(2^{g+1}+5\# L-3)\textbf{M}+(\# L+1)2^g \textbf{S}+(\# L+2)g 2^g \textbf{a}$\\
 \midrule
 \multirow{7}{*}{Dim. $4$} & \multirow{2}{*}{Doubling (Alg. \ref{alg: doubling})} & Precomp. & $\textcolor{gray}{\textbf{I}}+93\textbf{M}+16\textbf{S}+128\textbf{a}$\\
 & & Main & $32\textbf{M}+32\textbf{S}+128\textbf{a}$ \\
  \cmidrule{2-4}
  & \multicolumn{2}{c|}{Generic codomain (Alg. \ref{alg: dual theta null point})} & $\leq \textbf{I}+59\textbf{M}+1024\textbf{S}+256\textbf{a}$ \\
  \cmidrule{2-4}
  & \multirow{2}{*}{Generic eval. (Alg. \ref{alg: isogeny eval})} & Precomp. &  $\textcolor{gray}{\textbf{I}}+45\textbf{M}$\\
 & & Main & $16\textbf{M}+16\textbf{S}+128\textbf{a}$\\
 \cmidrule{2-4}
  & \multirow{2}{*}{Gluing eval. (Alg. \ref{alg: gluing isogeny eval})} & Precomp. & $\leq \textcolor{gray}{\textbf{I}}+45\textbf{M}$\\
 & & Main & $\leq \textbf{I}+39\textbf{M}+48\textbf{S}+256\textbf{a}$\\
\bottomrule
  \end{tabular}
  }
  \label{tab: operation counts}
\end{table}

\section{Cryptographic applications}\label{sec: application to SQIsignHD}

In this section, we apply the algorithms presented in \cref{sec: main section} to compute a $4$ dimensional $2^e$-isogeny between elliptic curve products. The main applications we have in mind is the verification procedure in SQIsignHD \cite{SQISignHD_autocite} and SIDH torsion attacks \cite{RobSIDH} but this could also be applied to other cryptographic constructions \cite{Leroux_VRFHD}, or more generally, an improvement of the Deuring correspondence \cite[Remark 2.9]{Clapotis}. In \cite[Appendix F]{SQISignHD_autocite}, algorithms for the verification procedure were briefly presented but they differ significantly from the real implementation\footnote{This implementation can be found here \url{https://github.com/Pierrick-Dartois/Theta_dim4}.} relying on the ideas of \cref{sec: main section} and do not include several optimizations and implementation details presented here.

Recall that in SQIsignHD, we compute the $2^e$-isogeny given by Kani's lemma \cite{Kani1997} as follows:
\begin{equation}F:=\left(\begin{matrix}
\alpha_1 & \widetilde{\Sigma}\\
-\Sigma & \widetilde{\alpha}_2
\end{matrix}\right)\in\End(E_1^2\times E_2^2),\label{eq: definition F}
\end{equation}
where $\Sigma:=\Diag(\sigma,\sigma): E_1^2\longrightarrow E_2^2$ with $\sigma: E_1\longrightarrow E_2$ a $q$-isogeny and for $i\in\{1,2\}$,
\[\alpha_i:=\left(\begin{matrix}
a_1 & a_2\\
-a_2 & a_1
\end{matrix}\right)\in \End(E_i^2),\]
with $a_1, a_2\in\Z$ such that $a_1^2+a_2^2+q=2^e$. Input data $a_1, a_2$ is given along with a basis $(P_1, P_2)$ of $E_1[2^f]$ (where $f\geq e/2+2$) and its image $(\sigma(P_1),\sigma(P_2))$. We have to compute $F$ and evaluate it on some points. We shall keep those notations in the following.

\subsection{Gluing isogenies}

As explained previously, $F$ will be computed as a chain of $2$-isogenies. Our goal here is to determine where gluing isogenies appear in the chain in order to optimize our computations (because gluing isogenies are computed differently than generic ones and more expensive).

Since $q$ is odd, either one of the $a_1$ or $a_2$ must be even. Without loss of generality, we can assume that $a_2$ is even (so that $a_1$ is odd). Then, we have:

\begin{Lemma}\label{lemma: first gluing isogenies}
Assume that $2|a_2$ and let $m:=v_2(a_2)$ be its $2$-adic valuation. Then $F:=G\circ f_{m+1}\circ f_m\circ\cdots \circ f_1$, with
\[E_1^2\times E_2^2\overset{f_1}{\relbar\joinrel\relbar\joinrel\longrightarrow} A_1^2 \quad \cdots \quad A_{m-1}^2 \overset{f_m}{\relbar\joinrel\relbar\joinrel\longrightarrow} A_m^2\overset{f_{m+1}}{\relbar\joinrel\relbar\joinrel\longrightarrow}B,\]
a chain of $2$-isogenies, where the $A_i$ are abelian surfaces and $B$ is an abelian variety of dimension 4. For all $i\in\Int{2}{m}$, $f_i:=(\varphi_i,\varphi_i)$, with $\varphi_i: A_{i-1}\longrightarrow A_i$ and $f_1: (R_1,S_1,R_2,S_2)\longmapsto (\varphi_1(R_1,R_2),\varphi_1(S_1,S_2))$, with $\varphi_1: E_1\times E_2\longrightarrow A_1$. Additionally,
\[\ker(\varphi_m\circ\cdots\circ \varphi_1)=\{([a_1]P,\sigma(P))\mid P\in E_1[2^m]\}.\]
\end{Lemma}

\begin{proof}
Kani's lemma \cite{Kani1997} ensures that 
\[\ker(F)=\{([a_1]P-[a_2]Q,[a_2]P+[a_1]Q,\sigma(P),\sigma(Q))\mid P, Q\in E_1[2^e]\}.\]
Let $f_1, \cdots, f_{m+1}$ be the $m+1$ first elements of the $2$-isogeny chain $F$. Then, since $a_2\equiv 0 \mod 2^m$, we have
\[\ker(f_m\circ\cdots\circ f_1)=[2^{e-m}]\ker(F)=K_1\oplus K_2,\]
where $K_1:=\{([a_1]P,0,\sigma(P),0)\mid P\in E_1[2^m]\}$ and $K_2:=\{(0,[a_1]P,0,\sigma(P))\mid P\in E_1[2^m]\}$. This proves the chain $f_m\circ\cdots\circ f_1$ has the desired form. 
This completes the proof.
\end{proof}

The above lemma indicates we should compute the first $m:=v_2(a_2)$ isogenies of the chain in dimension $2$ and treat $f_{m+1}$ as a gluing isogeny. This is how we proceed in the following.

\subsection{Computing a 4 dimensional endomorphism derived from Kani's lemma with full available torsion}\label{sec: method full torsion}

In this paragraph, we assume we can access to $2^{e+2}$-torsion points of supersingular elliptic curves. This way, we can compute at once the isogeny $F\in\End(E_1^2\times E_2^2)$ as a chain of $2$-isogenies. Using the notations of \cref{lemma: first gluing isogenies}, we propose the following strategy:

\begin{enumerate}
\item  We compute the first $m=v_2(a_2)$ isogenies by computing the $2$-isogeny chain $\Phi:=\varphi_m\circ\cdots\circ \varphi_1$ in dimension $2$. Our implementation of this step relies on \cite{Theta_dim2}.\label{step 1} 
\item We then compute the isogeny $f_{m+1}: A_m^2\longrightarrow B$ assuming it is a gluing isogeny, so using Algorithm \ref{alg: dual theta null point} with $5$ points on entry instead of $4$ (\cref{rem: additional kernel points gluing}). 
\item We can compute a maximal isotropic subgroup $K''\subset B[2^{e-m+1}]$ such that $[4]K''=\ker(G)$ and we can finally compute the $2^{e-m-1}$-isogeny $G: B\longrightarrow E_1^2\times E_2^2$.\label{step 3} 
\item We compute a change of theta coordinates to express image points by $G$ in $(x:z)$-Montgomery coordinates on $E_1^2\times E_2^2$.\label{step 4}
\end{enumerate}

Prior to the computation of (gluing) isogenies $\varphi_1$ and $f_{m+1}$, we have to compute changes of theta coordinates. These changes of coordinates are described in full detail in \cref{sec: change of basis dim 2,sec: change of basis dim 4 full}. Since $m$ can be significantly bigger than $1$ in some cases, the computation of $\Phi$ in step \ref{step 1} above uses optimal strategies that can be computed as in dimension 1 \cite{SIDH,Jesus2023}. 

However, for the computation of $G:=f_e\circ\cdots\circ f_{m+2}$ in step \ref{step 3} the optimal strategy has to satisfy two requirements: 

\begin{itemize}
\item First, we select a strategy of depth $e-m$ instead of $e-m-1$ that integrates the first $m+1$ isogenies $f_{m+1}\circ\cdots\circ f_1$ as one "first step".
\item This "first step" should account for the relative cost of evaluating $f_{m+1}\circ\cdots\circ f_1$ compared to a generic $2$-isogeny.
\end{itemize}
Indeed, if we started the strategy at $f_{m+2}$, we would need to compute $[2^{e-m-2}]K''$ to obtain $f_{m+2}$. Doubling $e-m-2$ times a basis of $K''$ may be more costly than reusing some point doublings we already have computed on $E_1^2\times E_2^2$ to compute the first $m$ isogenies and pushing them through $f_{m+1}\circ\cdots\circ f_1$. We refer to \cref{sec: constrained optimal strategies} for the construction of such optimal strategies.

Algorithm \ref{alg: Kani endo full torsion} summarizes steps 1-3 above to compute $F$ as a $2$-isogeny chain. The output is used in Algorithm \ref{alg: Kani endo full torsion eval} to evaluate $F$ on a point. The evaluation procedure requires a change of theta coordinates on the codomain $E_1^2\times E_2^2$ to recover the product theta structure (step \ref{step 4}) as explained in \cref{sec: splitting change of basis}. Points can then be converted into $(x:z)$-Montgomery coordinates with Algorithm \ref{alg: product to Montgomery dim 4}.

\begin{algorithm}
\newcommand\mycommfont[1]{\textcolor{blue}{#1}}
\SetCommentSty{mycommfont}
\SetAlgoLined
\KwData{\justifying $a_1, a_2, q$ such that $a_2$ is even, $q$ is odd and $a_1^2+a_2^2+q=2^e$, two supersingular elliptic curves $E_1$ and $E_2$ defined over $\F_{p^2}$, $(P''_1,Q''_1)$ a basis of $E_1[2^{e+2}]$, $(\sigma(P''_1),\sigma(Q''_1))$ for some $q$-isogeny $\sigma: E_1\longrightarrow E_2$.}
\KwResult{\justifying A chain representation of the isogeny $F\in\End(E_1^2\times E_2^2)$ given by~\cref{eq: definition F}.}
\justifying
$m\longleftarrow v_2(a_2)$, $r\longleftarrow 1/q \mod 4$, $\mu\longleftarrow 1/a_1 \mod 2^{e+2}$\;
(Pre)compute an optimal strategy $S$ with $m$ leaves \cite[Algorithm~60]{SIKE_spec}\;
(Pre)compute an optimal strategy $S'$ with $e-m$ leaves with more weight at the beginning (see \cref{sec: constrained optimal strategies})\;

\tcc{Step 1: First $m$ isogenies in dimension 2}

$c\longleftarrow e-m-1$, $P'_1, Q'_1, P'_2, R'_2\longleftarrow [2^c]P''_1, [2^c]Q''_1, [2^c]\sigma(P''_1), [2^c]\sigma(Q''_1)$\;

$P_1, Q_1, P_2, Q_2\longleftarrow [2^{m+1}]P'_1, [2^{m+1}]Q'_1, [2^{m+1}]P'_2, [r 2^{m+1}]R'_2$\;

$\zeta_4\longleftarrow e_4(P_1,Q_1)$\;

$T_1, T_2\longleftarrow [2]([a_1]P'_1-[a_2]Q'_1,P'_2), [2]([a_2]P'_1+[a_1]Q'_1,R'_2)$\;

For $i\in\{1,2\}$, compute a basis $(\alpha_i,\beta_i)$ of $E_i[4]$ such that $\beta_i=(-1:1)$ in $(x:z)$-Montgomery coordinates and $e_4(\alpha_i,\beta_i)=\zeta_4$\;

Compute the change of basis matrices $M_i$ from $(\alpha_i,\beta_i)$ to $(P_i,Q_i)$ for $i\in\{1,2\}$\;

Find a theta structure $\Theta'_\mL$ on $(E_1\times E_2,\mL)$ such that $K_2(\Theta'_\mL)=[2^{m+1}]\langle T_1, T_2\rangle$ and compute the change of coordinates matrix $N_{12}$ from $(x:z)$ to $\Theta'_\mL$ (using Algorithm \ref{alg: change of basis dim 2} with input $a_1, a_2, q, (\alpha_i:\beta_i), M_i, \zeta_4$)\;

$({\theta'}_\textbf{j}^{\mL}(T_i))_\textbf{j} \longleftarrow N_{12}\cdot \,^t (x_1(T_i)x_2(T_i), x_1(T_i)z_2(T_i), z_1(T_i)x_2(T_i), z_1(T_i)z_2(T_i))$ for $i\in\{1,2\}$\;

Use the coordinates $({\theta'}_\textbf{j}^{\mL}(T_i))_\textbf{j}$ and strategy $S$ to compute a $2$-dimensional $2$-isogeny chain $\Phi:=\varphi_m\circ\cdots\circ\varphi_1$ of kernel $\ker(\Phi)=[4]\langle T_1, T_2\rangle$ (see \cite{Theta_dim2})\;

\tcc{Step 2: Gluing isogeny $f_{m+1}$ in dimension 4}

$V_1 \longleftarrow (\Phi([a_1]P'_1,P'_2),\Phi([a_2]P'_1,0))$\;
$V_2 \longleftarrow (\Phi([a_1]Q'_1,R'_2),\Phi([a_2]Q'_1,0))$\;
$V_3\longleftarrow (\Phi(-[a_2]P'_1,0),\Phi([a_1]P'_1,P'_2))$\;
$V_4\longleftarrow (\Phi(-[a_2]Q'_1,0),\Phi([a_1]Q'_1,R'_2))$\;
$V_5\longleftarrow (\Phi([a_1](P'_1+Q'_1),P'_2+R'_2),\Phi([a_2](P'_1+Q'_1),0))=V_1+V_2$\;

Let $\Theta_{\mL_m}$ be the level $2$ theta-structure on the codomain $(A_m,\mL_m)$ of $\Phi$ and $\Theta_{\mM_m}:=\Theta_{\mL_m}\times\Theta_{\mL_m}$\;

Find a theta structure $\Theta'_{\mM_m}$ on $(A_m^2,\mM_m)$ such that $K_2(\Theta'_{\mM_m})=[4]\langle V_1, \cdots, V_4\rangle$ and compute the change of coordinates matrix $N_{24}$ from $\Theta_{\mM_m}$ to $\Theta'_{\mM_m}$-coordinates (using Algorithm \ref{alg: change of basis dim 4 full torsion} with input $a_1, a_2, q, m, \zeta_4$)\;

$({\theta'}_\textbf{j}^{\mM_m}(V_i))_\textbf{j}\longleftarrow N_{24}\cdot (\theta_\textbf{j}^{\mM_m}(V_i))_\textbf{j}$ for $i\in\Int{1}{5}$\;

Using the $({\theta'}_\textbf{j}^{\mM_m}(V_i))_\textbf{j}$ for $i\in\Int{1}{5}$, compute $f_{m+1}$ of kernel $[4]\langle V_1,\cdots, V_4\rangle$ (Algorithm \ref{alg: dual theta null point} and \cref{rem: additional kernel points gluing})\;

\caption{Computation of a 4 dimensional endomorphism derived from Kani's lemma with full available torsion.}\label{alg: Kani endo full torsion}
\end{algorithm}

\begin{algorithm}
\newcommand\mycommfont[1]{\textcolor{blue}{#1}}
\SetCommentSty{mycommfont}
\LinesNumbered
\setcounter{AlgoLine}{23}

\tcc{Step 3: Last $e-m-1$ isogenies in dimension 4}

$V'_1\longleftarrow f_{m+1}(\Phi([a_1]P''_1-[\mu]P_1,\sigma(P''_1)),\Phi([a_2]P''_1,0))$\;
$V'_2\longleftarrow f_{m+1}(\Phi([a_1]Q''_1,\sigma(Q''_1)),\Phi([a_2]Q''_1,0))$\;
$V'_3\longleftarrow f_{m+1}(\Phi(-[a_2]P''_1,0),\Phi([a_1]P''_1,\sigma(P''_1)))$\;
$V'_4\longleftarrow f_{m+1}(\Phi(-[a_2]Q''_1,0),\Phi([a_1]Q''_1-[\mu]Q_1,\sigma(Q''_1)))$\;

Use Algorithm \ref{alg: apply strategy dim 4} with input $V'_1, \cdots, V'_4$ and strategy $S'$ to compute a $4$-dimensional $2$-isogeny chain $f_e\circ\cdots\circ f_{m+2}$ of kernel $[4]\langle V'_1,\cdots, V'_4\rangle$\;

\tcc{Step 4: Splitting change of theta coordinates}

Let $\Theta_{\mL_0}:=\Theta_{\mL_1}\times\Theta_{\mL_1}\times\Theta_{\mL_2}\times\Theta_{\mL_2}$ be the product theta structure on $(E_1^2\times E_2^2,\mL_0)$ induced by the $(\alpha_i, \beta_i)$ ($i\in\{1,2\}$)\;

Let $\Theta'_{\mL_0}$ be the (non-product) level $2$ theta structure on $(E_1^2\times E_2^2,\mL_0)$ generated when computing $f_e$\;

Compute the change of coordinates matrix $N_{41}$ from $\Theta'_{\mL_0}$ to $\Theta_{\mL_0}$-coordinates (using Algorithm \ref{alg: splitting change of basis dim 4} with input $a_1, a_2, q, m, M_1, M_2, \zeta_4$)\;

\KwRet $N_{12}, \varphi_1, \cdots, \varphi_m, N_{24}, f_{m+1},\cdots, f_e, N_{41}, (\alpha_1,\beta_1), (\alpha_2,\beta_2)$\;

\end{algorithm}

\begin{algorithm}
\newcommand\mycommfont[1]{\textcolor{blue}{#1}}
\SetCommentSty{mycommfont}
\SetAlgoLined
\KwData{\justifying A chain $C$ outputted by Algorithm \ref{alg: Kani endo full torsion} representing $F\in\End(E_1^2\times E_2^2)$ given by \cref{eq: definition F},  and a point $Q\in E_1^2\times E_2^2$.}
\KwResult{\justifying The Montgomery $(x:z)$-coordinates of $F(Q)$.}

\justifying
Parse $C$ as $N_{12}, \varphi_1, \cdots, \varphi_m, N_{24}, f_{m+1},\cdots, f_e, N_{41}, (\alpha_1,\beta_1), (\alpha_2,\beta_2)$\;
$v\longleftarrow \,^t(x(Q_1)x(Q_3), x(Q_1)z(Q_3), z(Q_1)x(Q_3), z(Q_1)z(Q_3))$\;
$({\theta'}_{\textbf{i}}^{\mL}(Q_1,Q_3))_{\textbf{i}}\longleftarrow N_{12}\cdot v$\;
$(\theta_{\textbf{i}}^{\mL_m}(R_1))_{\textbf{i}}\longleftarrow \varphi_m\circ\cdots\circ\varphi_1(({\theta'}_{\textbf{i}}^{\mL}(Q_1,Q_3))_{\textbf{i}})$\;
$w\longleftarrow \,^t(x(Q_2)x(Q_4), x(Q_2)z(Q_4), z(Q_2)x(Q_4), z(Q_2)z(Q_4))$\;
$({\theta'}_{\textbf{i}}^{\mL}(Q_2,Q_4))_{\textbf{i}}\longleftarrow N_{12}\cdot w$\;
$(\theta_{\textbf{i}}^{\mL_m}(R_2))_{\textbf{i}}\longleftarrow \varphi_m\circ\cdots\circ\varphi_1(({\theta'}_{\textbf{i}}^{\mL}(Q_1,Q_3))_{\textbf{i}})$\;
$\theta_{\textbf{i}_1,\textbf{i}_2}^{\mM_m}(R)\longleftarrow\theta_{\textbf{i}_1}^{\mL_m}(R_1)\cdot \theta_{\textbf{i}_2}^{\mL_m}(R_2)$ for $\textbf{i}_1, \textbf{i}_2\in (\Z/2\Z)^2$\;
$({\theta'}_\textbf{i}^{\mM_m}(R))_\textbf{i}\longleftarrow N_{24}\cdot (\theta_\textbf{i}^{\mM_m}(R))_\textbf{i}$\;
$({\theta'}_\textbf{i}^{\mL_0}(S))_\textbf{i}\longleftarrow f_e\circ\cdots\circ f_{m+1}(({\theta'}_\textbf{i}^{\mM_m}(R))_\textbf{i})$\;
$(\theta_\textbf{i}^{\mL_0}(S))_\textbf{i}\longleftarrow N_{41}\cdot ({\theta'}_\textbf{i}^{\mL_0}(S))_\textbf{i}$\;
Use Algorithm \ref{alg: product to Montgomery dim 4} with input $(\theta_\textbf{i}^{\mL_0}(S))_\textbf{i}$ and $(\alpha_i,\beta_i)$ for $i\in\{1,2\}$ to obtain $(x_1(F(Q)):z_1(F(Q))),\cdots, (x_4(F(Q)):z_4(F(Q)))$\;
\KwRet $(x_1(F(Q)):z_1(F(Q))),\cdots, (x_4(F(Q)):z_4(F(Q)))$\;

\caption{Evaluation of a 4 dimensional endomorphism derived from Kani's lemma with full available torsion given its representation.}\label{alg: Kani endo full torsion eval}
\end{algorithm}

\subsection{Cutting the endomorphism computation in two}\label{sec: cutting in two}

In this paragraph, we explain how to compute $F\in\End(E_1^2\times E_2^2)$ as defined in \cref{eq: definition F} when we cannot access the $2^{e+2}$-torsion of elliptic curves but only "half" of it. Namely, we can access the $2^{e'+2}$-torsion with $e'\geq e/2$. We follow the approach of \cite[§ 4.4]{SQISignHD_autocite}: we write $F:=F_2\circ F_1$ where $F_i$ is a $2^{e_i}$-isogeny with $e_i\leq e'$ for $i\in\{1,2\}$ and $e=e_1+e_2$. We compute $F_1$ and $\widetilde{F}_2$ whose kernels are respectively:
\[\ker(F_1)=\{([a_1]P-[a_2]Q,[a_2]P+[a_1]Q,\sigma(P),\sigma(Q))\mid P, Q\in E_1[2^{e_1}]\}\]
\[\ker(\widetilde{F}_2)=\{([a_1]P+[a_2]Q,-[a_2]P+[a_1]Q,-\sigma(P),-\sigma(Q))\mid P, Q\in E_1[2^{e_2}]\}.\]
And we compute the dual $F_2=\widetilde{\widetilde{F}}_2$ to obtain $F=F_2\circ F_1$. 

Since $\ker(F_1)=\ker(F)[2^{e_1}]$ and $\ker(\widetilde{F}_2)=\ker(\widetilde{F})[2^{e_2}]$, \cref{lemma: first gluing isogenies} applies to $F_1$ and an analogue of \cref{lemma: first gluing isogenies} applies to $\widetilde{F}_2$ (see \cref{lemma: first gluing isogenies F_2}), so we may assume $e_1, e_2\geq m$. Then the computation of $F_1$ and $\widetilde{F}_2$ starts by a chain of $m$ isogenies of dimension $2$ and a gluing isogeny in dimension $4$. 

Unlike previously, we do not expect any splitting in the chains of $2$-isogenies representing $F_1$ and $\widetilde{F}_2$ (except in very rare cases). However, we expect to be able to recover the same codomain $\mC$ for $F_1$ and $\widetilde{F}_2$, or more exactly, to identify the theta structures on $\mC$ induced by $F_1$ and $\widetilde{F}_2$. This identification is not automatic and depends on the choice of theta structures (\emph{i.e.} of symplectic basis of the $2^{e'}$-torsion) that we make on $E_1^2\times E_2^2$ prior to the computation of $F_1$ and $\widetilde{F}_2$. We explain how to make this choice in \cref{sec: change of basis dim 4 half}. This choice also affects the change of theta coordinates that we perform to compute the $(m+1)$-th gluing isogenies in the $2$-isogeny chains $F_1$ and $\widetilde{F}_2$. This is explained in \cref{sec: dim 4 gluing half torsion}.

To compute the dual $F_2=\widetilde{\widetilde{F}}_2$, we only have to compute the dual isogeny of every $2$-isogeny intervening in the chain $\widetilde{F}_2$. This can be done easily by \cref{lemma: dual isogeny computation}. However, note that \cref{lemma: first gluing isogenies} also applies to $\widetilde{F}_2$. As a consequence, the first $m$ isogenies of the chain $\widetilde{F}_2$ are computed in dimension $2$ and the $(m+1)$-th isogeny is a gluing isogeny $g_{m+1}: {A'}_m^2\longrightarrow B'$. After the computation of $\widetilde{g}_{m+1}$, the product theta structure on the codomain ${A'}_m^2$ has to be recovered before computing the dual of the $m$ $2$-dimensional $2$-isogenies. This is explained in \cref{sec: intermediate product theta structure}. We refer to \cref{alg: Kani endo half torsion} in \cref{sec: appendix half torsion} for a detailed overview of the computation of $F$ with half available torsion and to \cref{alg: Kani endo half torsion eval} for its evaluation.

\subsection{Performance}

The computation and evaluation algorithms of $F$ defined in \cref{eq: definition F} have been implemented in Python/Sagemath for the needs of SQIsignHD. This computation has been tested on various parameters on random supersingular elliptic curves $E_1$ defined over finite fields $\F_{p^2}$ of characteristic $p$ between 30 and 378 bits. Primes are of the form $p=c\cdot 2^f\ell^{f'}-1$ with $\ell=3$ or $7$, $f\geq e+2$ and $c$ small. The isogeny $\sigma: E_1\longrightarrow E_2$ "embedded" in $F\in\End(E_1^2\times E_2^2)$ in dimension $4$ as defined in \cref{eq: definition F} is always a random cyclic isogeny of degree $q|\ell^{f'}$ and integers $a_1, a_2\in\Z$ such that $q+a_1^2+a_2^2=2^e$ are precomputed. In SQIsignHD verification, $q$ is not smooth and may vary and $a_1, a_2$ are computed at runtime, however we have chosen $q|\ell^{f'}$ here to be able to verify that point images of $F$ are correct. For every set of parameters, we compared the computation and evaluation of a $2^e$-isogeny $F\in\End(E_1^2\times E_2^2)$ in dimension $4$ as defined in \cref{eq: definition F} with the computation and evaluation of a cyclic $2^e$-isogeny in dimension $1$ with domain $E_1$ (using $x$-only arithmetic code due to Giacomo Pope\footnote{\url{https://github.com/GiacomoPope/KummerIsogeny}}). To compute $F$, both full torsion algorithms (\cref{alg: Kani endo full torsion,alg: Kani endo full torsion eval}) and half torsion algorithms (\cref{alg: Kani endo half torsion,alg: Kani endo half torsion eval}) were tested\footnote{The $2^{e+2}$-torsion is always available but we only used "half" of it to test \cref{alg: Kani endo half torsion,alg: Kani endo half torsion eval}.}. Computations were repeated 100 times and averaged.

Results are displayed in \cref{tab:benchmarks compute,tab:benchmarks eval}. We found that computing a $2^e$-isogeny in dimension $4$ is $16-18$ times more costly than in dimension $1$ over a large base field $\F_{p^2}$, with a slight advantage to the half torsion algorithms (due to the quasilinear complexity of an isogeny chain computation). Timings for evaluation are $\approx 20$ times faster in dimension 1 than in dimension 4. This suggests that our algorithmic approach is promising and can be made cryptographically relevant with a low level implementation (e.g. in C or Rust).   


\begin{table}[!h]
  \centering
  \caption{Comparison of timings (in ms) for $2^e$-isogeny computations in dimension 4 with full available torsion (\cref{alg: Kani endo full torsion}), half available torsion (\cref{alg: Kani endo half torsion}) and in dimension 1 with G. Pope's code for various parameters in Python/Sagemath on a 2.7 GHz Intel Core i5 CPU.}
  \begin{tabular}{cccc|cc|c}
    \toprule
    & & & & \multicolumn{2}{c}{Dimension 4} &  Dimension 1 \\
    $e$ & $\log_2(p)$ & $p$ & $\deg(\sigma)$ & Full tors. & Half tors. & G. Pope \\
    \midrule
    16 & 33 & $2^{19}\cdot 3^{9}-1$ & $3^{9}$ & 139 & 164 & 6\\
    32 & 55 & $2^{34}\cdot 3^{13}-1$ & $3^{13}$ & 366 & 384 & 12 \\
    64 & 121 & $11\cdot 2^{68}\cdot 3^{31}-1$ & $3^{31}$ & 741 & 695 & 37 \\
    64 & 125 & $5\cdot 2^{66}\cdot 3^{36}-1$ & $3^{35}$ & 678 & 674 & 36\\
    128 & 254 & $2^{131}\cdot 3^{78}-1$ & $3^{75}$ & 1519 & 1428 & 83 \\
    128 & 261 & $5^2\cdot 2^{131}\cdot 3^{79}-1$ & $3^{79}$ & 1586 & 1484 & 87 \\
    192 & 365 & $2^{199}\cdot 3^{105}-1$ & $3^{105}$ & 2447 & 2320 & 137 \\
    192 & 371 & $239\cdot 2^{194}\cdot 3^{107}-1$ & $3^{107}$ & 2459 & 2309 & 137 \\
    17 & 30 & $3\cdot 2^{20}\cdot 7^{3}-1$ & $7^{3}$ & 142 & 168 & 6\\
    17 & 35 & $2^{21}\cdot 7^{5}-1$ & $7^{5}$ & 131 & 164 & 6 \\
    33 & 52 & $3^2\cdot 2^{35}\cdot 7^{5}-1$ & $7^{5}$ & 256 & 261 & 12 \\
    33 & 71 & $2^{37}\cdot 7^{12}-1$ & $7^{11}$ & 352 & 351 & 18 \\
    65 & 110 & $109\cdot 2^{67}\cdot 7^{13}-1$ & $7^{13}$ & 691 & 685 & 37 \\
    65 & 137 & $5\cdot 2^{70}\cdot 7^{23}-1$ & $7^{23}$ & 723 & 708 & 39\\
    129 & 249 & $261\cdot 2^{131}\cdot 7^{39}-1$ & $7^{39}$ & 1559 & 1449 & 86 \\
    129 & 257 & $15\cdot 2^{132}\cdot 7^{43}-1$ & $7^{43}$ & 1612 & 1517 & 91 \\
    193 & 359 & $3^2\cdot 2^{196}\cdot 7^{57}-1$ & $7^{57}$ & 2499 & 2354 & 137 \\
    193 & 378 & $97\cdot 2^{195}\cdot 7^{63}-1$ & $7^{63}$ & 2488 & 2370 & 142 \\
    \bottomrule
  \end{tabular}
  \label{tab:benchmarks compute}
\end{table}\begin{table}[!h]
  \centering
  \caption{Comparison of timings (in ms) for $2^e$-isogeny evaluations in dimension 4 with full available torsion (\cref{alg: Kani endo full torsion eval}), half available torsion (\cref{alg: Kani endo half torsion eval}) and in dimension 1 with G. Pope's code for various parameters in Python/Sagemath on a 2.7 GHz Intel Core i5 CPU.}
  \begin{tabular}{cccc|cc|c}
    \toprule
     & & & & \multicolumn{2}{c}{Dimension 4} &  Dimension 1 \\
    $e$ & $\log_2(p)$ & $p$ & $\deg(\sigma)$ & Full tors. & Half tors. & G. Pope \\
    \midrule
    16 & 33 & $2^{19}\cdot 3^{9}-1$ & $3^{9}$ & 7.1 & 6.8 & 0.6 \\
    32 & 55 & $2^{34}\cdot 3^{13}-1$ & $3^{13}$ & 14.2 & 13.9 & 0.8 \\
    64 & 121 & $11\cdot 2^{68}\cdot 3^{31}-1$ & $3^{31}$ &   27.5 & 26.8 & 1.8 \\
    64 & 125 & $5\cdot 2^{66}\cdot 3^{36}-1$ & $3^{35}$ &   25.9 & 26.1 & 1.8 \\
    128 & 254 & $2^{131}\cdot 3^{78}-1$ & $3^{75}$ &   59.3 & 59.4 & 3.5 \\
    128 & 261 & $5^2\cdot 2^{131}\cdot 3^{79}-1$ & $3^{79}$ & 64.1 & 64.2 & 3.7 \\
    192 & 365 & $2^{199}\cdot 3^{105}-1$ & $3^{105}$ &  107.7 & 109.9  & 5.4 \\
    192 & 371 & $239\cdot 2^{194}\cdot 3^{107}-1$ & $3^{107}$ & 106.6 & 106.9 & 5.4 \\
    17 & 30 & $3\cdot 2^{20}\cdot 7^{3}-1$ & $7^{3}$ &  7.1 & 6.9 & 0.6 \\
    17 & 35 & $2^{21}\cdot 7^{5}-1$ & $7^{5}$ &  7.2 & 6.9 & 0.6 \\
    33 & 52 & $3^2\cdot 2^{35}\cdot 7^{5}-1$ & $7^{5}$ &  10.0 & 9.7 & 0.8 \\
    33 & 71 & $2^{37}\cdot 7^{12}-1$ & $7^{11}$ &  15.9 & 15.5 & 1.2 \\
    65 & 110 & $109\cdot 2^{67}\cdot 7^{13}-1$ & $7^{13}$ &  26.4 & 26.3 & 1.8 \\
    65 & 137 & $5\cdot 2^{70}\cdot 7^{23}-1$ & $7^{23}$ &  29.0 & 28.8 & 1.9 \\
    129 & 249 & $261\cdot 2^{131}\cdot 7^{39}-1$ & $7^{39}$ &  60.2 & 59.3 & 3.6 \\
    129 & 257 & $15\cdot 2^{132}\cdot 7^{43}-1$ & $7^{43}$ &  66.3 & 65.2 & 3.8 \\
    193 & 359 & $3^2\cdot 2^{196}\cdot 7^{57}-1$ & $7^{57}$ & 108.5 & 107.4 & 5.4 \\
    193 & 378 & $97\cdot 2^{195}\cdot 7^{63}-1$ & $7^{63}$ & 108.1 & 108.9 & 5.6 \\
    \bottomrule
  \end{tabular}
  \label{tab:benchmarks eval}
\end{table}

\subsection{Application to SIDH attacks}

In SIDH, Alice and Bob are given a starting supersingular elliptic curve $E_0$ defined over $\F_{p^2}$ where $p$ is a prime of the form $p=2^{e_2}3^{e_3}-1$ along with two basis $(P_A,Q_A)$ and $(P_B,Q_B)$ of $E_0[2^{e_2}]$ and $E_0[3^{e_3}]$ respectively. Alice and Bob sample secret integers $s_A\in\Z/2^{e_2}\Z$ and $s_B\in\Z/3^{e_3}\Z$ respectively. Then Alice computes a $2^{e_2}$-isogeny $\varphi_A: E_0\longrightarrow E_A$ of kernel $\ker(\varphi_A)=\langle P_A+[s_A]Q_A\rangle$ and Bob computes a $3^{e_3}$-isogeny $\varphi_B: E_0\longrightarrow E_B$ of kernel $\ker(\varphi_B)=\langle P_B+[s_B]Q_B\rangle$. Alice sends $(E_A,\varphi_A(P_B),\varphi_A(Q_B))$ to Bob and Bob sends $(E_B,\varphi_B(P_A),\varphi_B(Q_A))$ to Alice. Alice can then compute $\psi_A:=[\varphi_B]_*\varphi_A: E_B\longrightarrow E_{BA}$ of kernel $\ker(\psi_A)=\langle \varphi_B(P_A)+[s_A]\varphi_B(Q_A)\rangle$ and Bob computes $\psi_B:=[\varphi_A]_*\varphi_B: E_A\longrightarrow E_{AB}$ of kernel $\ker(\psi_B)=\langle \varphi_A(P_B)+[s_B]\varphi_A(Q_B)\rangle$. Then, they share knowledge of a secret elliptic curve $E_{AB}\simeq E_{BA}$.

\begin{figure}[h]
\centering
\begin{tikzpicture}[line cap=round,line join=round,>=triangle 45,x=1cm,y=1cm,scale=1]
\clip(0,0.5) rectangle (4,3.5);

\draw (1,1) node {$E_0$};
\draw (3,1) node {$E_A$};
\draw (1,3) node {$E_B$};
\draw (3,3) node {$E_{AB}$};

\draw [-latex,line width=0.5pt] (1.3,1)--(2.7,1);
\draw [-latex,line width=0.5pt] (1,1.3)--(1,2.7);
\draw [-latex,line width=0.5pt] (1.3,3)--(2.6,3);
\draw [-latex,line width=0.5pt] (3,1.3)--(3,2.7);

\draw (2,1.3) node {$\varphi_A$};
\draw (1.35,2) node {$\varphi_B$};
\draw (2,3.3) node {$\psi_A$};
\draw (3.35,2) node {$\psi_B$};
\end{tikzpicture}
\end{figure}

However, given $(E_B,\varphi_B(P_A),\varphi_B(Q_A))$, an attacker is able to recover $\varphi_B$ (hence~$s_B$) in polynomial time. Knowing $s_B$, they can then compute $\psi_B$ and find the secret $E_{AB}$. To compute $\varphi_B$, the attacker embeds $\varphi_B$ into a dimension $4$ isogeny:
\[F_B:=\left(\begin{array}{cc} \alpha_0 & \widetilde{\Phi}_B\\
-\Phi_B & \widetilde{\alpha}_B\end{array}\right)\in\End(E_0^2\times E_B^2),\]
where $\Phi_B:=\Diag(\varphi_B,\varphi_B): E_0^2\longrightarrow E_B^2$ and for $i=0, B$,
\[\alpha_i:=\left(\begin{matrix}
a_1 & a_2\\
-a_2 & a_1
\end{matrix}\right)\in \End(E_i^2),\]
with $a_1, a_2\in\Z$ such that $a_1^2+a_2^2+3^{e_3}=2^e$ and $\lceil e/2\rceil+2\leq e_2$. Knowing the $2^{e_2}$-torsion point images $\varphi_B(P_A)$ and $\varphi_B(Q_A)$, the attacker can compute $F_B=F_2\circ F_1$ in two parts as in \cref{sec: cutting in two}. 

The parameters $a_1, a_2$ and $e$ such that $a_1^2+a_2^2+3^{e_3}=2^e$ are precomputed before the attack. In practice, we increment $e$ until $2^e-3^{e_3}$ is easy to factor in the form $\alpha^2\beta N$, where $\alpha$ is smooth, $\beta$ is smooth and all its prime factors are congruent to $1$~mod~$4$ and $N$ is a big prime congruent to $1$ mod $4$. Then $2^e-3^{e_3}$ can be easily decomposed as a sum of two squares using Cornacchia's algorithm \cite{Cornacchia}. This procedure terminates and outputs a small value of $e$ (satisfying $\lceil e/2\rceil+2\leq e_2$) only when $e_3$ is odd. When $e_3$ is even (as in SIKE p610), we look for $a_1, a_2$ and $e$ such that $a_1^2+a_2^2+3^{e_3+1}=2^e$ and embed $\varphi'\circ\varphi_B$ instead of $\varphi_B$ in dimension $4$, where $\varphi': E_B\longrightarrow E'_B$ is a random $3$-isogeny. The parameter search took less than 1 s on a laptop even for the biggest prime (SIKE p751).

We performed our attack 100 times on all SIKE NIST primes (p434, p503, p610 and p751) with starting elliptic curves $E_0$ sampled at random in the supersingular isogeny graph with an isogeny walk from the elliptic curve of $j$-invariant $1728$. Timings and parameters are displayed in \cref{tab:SIDH attacks}. This attack runs in less than 15 s on a laptop for SIKE p751. This significantly improves previous attack implementations using $2$-dimensional isogenies. The implementation by W. Castryck, T. Decru, G. Pope and R. Oudompheng\footnote{\url{https://github.com/GiacomoPope/Castryck-Decru-SageMath}} only worked with a special starting curve of known endomorphism ring and broke SIKE p751 in 1 h. The implementation by L. Maino, L. Panny, G. Pope and B. Wesolowski\footnote{\url{https://github.com/Breaking-SIDH/direct-attack}} worked with any starting curve but only for small parameters. 

\begin{table}[!h]
  \centering
  \caption{Timings (in s) and parameters of the complete SIDH key recovery attack with a random starting curve in Python/Sagemath for various NIST SIKE primes on a 2,7 GHz Intel Core i5 CPU.}
  \begin{tabular}{c|ccc|c}
    \toprule
    SIKE prime & $e_2$ & $e_3$ & $e$ & Attack timing (s) \\
    \midrule
    p434 & 216 & 137 & 225 & 3.82 \\
    p503 & 250 & 159 & 290 & 5.47 \\
    p610 & 305 & 192 & 407 & 8.61 \\
    p751 & 372 & 239 & 589 & 14.02 \\
    \bottomrule
  \end{tabular}
  \label{tab:SIDH attacks}
\end{table}

\newpage

\appendix

\section{Doubling algorithm}\label{sec: doubling}

\subsection{The generic case}

Let $\Theta_\mL$ be a level $2$ theta structure on a polarised abelian variety $(A,\mL)$. We explain here how to compute $(\theta_\textbf{i}^{\mL}(2x))_\textbf{i}$ when $(\theta_\textbf{i}^{\mL}(x))_\textbf{i}$ is given using the formulas of \cref{thm: duplication formulas}. This is described in \cref{alg: doubling}, which is derived from \cite[Algorithm 4.4.10]{DRphd}. \cref{alg: doubling} requires that: 
\begin{equation}\forall \textbf{i}\in(\Z/2\Z)^g, \quad \theta_\textbf{i}^{\mL}(0_A)\neq 0 \quad \mbox{and} \quad \forall \chi\in\widehat{(\Z/2\Z)^g}, \quad U_{\chi}^{\mL^2}(0_A)\neq 0.\label{eq: non-zero theta constants}
\end{equation}
We explain in the following how to treat special cases with vanishing theta constants in the following. 

\begin{algorithm}
\SetAlgoLined
\KwData{\justifying A theta point $(\theta_\textbf{i}^\mL(x))_{\textbf{i}}$ of $A$ and the dual theta null point $(U_{\chi}^{\mL}(0_A))_{\chi}$ of $A$ with non-vanishing coordinates.}
\KwResult{\justifying $(\theta_\textbf{i}^\mL(2x))_{\textbf{i}}$.}
\justifying 
Precompute ${\theta_\textbf{i}^\mL(0_A)}^{-1}\longleftarrow H((U_{\chi}^{\mL}(0_A))_{\chi})_\textbf{i}$ for all $\textbf{i}\in (\Z/2\Z)^g$\;
Precompute ${U_{\chi}^{\mL^2}(0_A)}^{-2}\longleftarrow 1/H\circ S((\theta_\textbf{i}^\mL(0_A))_{\textbf{i}})_{\chi}$ for all $\chi\in \widehat{(\Z/2\Z)^g}$\;
$(Z_\chi)_\chi\longleftarrow S\circ H\circ S((\theta_\textbf{i}^\mL(x))_\textbf{i})$\;
$(Y_\chi)_\chi\longleftarrow ({U_{\chi}^{\mL^2}(0_A)}^{-2}\cdot Z_\chi)_\chi$\;
$(X_\textbf{i})_\textbf{i}\longleftarrow H((Y_\chi)_{\chi})$\;
$(W_\textbf{i})_\textbf{i}\longleftarrow ({\theta_\textbf{i}^\mL(0_A)}^{-1}\cdot X_\textbf{i})_\textbf{i}$\;
\KwRet $(W_\textbf{i})_\textbf{i}$\;
\caption{Generic doubling algorithm.}\label{alg: doubling}
\end{algorithm}

\subsection{The vanishing case}

The $U_{\chi}^{\mL^2}(0_A)$ may vanish when $(A,\mL)$ is $2$-isogenous to a product, \emph{e.g.}\ if we are doubling points on the domain of a splitting $2$-isogeny $f: (A, \mL^2)\longrightarrow (B,\mM)$ during an isogeny chain computation (as in \cref{sec: algo overview}). Indeed, as a consequence of \cref{lemma: f change of level}, $(U_{\chi}^{\mL^2}(0_A))_{\chi}$ is the dual theta null point $(U_{\chi}^{\mM}(0_B))_{\chi}$ of the codomain of the $2$-isogeny $f$ of kernel $\ker(f)=K_2(\Theta_\mL)$. And when $f$ is a splitting, $B$ is a product, so the dual theta constants $U_{\chi}^{\mM}(0_B)$ may vanish and \cref{alg: doubling} may not be applicable. 

In this case, to perform doublings on $A$, we may apply a Hadamard transform on $\Theta_\mL$ (as in the proof of \cref{lemma: dual isogeny computation}) to obtain a theta structure $\Theta'_\mL$ with associated theta coordinates $({\theta'}_\textbf{i}^{\mL})_\textbf{i}=H((\theta_\textbf{i}^{\mL})_{\textbf{i}})$. In that case, to perform doublings, we need to invert the squared dual theta constants 
\[{U'}_{\chi}^{\mL^2}(0_A)^2=\sum_{\textbf{t}\in (\Z/2\Z)^g}\chi(\textbf{t}){\theta'}_{\textbf{t}}^{\mL}(0_A)^2,\]
where $({U'}_{\chi}^{\mL^2}(0_A))_{\chi}$ is the dual theta null point of the codomain of the $2$-isogeny $f': (A,\mL^2)\longrightarrow (B',\mM')$ of kernel $\ker(f')=K_2(\overline{\Theta}'_\mL)=K_1(\overline{\Theta}_\mL)$. In an extreme majority of cases, we do not expect $f'$ to be a splitting isogeny so doublings may be feasible. Once we have computed dual theta coordinates $({\theta'}_\textbf{i}^\mL([2]x))_{\textbf{i}}$, we obtain the desired theta coordinates $(\theta_\textbf{i}^\mL([2]x))_{\textbf{i}}$ by applying another Hadamard transform.

Algorithm \ref{alg: doubling} cannot be applied either when some theta constants $\theta_\textbf{i}^\mL(0_A)$ are zero. Then the Hadamard transform may also be a solution as above. If Algorithm \ref{alg: doubling} still cannot be applied because some theta constants ${\theta'}_\textbf{i}^\mL(0_A)$ and ${U'}_{\chi}^{\mL^2}(0_A)$ are still zero, we may apply a random change of theta coordinates obtained from \cref{thm: symplectic change of basis}. This would be costly but so far we have not encountered this case in practical algorithmic applications.

\section{Explicit change of coordinates computations for 4 dimensional isogenies derived from Kani's lemma with full available torsion}\label{sec: basis change with full torsion}

Throughout this section, we keep the notations of \cref{sec: application to SQIsignHD}. We assume we have full available torsion $(E_1^2\times E_2^2)[2^{e+2}]$ to compute $F\in\End(E_1^2\times E_2^2)$ as defined in~\cref{eq: definition F}.

\subsection{Change of coordinates in dimension 2}\label{sec: change of basis dim 2}

In this paragraph, we explain how to perform the change of coordinates prior to the computation of the first (gluing) $2$ dimensional $2$-isogeny $\varphi_1$ (see \cref{lemma: first gluing isogenies}).

For $i\in\{1,2\}$, consider a basis $(\alpha_i, \beta_i)$ of $E_i[4]$ such that $\beta_i=(-1:1)$ in Montgomery $(x:z)$-coordinates and the associated level $2$ theta structure $\Theta_{\mL_i}$ on $(E_i,\mL_i)$. We assume that $e_4(\alpha_1,\beta_1)=e_4(\alpha_2,\beta_2)$ and we denote by $\zeta_4$ this quantity. Let $\mL:=\pi_1^*\mL_1\otimes\pi_2^*\mL_2$, where the $\pi_i$ are the projections $E_1\times E_2\longrightarrow E_i$ and consider the product theta structure $\Theta_\mL:=\Theta_{\mL_1}\times\Theta_{\mL_2}$. Then, $\Theta_\mL$ is associated to the $\zeta_4$-symplectic basis of $(E_1\times E_2)[4]$: $\mathscr{B}_0:=((\alpha_1,0),(0,\alpha_2),(\beta_1,0),(0,\beta_2))$ (see \cref{sec: product}). 

\begin{Lemma}\label{lemma: change of basis dim 2}
Let $(P_1, Q_1)$ be a basis of $E_1[4]$ such that $e_4(P_1, Q_1)=\zeta_4$ and $(P_2, Q_2):=(\sigma(P_1),[r]\sigma(Q_1))$, where $rq \equiv 1 \mod 4$. Let $M_i$ be the change of basis matrices (in columns convention) from $(\alpha_i, \beta_i)$ to $(P_i,Q_i)$ for $i\in\{1,2\}$. 

Let $\mu$ be a modular inverse of $a_1$ modulo $4$. Consider $\mathscr{B}_1$ given by:
\[((0,-P_2),([\mu]P_1,[\mu a_2]P_2),([a_1]P_1-[a_2]Q_1,P_2),([a_2]P_1+[a_1]Q_1,[q]Q_2)).\]
Then $\mathscr{B}_1$ is a $\zeta_4$-symplectic basis of $(E_1\times E_2)[4]$ which induces a level $2$ theta structure $\Theta'_\mL$ on $E_1\times E_2$ such that $K_2(\Theta'_\mL)=\ker(\varphi_1)$. 

In addition, the change of basis matrix (in columns convention) from $\mathscr{B}_0$ to $\mathscr{B}_1$ is $M:=M_l\cdot M_r$, where:
\[M_l:=\left(\begin{array}{cccc}
M_{1,1,1} & 0 & M_{1,1,2} & 0\\
0 & M_{2,1,1} & 0 & M_{2,1,2} \\
M_{1,2,1} & 0 & M_{1,2,2} & 0\\
0 & M_{2,2,1} & 0 & M_{2,2,2}
\end{array}\right)\quad \mbox{and} \quad M_r:=\left(\begin{array}{cccc}
0 & \mu & a_1 & a_2\\
0 & 0 & 1 & 0\\
0 & 0 & -a_2 & a_1 \\
-1 & \mu a_2 & 0 & q
\end{array}\right)\]
\end{Lemma}

\begin{proof}
The left factor $M_l$ is the change of basis matrix from $\mathscr{B}_0$ to the $\zeta_4$-symplectic basis $\mathscr{B}'_0:=((P_1,0),(0,P_2),(Q_1,0),(0,Q_2))$. The right factor $M_r$ is the change of basis matrix from $\mathscr{B}'_0$ to $\mathscr{B}_1$. Since $\mathscr{B}_0$ and $\mathscr{B}'_0$ are both $\zeta_4$-symplectic basis by construction, we have $M_l\in\Sp_4(\Z/4\Z)$. Furthermore, we easily check that $M_r\in\Sp_4(\Z/4\Z)$ by decomposing $M_r$ into $2\times 2$-blocks:
\[M_r=\left(\begin{array}{cc}
A & C\\
B & D
\end{array}\right)\]
and verifying that $\,^t B A\equiv \,^t A B$, $\,^t D C\equiv\,^t C D$ and $\,^t A D-\,^t B C\equiv I_2$ mod $4$, as in \cref{lemma: symplectic matrix}. Hence, the change of basis matrix from $\mathscr{B}_0$ to $\mathscr{B}_1$ $M=M_l\cdot M_r$ belongs to $\Sp_4(\Z/4\Z)$ and $\mathscr{B}_1$ is a $\zeta_4$-symplectic basis. 

Since $2|a_2$, we have:
\[[2]\langle \mathscr{B}_{1,3}, \mathscr{B}_{1,4}\rangle=[2]\langle ([a_1]P_1,\sigma(P_1)),([a_1]Q_1,\sigma(Q_1))\rangle=\ker(\varphi_1),\]
so the level $2$ theta structure $\Theta'_\mL$ associated to $\mathscr{B}_1$ by \cref{thm: compatible symmetric theta structures}.(ii) satisfies $K_2(\Theta'_\mL)=\ker(\varphi_1)$.
\end{proof}

If we want to compute the $\Theta'_{\mL}$-coordinates of a point $(R_1, R_2)\in E_1\times E_2$ expressed in Montgomery $(x:z)$-coordinates, we first have to translate $(x(R_i):z(R_i))$ into $(\theta_0^{\mL_i}(R_i):\theta_1^{\mL_i}(R_i))$ for $i\in\{1,2\}$, using the formulas of \cref{sec: Montgomery theta}, then we compute the product $\Theta_\mL$-coordinates $\theta_{i,j}^{\mL}(R_1,R_2)=\theta_i^{\mL_1}(R_1)\cdot \theta_j^{\mL_2}(R_2)$ for all $i,j \in\Z/2\Z$ (by \cref{lemma: product theta structure}) and then act on $(\theta_{i,j}^{\mL}(R_1,R_2))_{i,j}$ with the change of coordinates matrix $N$ from $\Theta_{\mL}$ to $\Theta'_{\mL}$-coordinates obtained by \cref{lemma: change of basis dim 2} and \cref{thm: symplectic change of basis}. Assuming $N$ has been precomputed, each conversion of $(x:z)$-coordinates into $\Theta'_{\mL}$-coordinates costs $2\times 2$ multiplication to translate $(x:z)$ into $\Theta_{\mL_i}$-coordinates, $4$ multiplications to compute the product $\Theta_{\mL}$-coordinates and $16$ multiplication to apply the matrix $N$, for a total of $24$ multiplications. 

Instead, we can translate directly $(x:z)$-coordinates to $\Theta'_{\mL}$-coordinates without intermediate $\Theta_{\mL_i}$-coordinates computation. We first compute the product $(x:z)$-coordinates $(x(R_1)x(R_2):x(R_1)z(R_2):z(R_1)x(R_2):z(R_1)z(R_2))$. Then, we act with the \emph{change of ccordinates matrix from $(x:z)$ to $\Theta'_{\mL}$} $N'':=N\cdot N'$, where $N'$ translates product $(x:z)$-coordinates into $\Theta_\mL$-coordinates and is given by: 
\[N':=\left(\begin{array}{cccc}
a_1 a_2 & - a_1 a_2 & -a_1 a_2 & a_1 a_2\\
b_1 a_2 & - b_1 a_2 & b_1 a_2 & -b_1 a_2\\
a_1 b_2 & a_1 b_2 & -a_1 b_2 & -a_1 b_2\\
b_1 b_2 & b_1 b_2 & b_1 b_2 & b_1 b_2
\end{array}\right),\]
with $(a_i:b_i)$ the theta null point of $(E_i,\mL_i,\Theta_{\mL_i})$ for $i\in\{1,2\}$. Assuming $N''$ has been precomputed, translating $(x:z)$-coordinates to $\Theta'_{\mL}$-coordinates costs 20 multiplications, saving 4 multiplications compared to the previous method. The precomputation of the product $N''=N\cdot N'$ is not too costly compared to the computation of $N$ alone since the product $N\cdot N'$ can be computed with $16$ multiplications only instead of $64$ (see \cref{alg: change of basis dim 2}).

\begin{algorithm}[h]
\SetAlgoLined
\KwData{\justifying Integer parameters $a_1, a_2, q$, the basis $(\alpha_i,\beta_i)$ of $E_i[4]$, the change of basis matrices $M_i$ from $(\alpha_i,\beta_i)$ to $(P_i,Q_i)$ (as defined in \cref{lemma: change of basis dim 2} for $i\in\{1,2\}$) and the Weil pairing $\zeta_4=e_4(P_1,Q_1)$.}
\KwResult{\justifying A change of coordinates matrix from $(x:z)$ to $\Theta'_{\mL}$, where $\Theta'_{\mL}$ is the theta structure induced by $\mathscr{B}_1$ (as defined in \cref{lemma: change of basis dim 2}) satisfying $K_2(\Theta'_{\mL})=\ker(\varphi_1)$.}

$\mu\longleftarrow 1/a_1 \mod 4$\;
Using $a_1, a_2, \mu$, compute the matrix $M_r$ of \cref{lemma: change of basis dim 2}\;
Using $M_1$ and $M_2$, compute the matrix $M_l$ of \cref{lemma: change of basis dim 2}\;
$M\longleftarrow M_l\cdot M_r$\;
Let $\Theta_{\mL}=\Theta_{\mL_1}\times\Theta_{\mL_2}$ be the product theta structure of $E_1\times E_2$ induced by $(\alpha_1,\beta_1)$ and $(\alpha_2,\beta_2)$\;
Using the formulas of \cref{thm: symplectic change of basis} with $M$ and $\zeta_4$ compute the change of coordinates matrix $N$ from $\Theta_{\mL}$ to $\Theta'_{\mL}$-coordinates\;
$a_i\longleftarrow x(\alpha_i)+z(\alpha_i)$, $b_i\longleftarrow x(\alpha_i)-z(\alpha_i)$ for $i\in\{1,2\}$\;
$c_1, c_2, c_3, c_4\longleftarrow a_1 a_2, b_1 a_2, a_1 b_2, b_1 b_2$\;
$N'_{i,j}\longleftarrow N_{i,j}\cdot c_j$ for all $i, j\in\Int{1}{4}$\;
\For{$i=1$ \KwTo $4$}{
$N''_{i,1}\longleftarrow N'_{i,1}+N'_{i,2}+N'_{i,3}+N'_{i,4}$\;
$N''_{i,2}\longleftarrow -N'_{i,1}-N'_{i,2}+N'_{i,3}+N'_{i,4}$\;
$N''_{i,3}\longleftarrow -N'_{i,1}+N'_{i,2}-N'_{i,3}+N'_{i,4}$\;
$N''_{i,4}\longleftarrow N'_{i,1}-N'_{i,2}-N'_{i,3}+N'_{i,4}$\;
}
\KwRet $N''$\;

\caption{Change of coordinates in dimension 2.}\label{alg: change of basis dim 2}
\end{algorithm}

\subsection{Change of coordinates in dimension 4 with full available torsion}\label{sec: change of basis dim 4 full}

In this paragraph, we explain how to perform the change of coordinates prior to the computation of the first (gluing) $4$ dimensional $2$-isogeny $f_{m+1}: A_m^2\longrightarrow B$ (see \cref{lemma: first gluing isogenies}) when we can access the $2^{e+2}$-torsion of $E_1$. We first determine the product theta structure $\Theta_{\mM_m}$ on $(A_m^2,\mM_m)$ and then explain how to compute a new theta structure such that $K_2(\Theta'_{\mM_m})=\ker(f_{m+1})$.

\subsubsection{Computing the theta structure induced by the 2-dimensional chain}

After the computation of $\Phi:=\varphi_m\circ\cdots\circ\varphi_1: E_1\times E_2\longrightarrow A_m$, we obtain a level $2$ theta structure $\Theta_{\mL_m}$ on the polarised abelian surface $(A_m,\mL_m)$. This theta structure $\Theta_{\mL_m}$ is induced by the symplectic basis of $A_m[4]$ given by 
\[\mathscr{C}_1:=([2^m]\Phi(S_1),[2^m]\Phi(S_2),\Phi(T_1),\Phi(T_2)),\] 
where $\widetilde{\mathscr{B}}_1:=(S_1,S_2,T_1,T_2)$ is a symplectic basis of $(E_1\times E_2)[2^{m+2}]$ such that:

\begin{enumerate}[label=(\roman*)]
\item $[2^m]\widetilde{\mathscr{B}}_1$ is the basis $\mathscr{B}_1$ of \cref{lemma: change of basis dim 2};
\item For all $i\in\Int{1}{m}$, the $8$-torsion points $[2^{m-i-1}]\varphi_{i-1}\circ\cdots\circ\varphi_1(T_j)$ where $j\in\{1,2\}$ lye above $\ker(\varphi_i)$ and have been used to compute $\varphi_i$ (on entry of \cite[Algorithm 5 or 7]{Theta_dim2} or Algorithm \ref{alg: dual theta null point}).
\end{enumerate}

To satisfy point (ii) above, it is sufficient that $\widetilde{\mathscr{B}}_1$ satisfies the following:

\begin{enumerate}[label=(ii)']
\item $\langle T_1,T_2\rangle$ is a maximal isotropic subgroup of $(E_1\times E_2)[2^{m+2}]$ such that:
\[\langle [4]T_1,[4]T_2\rangle=\ker(\Phi)=\{([a_1]P,\sigma(P))\mid P\in E_1[2^m]\}.\]
\end{enumerate}

\begin{Lemma}\label{lemma: theta structure dim 2 end chain}
Let $(P'_1,Q'_1)$ be a basis of $E_1[2^{m+2}]$ such that $P_1=[2^m]P'_1$ and $Q_1=[2^m]Q'_1$, where $(P_1, Q_1)$ is the basis of $E_1[4]$ of \cref{lemma: change of basis dim 2}. Let $\zeta:=e_{2^{m+2}}(P'_1,Q'_1)$. Consider $\widetilde{\mathscr{B}}_1:=(S_1,S_2,T_1,T_2)$, with:
\[S_1:=([2^{m+1}]Q'_1,[a]\sigma(P'_1)+[b]\sigma(Q'_1)), \quad S_2:=([\mu]P'_1,[c]\sigma(P'_1)+[d]\sigma(Q'_1)),\]
\[T_1:=([a_1]P'_1-[a_2]Q'_1,\sigma(P'_1)) \quad \mbox{and} \quad T_2:=([a_2]P'_1+[a_1]Q'_1,\sigma(Q'_1)),\]
$a\equiv 2^{m+1}a_2/q$, $b\equiv -(1+2^{m+1}a_1)/q$, $\mu\equiv (1-2^{m+1}q)/a_1$, $d\equiv -\mu a_2/q \mod 2^{m+2}$ and $c=2^{m+1}$. Then $\widetilde{\mathscr{B}}_1$ is a $\zeta$-symplectic basis of $(E_1\times E_2)[2^{m+2}]$ satisfying (i) and (ii)'.
\end{Lemma}

\begin{proof}
Let $r$ be the modular inverse of $q$ modulo $2^{m+2}$, $P'_2:=\sigma(P_1)$ and $Q'_2:=[r]\sigma(Q'_1)$. Consider the $\zeta$-symplectic basis of $(E_1\times E_2)[2^{m+2}]$ given by $\widetilde{\mathscr{B}}_0:=((P'_1,0),(0,P'_1),(Q'_1,0),(0,Q'_2))$. Then, the change of basis matrix from $\widetilde{\mathscr{B}}_0$ to $\widetilde{\mathscr{B}}_1$ is:
\[M:=\left(\begin{array}{cccc}
0 & \mu & a_1 & a_2\\
a & c & 1 & 0\\
2^{m+1} & 0 & -a_2 & a_1\\
bq & dq & 0 & q
\end{array}\right)\]
As in the proof of \cref{lemma: change of basis dim 2}, we easily check that $M\in \Sp_4(\Z/2^{m+2}\Z)$, which proves that $\widetilde{\mathscr{B}}_1$ is a $\zeta$-symplectic basis of $(E_1\times E_2)[2^{m+2}]$.

We also see that the reduction of $M$ modulo $4$ is the matrix $M_r$ introduced in the proof of \cref{lemma: change of basis dim 2}, which is the change of basis matrix from $\mathscr{B}'_0$ to $\mathscr{B}_1$. Furthermore, $[2^m]\widetilde{\mathscr{B}}_0=\mathscr{B}'_0$ so we obtain $[2^m]\widetilde{\mathscr{B}}_1=\mathscr{B}_1$, which proves (i). 

Finally, since $2^m|a_2$, we see that:
\begin{align*}\langle [4]T_1,[4]T_2\rangle &=\langle ([4a_1]P'_1,\sigma([4]P'_1)),([4a_1]Q'_1,\sigma([4]Q'_1))\rangle\\
&=\{([a_1]P,\sigma(P))\mid P\in E_1[2^m]\}=\ker(\Phi),
\end{align*}
which proves (ii)' and completes the proof. 
\end{proof}

\subsubsection{Computing the change of theta coordinates before the 4-dimensional gluing}\label{sec: dim 4 gluing full torsion}

Consider the product $(A_m^2,\mM_m)$ with $\mM_m:=\pi_1^*\mL_m\otimes\pi_2^*\mL_m$, where the $\pi_i$ are the projections $A_m^2\longrightarrow A_m$ on the $i$-th component and the product theta structure $\Theta_{\mM_m}:=\Theta_{\mL_m}\times\Theta_{\mL_m}$. Then $\Theta_{\mM_m}$ is associated to the $4$-torsion symplectic basis:
\begin{align*}
\mathscr{C}_1\times\mathscr{C}_1&:=(([2^m]\Phi(S_1),0),([2^m]\Phi(S_2),0),(0,[2^m]\Phi(S_1)),(0,[2^m]\Phi(S_2)),\\
&\qquad (\Phi(T_1),0),(\Phi(T_2),0),(0,\Phi(T_1)),(0,\Phi(T_2)))
\end{align*}
We want to compute a theta structure $\Theta'_{\mM_m}$ on $(A_m^2,\mM_m)$ such that $K_2(\Theta'_{\mM_m})=\ker(f_{m+1})$. In order to do that, we compute a change of basis from $\mathscr{C}_1\times\mathscr{C}_1$ to a symplectic basis $\mathscr{C}:=(U_1,\cdots, U_4,V_1,\cdots, V_4)$ such that:
\[V_1=(\Phi([a_1]P'_1,\sigma(P'_1)),\Phi([a_2]P'_1,0)), \quad V_2=(\Phi([a_1]Q'_1,\sigma(Q'_1)),\Phi([a_2]Q'_1,0)),\]
\[V_3=(\Phi(-[a_2]P'_1,0),\Phi([a_1]P'_1,\sigma(P'_1))), \quad V_4=(\Phi(-[a_2]Q'_1,0),\Phi([a_1]Q'_1,\sigma(Q'_1))),\]
so that $\langle V_1,\cdots, V_4\rangle=[2^{e-m}]f_m\circ\cdots\circ f_1(K'')$. Any symplectic complement $(U_1,\cdots, U_4)$ of $(V_1,\cdots, V_4)$ is acceptable. When we cannot access the full $2^{e+2}$-torsion, we may choose a different basis $\mathscr{C}$ because we have to satisfy more constraints. Here, to compute $\mathscr{C}$ we first compute the matrix of $(V_1,\cdots, V_4)$ in the basis $\mathscr{C}_1\times\mathscr{C}_1$ and complete it to obtain a symplectic matrix of $\Sp_8(\Z/4\Z)$ (which can be done with easy linear algebra over $\Z/4\Z$).

\begin{Lemma}\label{lemma: matrix V_i}
The matrix of $(V_1,\cdots, V_4)$ in $\mathscr{C}_1\times\mathscr{C}_1$ is:
\[\left(\begin{array}{cccc}
-a_1a_2/2^m & a_2^2/2^m & a_2^2/2^m & a_1a_2/2^m\\
-a_2^2/2^m & -a_1a_2/2^m & -a_1a_2/2^m & a_2^2/2^m\\
-a_2^2/2^m & -a_1a_2/2^m & -a_1a_2/2^m & a_2^2/2^m\\
a_1a_2/2^m & -a_2^2/2^m & -a_2^2/2^m & -a_1a_2/2^m\\
1 & 0 & 0 & 0\\
\mu a_2 & 1 & 0 & -\mu a_2\\
0 & 0 & 1 & 0 \\
0 & \mu a_2 & \mu a_2 & 1
\end{array}\right),\]
where $\mu$ has been defined in \cref{lemma: theta structure dim 2 end chain}. 
\end{Lemma}

\begin{proof}
Let $W_1:=\Phi([a_1]P'_1,\sigma(P'_1))$, $W_2:=\Phi([a_1]Q'_1,\sigma(Q'_1))$, $W_3:=\Phi([a_2]P'_1,0)$ and $W_4:=\Phi([a_2]Q'_1,0)$. To conclude, it suffices to compute the matrix of $(W_1,\cdots, W_4)$ in $\mathscr{C}_1$.
Let us write:
\[W_j:=\sum_{i=1}^2 [c_{i,j} 2^m]\Phi(S_i)+\sum_{i=1}^2 [d_{i,j}]\Phi(T_i)\]
for all $j\in\Int{1}{4}$. Then, we have for all $i\in\{1,2\}$ and $j\in\Int{1}{4}$,
\[e_4(W_j,[2^m]\Phi(S_i))=e_4([2^m]\Phi(S_i),\Phi(T_i))^{-d_{i,j}}=\zeta_4^{-d_{i,j}}\]
\[\mbox{and} \quad e_4(W_j,\Phi(T_i))=e_4([2^m]\Phi(S_i),\Phi(T_i))^{c_{i,j}}=\zeta_4^{c_{i,j}},\]
where $\zeta_4=\zeta^{2^m}=e_{2^{m+2}}(P'_1,Q'_1)^{2^m}$. Hence, we can obtain the coefficients $c_{i,j}$ and $d_{i,j}$ that we are looking for by computing Weil pairings.

We have:
\begin{align*}
e_4(W_1,[2^m]\Phi(S_1))&=e_4(\Phi([a_1]P'_1,\sigma(P'_1)),[2^m]\Phi(S_1))\\
&=e_4(([a_1]P'_1,\sigma(P'_1)),[2^m]\widetilde{\Phi}\circ\Phi(S_1)) \quad \mbox{\cite[Lemma 16.2.(a)]{Milne1986}}\\
&=e_4(([a_1]P'_1,\sigma(P'_1)),[2^{2m}]S_1)\\
&=e_4([2^m]([a_1]P'_1,\sigma(P'_1)),[2^{m}]S_1)\\
&=e_{2^{m+2}}(([a_1]P'_1,\sigma(P'_1)),S_1)^{2^m} \quad \mbox{\cite[Lemma 16.1]{Milne1986}}\\
&=e_{2^{m+2}}(([a_1]P'_1,\sigma(P'_1)),([2^{m+1}]Q'_1,[a]\sigma(P'_1)+[b]\sigma(Q'_1)))^{2^m}\\
&=e_{2^{m+2}}(P'_1,Q'_1)^{2^m(a_1 2^{m+1}+bq)}=\zeta_4^{a_1 2^{m+1}+bq}=\zeta_4^{bq}=\zeta_4^{-1},
\end{align*}
so that $d_{1,1}=1$. Similarly, we obtain:
\begin{align*}
e_4(W_1,[2^m]\Phi(S_2))&=e_{2^{m+2}}(([a_1]P'_1,\sigma(P'_1)),S_2)^{2^m}\\
&=e_{2^{m+2}}(([a_1]P'_1,\sigma(P'_1)),([\mu]P'_1,[c]\sigma(P'_1)+[d]\sigma(Q'_1)))^{2^m}\\
&=e_{2^{m+2}}(P'_1,Q'_1)^{2^m dq}=\zeta_4^{dq}=\zeta_4^{-\mu a_2},
\end{align*}
\begin{align*}
e_4(W_1,\Phi(T_1))&=e_{4}(\Phi([a_1]P'_1,\sigma(P'_1)),\Phi([a_1]P'_1-[a_2]Q'_1,\sigma(P'_1)))\\
&=e_{4}(\Phi([a_1]P'_1,\sigma(P'_1)),\Phi(-[a_2]Q'_1,0))\\
& \quad \times e_{4}(\Phi([a_1]P'_1,\sigma(P'_1)),\Phi([a_1]P'_1,\sigma(P'_1)))\\
&=e_{4}(\Phi([a_1]P'_1,\sigma(P'_1)),\Phi(-[a_2]Q'_1,0))\\
&=e_{4}(\Phi([a_1]P'_1,\sigma(P'_1)),[2^m]\Phi(-[a_2/2^m]Q'_1,0))\\
&=e_{2^{m+2}}(([a_1]P'_1,\sigma(P'_1)),(-[a_2/2^m]Q'_1,0))^{2^m}\\
&=e_{2^{m+2}}(P'_1,Q'_1)^{-a_1a_2}=\zeta^{-a_1a_2}=\zeta_4^{-a_1a_2/2^m},
\end{align*}
so that $d_{2,1}=\mu a_2$ and $c_{1,1}=-a_1a_2/2^m$. Let $W'_1\in A_m[2^{m+2}]$ and $T'_2\in (E_1\times E_2)[2^{2m+2}]$ such that $[2^m]W'_1=W_1$ and $[2^m]T'_2=T_2$. Then:
\begin{align*}
e_4(W_1,\Phi(T_2))&=e_{4}([2^m]W'_1,[2^m]\Phi(T'_2))=e_{2^{m+2}}(W'_1,\Phi(T'_2))^{2^m}\\
&=e_{2^{m+2}}([2^m]W'_1,\Phi(T'_2))=e_{2^{m+2}}(\Phi([a_1]P'_1,\sigma(P'_1)),\Phi(T'_2))\\
&=e_{2^{m+2}}(([a_1]P'_1,\sigma(P'_1)),\widetilde{\Phi}\circ\Phi(T'_2))\\
&=e_{2^{m+2}}(([a_1]P'_1,\sigma(P'_1)),[2^m]T'_2)\\
&=e_{2^{m+2}}(([a_1]P'_1,\sigma(P'_1)),([a_2]P'_1+[a_1]Q'_1,\sigma(Q'_1)))\\
&=e_{2^{m+2}}(P'_1,Q'_1)^{a_1^2+q}=\zeta^{2^e-a_2^2}=\zeta^{-a_2^2}=\zeta_4^{-a_2^2/2^m},
\end{align*}
so that $c_{2,1}=-a_2^2/2^m$. It follows that:
\[W_1=-\left[\frac{a_1a_2}{2^m}\right][2^m]\Phi(S_1)-\left[\frac{a_2^2}{2^m}\right][2^m]\Phi(S_2)+\Phi(T_1)+[\mu a_2]\Phi(T_2).\]
Similarly, we obtain the coordinates of $W_2, W_3$ and $W_4$ to finally get the matrix of $(W_1, \cdots, W_4)$ in $\mathscr{C}_1$:
\[\left(\begin{array}{cccc}
-a_1a_2/2^m & a_2^2 & -a_2^2/2^m & -a_1a_2/2^m\\
-a_2^2/2^m & -a_1a_2/2^m & a_1a_2/2^m & -a_2^2/2^m\\
1 & 0 & 0 & 0\\
\mu a_2 & 1 & 0 & \mu a_2
\end{array}\right).\]
Since $V_1=(W_1,W_3)$, $V_2=(W_2,W_4)$, $V_3=(-W_3,W_1)$ and $V_4=(-W_4,W_2)$, we easily obtain the matrix of $(V_1, \cdots, V_4)$ in $\mathscr{C}_1\times\mathscr{C}_1$.
\end{proof}

We summarize the change of theta coordinates computation prior to the gluing $f_{m+1}: A_m^2\longrightarrow B$ in Algorithm \ref{alg: change of basis dim 4 full torsion}.

\begin{algorithm}[h]
\SetAlgoLined
\KwData{\justifying Integer parameters $a_1, a_2, q, m=v_2(a_2)$ and the Weil pairing $\zeta_4=e_4(P_1,Q_1)=e_{2^{m+2}}(P'_1,Q'_1)^{2^m}$.}
\KwResult{\justifying A change of coordinates matrix from $\Theta_{\mM_m}$ to $\Theta'_{\mM_m}$-coordinates, where $\Theta_{\mM_m}$ is the level $2$ product theta structure on $(A_m^2,\mM_m)$ and $\Theta'_{\mM_m}$ satisfies $K_2(\Theta'_{\mM_m})=\ker(f_{m+1})$.}

$\mu\longleftarrow (1-2^{m+1}q)/a_1 \mod 2^{m+2}$\;
Using $\mu$, compute the matrix $M$ of \cref{lemma: matrix V_i}\;
Decompose $M$ in $4\times 4$-blocks $M:=\left(\begin{array}{c}
C\\
D
\end{array}\right)$\;
Find $A, B\in M_4(\Z/4\Z)$ such that $\,^t B A =\,^t A B$, $\,^t D C=\,^t C D$ and $\,^t A D-\,^t B C=I_4$\;
$M'\longleftarrow\left(\begin{array}{cc}
A & C\\
B & D
\end{array}\right)$\;
Let $\Theta'_{\mM_m}$ be the theta structure given by the action of $M'$ on $\Theta_{\mM_m}$\;
Using \cref{thm: symplectic change of basis}, compute the change of coordinates matrix $N$ from $\Theta_{\mM_m}$ to $\Theta'_{\mM_m}$-coordinates\;
\KwRet $N$\;

\caption{Change of coordinates in dimension 4 (full torsion case).}\label{alg: change of basis dim 4 full torsion}
\end{algorithm}

\subsection{Recovering the product theta structure on the codomain with full available torsion}\label{sec: splitting change of basis}

When we can access the full $2^{e+2}$-torsion and compute $F\in\End(E_1^2\times E_2^2)$ as a $2$-isogeny chain at once, we have to recover the product theta structure on the codomain $E_1^2\times E_2^2$ at the end in order to be able to evaluate $F$ in Montgomery coordinates.

Once we have computed the whole $2$-isogeny chain $F$, the level $2$ theta structure $\Theta'_{\mL_0}$ naturally induced on the codomain $(E_1^2\times E_2^2,\mL_0)$ is associated to the symplectic $4$-torsion basis: 
\begin{equation}\mathscr{D}:=([2^{e-m}]H(U'_1),\cdots, [2^{e-m}]H(U'_4), H(V'_1), \cdots, H(V'_4)), \label{eq: D}
\end{equation}
where $H:=f_e\circ\cdots\circ f_{m+1}$, so that $F:=H\circ f_m\circ\cdots\circ f_1$ and $\widetilde{\mathscr{C}}_1:=(U'_1, \cdots, U'_4, V'_1, \\ \cdots, V'_4)$ is a symplectic basis of $A_m^2[2^{e-m+2}]$ such that:

\begin{enumerate}[label=(\roman*)]
\item $[2^{e-m}]\widetilde{\mathscr{C}}$ is the basis $\mathscr{C}$ of $A_m^2[4]$ introduced in \cref{sec: change of basis dim 4 full}.
\item For all $i\in\Int{m+1}{e}$, the $8$-torsion points $[2^{e-i-1}]f_{i-1}\circ\cdots\circ f_{m+1}(V'_j)$ where $j\in\Int{1}{4}$ lye above $\ker(f_i)$ and have been used to compute $f_i$ on entry of Algorithm \ref{alg: dual theta null point}.
\end{enumerate}

As in \cref{sec: change of basis dim 4 full}, to satisfy point (ii) above, it is sufficient that $\widetilde{\mathscr{C}}$ satisfies the following:

\begin{enumerate}[label=(ii)']
\item $\langle V'_1,\cdots, V'_4\rangle$ is a maximal isotropic subgroup of $A_m^2[2^{e-m+2}]$ such that $\langle [4]V'_1, \cdots, [4]V'_4\rangle=\ker(H)$.
\end{enumerate}

In the following lemma, we give an explicit construction of $\mathscr{D}$ induced by a basis $\widetilde{\mathscr{C}}$ satisfying (i) and (ii)'.

\begin{Lemma}\label{lemma: splitting dim 4}
Let:
\begin{itemize}
\item $(P_1, Q_1)$ and $(P_2, Q_2):=(\sigma(P_1),[r]\sigma(Q_1))$ be the $\zeta_4$-symplectic basis of $E_1[4]$ and $E_2[4]$ introduced in \cref{lemma: change of basis dim 2}.
\item $\mathscr{B}'_0\times\mathscr{B}'_0$ be the $\zeta_4$-symplectic basis of $(E_1^2\times E_2^2)[4]$ given by:
\begin{align*}\mathscr{B}'_0\times \mathscr{B}'_0 &:=((P_1,0,0,0), (0,P_1,0,0), (0,0,P_2,0), (0,0,0,P_2),\\
& \qquad (Q_1,0,0,0), (0,Q_1,0,0), (0,0,Q_2,0), (0,0,0,Q_2)).
\end{align*}
\item $\mathscr{D}'$ be the basis:
\begin{align*}\mathscr{D}'&:=([2^{e-m}]H(\Phi(S_1),0),[2^{e-m}]H(\Phi(S_2),0),[2^{e-m}]H(0,\Phi(S_1)),[2^{e-m}]H(0,\Phi(S_2)),\\
&\qquad H(\Phi(T_1),0),H(\Phi(T_2),0),H(0,\Phi(T_1)),H(0,\Phi(T_2))),
\end{align*}
with $S_1, S_2, T_1, T_2$ as in \cref{lemma: theta structure dim 2 end chain}.
\item $M_{\mathscr{B}'_0\times\mathscr{B}'_0 , \mathscr{D}'}$ be the change of basis matrix from $\mathscr{B}'_0\times\mathscr{B}'_0$ to $\mathscr{D}'$.
\item $M_{\mathscr{C}_1\times \mathscr{C}_1, \mathscr{C}}$ be the change of basis matrix from $\mathscr{C}_1\times \mathscr{C}_1$ to $\mathscr{C}$ (introduced in \cref{sec: change of basis dim 4 full}) decomposed into two $8\times 4$ blocks $M_{\mathscr{C}_1\times \mathscr{C}_1, \mathscr{C}}:=(L_{\mathscr{C}_1\times \mathscr{C}_1, \mathscr{C}}|R_{\mathscr{C}_1\times \mathscr{C}_1, \mathscr{C}})$.
\end{itemize}

Then, there exists a symplectic basis of $A_m[2^{e-m+2}]$, $\widetilde{\mathscr{C}}:=(U'_1, \cdots, U'_4, V'_1, \cdots, V'_4)$ satisfying conditions (i) and (ii)' such that the change of basis matrix from $\mathscr{B}'_0\times \mathscr{B}'_0$ to the basis $\mathscr{D}$ related to $\widetilde{\mathscr{C}}$ by \cref{eq: D} is given by:
\[M_{\mathscr{B}'_0\times\mathscr{B}'_0,\mathscr{D}}=(M_{\mathscr{B}'_0\times\mathscr{B}'_0, \mathscr{D}'}\cdot L_{\mathscr{C}_1\times \mathscr{C}_1, \mathscr{C}}|R_{\mathscr{B}'_0\times\mathscr{B}'_0,\mathscr{D}}),\]
with $M_{\mathscr{B}'_0\times\mathscr{B}'_0, \mathscr{D}'}$, the matrix:
\[\left(\begin{array}{ccccccccc}
0 & 1 & 0 & \mu a_2 & (a_1^2+q)/2^m  & a_1 a_2/2^m & a_1 a_2/2^m & a_2^2/2^m \\
0 & -\mu a_2 & 0 & 1 & -a_1 a_2/2^m & -a_2^2/2^m & (a_1^2+q)/2^m & a_1 a_2/2^m \\
0 & -\mu & 0 & 0 & 0 & -a_2/2^m & -a_2/2^m & 0\\
0 & 0 & 0 & -\mu & a_2/2^m & 0 & 0 & -a2/2^m \\
-1 & -\mu a_2 & 0 & 0 & -a_1 a_2/2^m & (a_1^2+q)/2^m & -a_2^2/2^m & a_1 a_2/2^m \\
0 & 0 & -1 & -\mu a_2 & a_2^2/2^m & -a_1 a_2/2^m & -a_1 a_2/2^m & (a_1^2+q)/2^m\\
-a_1 & -\mu a_1 a_2 & a_2 & \mu a_2^2 & a_2q/2^m & 0 & 0 & -a_2 q/2^m\\
 -a_2 & -\mu a_2^2 & -a_1 & -\mu a_1 a_2 & 0 & a_2 q/2^m & a_2 q/2^m & 0
\end{array}\right)\]
and:
\[R_{\mathscr{B}'_0\times\mathscr{B}'_0,\mathscr{D}}:=\left(\begin{array}{cccc}
0 & 0 & 0 & 0\\
\mu a_2 & 0 & 1 & 0\\
\mu & 0 & 0 & 0\\
0 & 0 & 0 & 0\\
0 & 1 & 0 & -\mu a_2 \\
0 & 0 & 0 & 0\\
0 & 0 & 0 & 0\\
0 & 0 & 0 & \mu q 
\end{array}\right),\]
where $\mu \equiv 1/a_1 \mod 4$. 
\end{Lemma}

\begin{proof}
Let $(P''_1, Q''_1)$ be a basis of $E_1[2^{e+2}]$ such that $[2^{e-m}]P''_1=P'_1$ and $[2^{e-m}]Q''_1=Q'_1$, where $(P'_1,Q'_1)$ has been introduced in \cref{lemma: theta structure dim 2 end chain}, so that $[2^e]P''_1=P_1$ and $[2^e]Q''_1=Q_1$. Then, we may choose:
\[V'_1=(\Phi([a_1]P''_1-[\mu]P_1,\sigma(P''_1)),\Phi([a_2]P''_1,0)),\]
\[V'_2=(\Phi([a_1]Q''_1,\sigma(Q''_1)),\Phi([a_2]Q''_1,0)),\]
\[V'_3=(\Phi(-[a_2]P''_1,0),\Phi([a_1]P''_1,\sigma(P''_1))),\]
\[V'_4=(\Phi(-[a_2]Q''_1,0),\Phi([a_1]Q''_1-[\mu]Q_1,\sigma(Q''_1))),\]
and the $U'_i$ in a symplectic complement, so that $[2^{e-m}]U'_i=U_i$ for all $i\in\Int{1}{4}$ and (i) is satisfied. By construction of the $V'_i$, we have  $\langle [4]V'_1, \cdots, [4]V'_4\rangle=\ker(H)$ and we can also easily check that $e_{2^{e-m+2}}(V'_i, V'_j)=1$ for all $i, j\in\Int{1}{4}$. Hence, (ii)' is also satisfied.  

In addition, we have:
\[H(V'_1)=F([a_1]P''_1-[\mu]P_1,[a_2]P''_1, \sigma(P''_1),0)=(0,[\mu a_2 ]P_1,[\mu]\sigma(P_1),0)\]
\[H(V'_2)=F([a_1]Q''_1,[a_2]Q''_1, \sigma(Q''_1),0)=(Q_1,0,0,0)\]
\[H(V'_3)=F(-[a_2]P''_1,[a_1]P''_1, 0,\sigma(P''_1))=(0,P_1,0,0)\]
\[H(V'_4)=F(-[a_2]Q''_1,[a_1]Q''_1-[\mu]Q_1, 0,\sigma(P''_1))=(-[\mu a_2 ]Q_1,0,0,[\mu]Q_1).\]
The expression of $R_{\mathscr{B}'_0\times\mathscr{B}'_0,\mathscr{D}}$ follows.

Let us write $M_{\mathscr{C}_1\times \mathscr{C}_1, \mathscr{C}}:=(M_{i,j})_{1\leq i, j\leq 8}$, $M_{\mathscr{B}'_0\times\mathscr{B}'_0, \mathscr{D}'}:=(M'_{i,j})_{1\leq i, j\leq 8}$ and $\mathscr{B}'_0\times\mathscr{B}'_0:=(x_1,\cdots, x_8)$. We have chosen $\widetilde{\mathscr{C}}$ such that for all $j\in\Int{1}{4}$,
\begin{align*} [2^{e-m}]U'_j&=U_j=\sum_{i=1}^2 [M_{i,j}]([2^m]\Phi(S_i),0)+\sum_{i=1}^2 [M_{i+2,j}](0,[2^m]\Phi(S_i))\\
&\quad +\sum_{i=1}^2 [M_{i+4,j}](\Phi(T_i),0)+\sum_{i=1}^2 [M_{i+6,j}](0,\Phi(T_i))
\end{align*}
Hence, 
\begin{align*} [2^{e-m}]H(U'_j)&=\sum_{i=1}^2 [M_{i,j}]H([2^m]\Phi(S_i),0)+\sum_{i=1}^2 [M_{i+2,j}]H(0,[2^m]\Phi(S_i))\\
&\quad +\sum_{i=1}^2 [M_{i+4,j}]H(\Phi(T_i),0)+\sum_{i=1}^2 [M_{i+6,j}]H(0,\Phi(T_i))\\
&=\sum_{i=1}^8 [M_{i,j}]\mathscr{D}'_i=\sum_{i=1}^8 [M_{i,j}]\sum_{l=1}^8 [M'_{l,i}]x_l\\
&= \sum_{l=1}^8 \left[\sum_{i=1}^8 M'_{l,i}M_{i,j}\right]x_l=\sum_{l=1}^8 [(M_{\mathscr{B}'_0\times\mathscr{B}'_0, \mathscr{D}'}\cdot L_{\mathscr{C}_1\times \mathscr{C}_1, \mathscr{C}})_{l,j}]x_l
\end{align*}
The expression of $M_{\mathscr{B}'_0\times\mathscr{B}'_0,\mathscr{D}}$ follows.

Finally, to obtain the expression of $M_{\mathscr{B}'_0\times\mathscr{B}'_0, \mathscr{D}'}$ it suffices to evaluate $F$ on several points, given that $F$ is determined by $H$ and $\Phi$ and given the expression of $S_1, S_2, T_1, T_2$ in \cref{lemma: theta structure dim 2 end chain}.
\end{proof}

\begin{algorithm}[tbh]
\SetAlgoLined
\KwData{\justifying Integer parameters $a_1, a_2, q, m=v_2(a_2)$, the left $4\times 8$ block $L_{\mathscr{C}_1\times \mathscr{C}_1, \mathscr{C}}$ of the change of basis matrix from $\mathscr{C}_1\times \mathscr{C}_1$ to $\mathscr{C}$ (as defined in \cref{sec: change of basis dim 4 full}), the change of basis matrices $M_i$ from $(\alpha_i,\beta_i)$ to $(P_i,Q_i)$ (as defined in \cref{lemma: change of basis dim 2} for $i\in\{1,2\}$) and the Weil pairing $\zeta_4=e_4(P_1,Q_1)$.}
\KwResult{\justifying A change of coordinates matrix from $\Theta'_{\mL_0}$ to $\Theta_{\mL_0}$-coordinates, where  $\Theta'_{\mL_0}$ is the theta structure induced by $\mathscr{D}$ \cref{eq: D} on $(E_1^2\times E_2^2,\mL_0)$ and $\Theta_{\mL_0}$ is the product theta structure on $(E_1^2\times E_2^2,\mL_0)$ induced by $(\alpha_i, \beta_i)$ ($i\in\{1,2\}$).}

$\mu\longleftarrow 1/a_1 \mod 4$\;
Using $\mu, a_1, a_2, q, m$, compute the matrices $M_{\mathscr{B}'_0\times\mathscr{B}'_0, \mathscr{D}'}$ and $R_{\mathscr{B}'_0\times\mathscr{B}'_0,\mathscr{D}}$ of \cref{lemma: splitting dim 4}\;

$M_{\mathscr{B}'_0\times\mathscr{B}'_0,\mathscr{D}}\longleftarrow(M_{\mathscr{B}'_0\times\mathscr{B}'_0, \mathscr{D}'}\cdot L_{\mathscr{C}_1\times \mathscr{C}_1, \mathscr{C}}|R_{\mathscr{B}'_0\times\mathscr{B}'_0,\mathscr{D}})$\;

$M_0\longleftarrow \left(\begin{array}{cccccccc}
M_{1,1,1} & 0 & 0 & 0 & M_{1,1,2} & 0 & 0 & 0\\
0 & M_{1,1,1} & 0 & 0 & 0 & M_{1,1,2} & 0 & 0\\
0 & 0 & M_{2,1,1} & 0 & 0 & 0 & M_{2,1,2} & 0\\
0 & 0 & 0 & M_{2,1,1} & 0 & 0 & 0 & M_{2,1,2}\\
M_{1,2,1} & 0 & 0 & 0 & M_{1,2,2} & 0 & 0 & 0\\
0 & M_{1,2,1} & 0 & 0 & 0 & M_{1,2,2} & 0 & 0\\
0 & 0 & M_{2,2,1} & 0 & 0 & 0 & M_{2,2,2} & 0\\
0 & 0 & 0 & M_{2,2, 1} & 0 & 0 & 0 & M_{2,2,2}
\end{array}\right)$\;

$M\longleftarrow (M_0\cdot M_{\mathscr{B}'_0\times\mathscr{B}'_0,\mathscr{D}})^{-1}$\;

Apply the formulas of \cref{thm: symplectic change of basis} to $M$ and $\zeta_4$ to compute a change of coordinates matrix $N$ from $\Theta'_{\mL_0}$ to $\Theta_{\mL_0}$-coordinates\;

\KwRet $N$\;

\caption{Splitting change of theta coordinates (from dimension 4 to 1).}\label{alg: splitting change of basis dim 4}
\end{algorithm}

\begin{algorithm}[tbh]
\SetAlgoLined
\KwData{\justifying Product $\Theta_{\mL_0}$-coordinates $(\theta^{\mL_0}_\textbf{i}(Q))_{\textbf{i}}$ of a point $Q\in E_1^2\times E_2^2$, where $\Theta_{\mL_0}$ is the product theta structure on $(E_1^2\times E_2^2,\mL_0)$ induced by $(\alpha_l, \beta_l)$ (as defined in \cref{lemma: change of basis dim 2} for $l\in\{1,2\}$), and the Montgomery $(x:z)$-coordinates of $\alpha_l$ for $l\in\{1,2\}$.}
\KwResult{\justifying Montgomery $(x:z)$-coordinates of $Q$.}
Parse $\alpha_l:=(r_l:s_l)$ in Montgomery $(x:z)$-coordinates for $l\in\{1,2\}$\;
$a_l, b_l\longleftarrow r_l+s_l, r_l-s_l$ for $l\in\{1,2\}$\;
\For{$j=1$ \KwTo $4$}{
\eIf{$\exists \, \textbf{i}\in(\Z/2\Z)^4, \ i_j=1 \wedge \theta^{\mL_0}_{\textbf{i}}(Q)\neq 0$}{
Find such $\textbf{i}\in(\Z/2\Z)^4$\;
Let $\textbf{i}'\in(\Z/2\Z)^4$ such that $i'_l=i_l$ if $l\neq j$ and $i'_j=0$\;
$\theta_0(Q_j)\longleftarrow \theta^{\mL_0}_{\textbf{i}'}(Q)/\theta^{\mL_0}_{\textbf{i}}(Q)$, $\theta_1(Q_j)\longleftarrow 1$\;
}{
$\theta_0(Q_j)\longleftarrow 1$, $\theta_1(Q_j)\longleftarrow 0$\;
}
$x(Q_j), z(Q_j) \longleftarrow a_{\lceil j/2\rceil}\theta_1(Q_j)+b_{\lceil j/2\rceil}\theta_0(Q_j),a_{\lceil j/2\rceil}\theta_1(Q_j)-b_{\lceil j/2\rceil}\theta_0(Q_j)$\;
}

\KwRet $(x(Q_1):z(Q_1)), \cdots, (x(Q_4):z(Q_4))$\;

\caption{Product theta coordinates to Montgomery coordinates (in dimension 4).}\label{alg: product to Montgomery dim 4}
\end{algorithm}

From \cref{lemma: splitting dim 4}, we derive Algorithm \ref{alg: splitting change of basis dim 4} to compute the the change of coordinates matrix from the theta coordinates associated to the theta structure $\Theta'_{\mL_0}$ induced by $\mathscr{D}$ to the product $\Theta_{\mL_0}$-coordinates on $E_1^2\times E_2^2$. 

Then, we can evaluate $F$ in $(x:z)$-Montgomery coordinates. If $P\in E_1^2\times E_2^2$ and if we have computed the $\Theta'_{\mL_0}$-coordinates $(\theta'^{\mL_0}_\textbf{i}(F(P))))_{\textbf{i}}$, then we apply the change of coordinates matrix of theta coordinates $N$ outputted by Algorithm \ref{alg: splitting change of basis dim 4} to obtain the product $\Theta_{\mL_0}$-coordinates $(\theta^{\mL_0}_\textbf{i}(F(P))))_{\textbf{i}}$. Then, Algorithm \ref{alg: product to Montgomery dim 4} transforms these coordinates into the $(x:z)$-Montgomery coordinates of $F(P)$.

The underlying idea of Algorithm \ref{alg: product to Montgomery dim 4} is the following. Let $Q:=F(P)$, Then, we may write $Q:=(Q_1,\cdots, Q_4)$ and by \cref{lemma: product theta structure}, $\theta^{\mL_0}_\textbf{i}(x)=\prod_{j=1}^4 \theta_{i_j}(Q_j)$ for all $\textbf{i}:=(i_1,\cdots, i_4)\in (\Z/2\Z)^4$, where the theta coordinates $\theta_{i_j}(Q_j)$ are the level $2$ theta coordinates of $E_{\lceil j/2\rceil}$ (induced by the basis $(\alpha_{\lceil j/2\rceil},\beta_{\lceil j/2\rceil})$ from \cref{lemma: change of basis dim 2}). Assuming $\theta_{1}(Q_j)\neq 0$, then we can find $\textbf{i}\in (\Z/2\Z)^4$ such that $i_j=1$ and $\theta^{\mL_0}_\textbf{i}(Q_j)\neq 0$ and we have $\theta_{0}(Q_j)/\theta_{1}(Q_j)=\theta^{\mL_0}_{\textbf{i}'}(Q)/\theta^{\mL_0}_{\textbf{i}}(Q)$, where $\textbf{i}'\in (\Z/2\Z)^4$ satisfies $i'_l=i_l$ for all $l\neq j$ and $i'_j=0$. This way, we can compute $(\theta_{0}(Q_j):\theta_{1}(Q_j))$ for all $j\in\Int{1}{4}$. If we denote by $(a_l, b_l)$ the theta null point of $E_l$ for $l\in\{1,2\}$, we can then write $Q_j:=(a_{\lceil j/2\rceil}\theta_1(Q_j)+b_{\lceil j/2\rceil}\theta_0(Q_j):a_{\lceil j/2\rceil}\theta_1(Q_j)-b_{\lceil j/2\rceil}\theta_0(Q_j))$ in Montgomery $(x:z)$-coordinates for all $j\in\Int{1}{4}$.

\newpage

\section{Computing 4 dimensional isogenies derived from Kani's lemma with half available torsion}\label{sec: appendix half torsion}

Throughout this section, we keep the notations of \cref{sec: application to SQIsignHD}. Unlike in \cref{sec: basis change with full torsion}, we assume that all the $2^{e+2}$-torsion is not available. Namely, we can access the $2^{e'+2}$-torsion with $e'\geq e/2$. We proceed as explained in \cref{sec: cutting in two} to compute $F$ in two parts $F_1: E_1^2\times E_2^2\longrightarrow \mC$ and $\widetilde{F}_2: E_1^2\times E_2^2\longrightarrow \mC$ such that $F=F_2\circ F_1$, $F_1$ being a $2^{e_1}$-isogeny and $F_2$ being a $2^{e_2}$-isogeny with $e=e_1+e_2$ and $m\leq e_1, e_2\leq e'$.

Since $\ker(F_1)=\ker(F)[2^{e_1}]$ and $\ker(\widetilde{F}_2)=\ker(\widetilde{F})[2^{e_2}]$, \cref{lemma: first gluing isogenies} applies to $F_1$ and an analogue of \cref{lemma: first gluing isogenies} applies to $\widetilde{F}_2$.

\begin{Lemma}\label{lemma: first gluing isogenies F_2}
Assume that $2|a_2$ and let $m:=v_2(a_2)$ be its $2$-adic valuation. Then $\widetilde{F}_2=g_{e_2}\circ\cdots\circ g_1$, with
\[E_1^2\times E_2^2\overset{g_1}{\relbar\joinrel\relbar\joinrel\longrightarrow} {A'}_1^2 \quad \cdots \quad {A'}_{m-1}^2 \overset{g_m}{\relbar\joinrel\relbar\joinrel\longrightarrow} {A'}_m^2\overset{g_{m+1}}{\relbar\joinrel\relbar\joinrel\longrightarrow}B',\]
a chain of $2$-isogenies, where the $A'_i$ are abelian surfaces and $B'$ is an abelian variety of dimension 4. For all $i\in\Int{2}{m}$, $g_i:=(\psi_i,\psi_i)$, with $\psi_i: A'_{i-1}\longrightarrow A'_i$ and $g_1: (R_1,S_1,R_2,S_2)\longmapsto (\psi_1(R_1,R_2),\psi_1(S_1,S_2))$, with $\psi_1: E_1\times E_2\longrightarrow A'_1$. In addition,
\[\ker(\psi_m\circ\cdots\circ \psi_1)=\{([a_1]P,-\sigma(P))\mid P\in E_1[2^m]\}.\]
\end{Lemma}

\begin{proof}
We proceed as in the proof of \cref{lemma: first gluing isogenies}. We have:
\[\ker(\widetilde{F})=\{([a_1]P+[a_2]Q,-[a_2]P+[a_1]Q,-\sigma(P),-\sigma(Q))\mid P, Q\in E_1[2^e]\}.\]
Let $g_1, \cdots, g_{m+1}$ be the $m+1$ first elements of the $2$-isogeny chain $\widetilde{F}$. Then, since $a_2\equiv 0 \mod 2^m$, we have
\[\ker(g_m\circ\cdots\circ g_1)=[2^{e-m}]\ker(\widetilde{F})=K_1\oplus K_2,\]
where $K_1:=\{([a_1]P,0,-\sigma(P),0)\mid P\in E_1[2^m]\}$ and $K_2:=\{(0,[a_1]P,0,-\sigma(P))\mid P\in E_1[2^m]\}$. This proves the chain $g_m\circ\cdots\circ g_1$ has the desired form. 
\end{proof}

Hence, we can compute $\widetilde{F}_2$ exactly as we would compute $F_1$ (or $F$ with the full torsion available). We give a detailed algorithmic overview of the computation of $F$ in \cref{alg: Kani endo half torsion}. 

\begin{algorithm}
\newcommand\mycommfont[1]{\textcolor{blue}{#1}}
\SetCommentSty{mycommfont}
\SetAlgoLined
\KwData{\justifying $a_1, a_2, q$ such that $a_2$ is even, $q$ is odd and $a_1^2+a_2^2+q=2^e$, two supersingular elliptic curves $E_1$ and $E_2$ defined over $\F_{p^2}$, $(P''_1,Q''_1)$ a basis of $E_1[2^{e'+2}]$ with $e'\geq e/2$, $(\sigma(P''_1),\sigma(Q''_1))$ for some $q$-isogeny $\sigma: E_1\longrightarrow E_2$ and two $4$-torsion basis $(\alpha_i,\beta_i)$ of $E_i$ for $i\in\{1,2\}$.}
\KwResult{\justifying A chain representation of the isogeny $F\in\End(E_1^2\times E_2^2)$ given by \cref{eq: definition F}.}
\justifying
$m\longleftarrow v_2(a_2)$, $e_1\longleftarrow \lceil e/2\rceil$ and $e_2\longleftarrow e-e_1$, $r\longleftarrow 1/q \mod 2^{e'+2}$, $\mu\longleftarrow 1/a_1 \mod 2^{e'+2}$\;
(Pre)compute an optimal strategy $S$ with $m$ leaves \cite[Algorithm~60]{SIKE_spec}\;
(Pre)compute optimal strategies $S_i$ ($i\in\{1,2\}$) with $e_i-m$ leaves with more weight at the beginning (see \cref{sec: constrained optimal strategies})\;

\tcc{Step 1: Two symplectic basis inducing dual theta structures on $\mC$}

$\mathscr{B}_0\longleftarrow ((P''_1,0,0,0),(0,P''_1,0,0),(0,0,\sigma(P''_1),0),(0,0,0,\sigma(P''_1)), (Q''_1,0,0,0),$ $(0,Q''_1,0,0),(0,0,[r]\sigma(Q''_1),0),(0,0,0,[r]\sigma(Q''_1))$\;

Compute symplectic basis of $(E_1^2\times E_2^2)[2^{e'+2}]$, $\mathscr{B}_1$ and $\mathscr{B}_2$ satisfying the conditions of \cref{lemma: matching symplectic basis} and the symplectic change of basis matrices $M_1$ and $M_2$ from $\mathscr{B}_0$ to $\mathscr{B}_1$ and $\mathscr{B}_2$ using \cref{alg: meet in the middle basis}\;

\tcc{Step 2.a: First $m$ isogenies of $F_1$ in dimension 2}

$c\longleftarrow e'-m$, $P'_1, Q'_1, P'_2, R'_2\longleftarrow [2^c]P''_1, [2^c]Q''_1, [2^c]\sigma(P''_1), [2^c]\sigma(Q''_1)$\;
$P_1, Q_1, P_2, Q_2\longleftarrow [2^{m}]P'_1, [2^{m}]Q'_1, [2^{m}]P'_2, [r 2^{m}]R'_2$\;
$\zeta_4\longleftarrow e_4(P_1,Q_1)$\;
$T_1, T_2\longleftarrow [2]([a_1]P'_1-[a_2]Q'_1,P'_2), [2]([a_2]P'_1+[a_1]Q'_1,R'_2)$\;
For $i\in\{1,2\}$, compute a basis $(\alpha_i,\beta_i)$ of $E_i[4]$ such that $\beta_i=(-1:1)$ in $(x:z)$-Montgomery coordinates and $e_4(\alpha_i,\beta_i)=\zeta_4$\;

Compute the change of basis matrices $M_i$ from $(\alpha_i,\beta_i)$ to $(P_i,Q_i)$ for $i\in\{1,2\}$\;

Find a theta structure $\Theta'_\mL$ on $(E_1\times E_2,\mL)$ such that $K_2(\Theta'_\mL)=[2^{m+1}]\langle T_1, T_2\rangle$ and compute the change of coordinates matrix $N_{12}$ from $(x:z)$ to $\Theta'_\mL$-coordinates (using Algorithm \ref{alg: change of basis dim 2} with input $a_1, a_2, q, (\alpha_i:\beta_i), M_i, \zeta_4$)\;

$({\theta'}_\textbf{j}^{\mL}(T_i))_\textbf{j} \longleftarrow N_{12}\cdot \,^t (x_1(T_i)x_2(T_i), x_1(T_i)z_2(T_i), z_1(T_i)x_2(T_i), z_1(T_i)z_2(T_i))$ for $i\in\{1,2\}$\;

Use the coordinates $({\theta'}_{\textbf{j}}^{\mL}(T_i))_{\textbf{j}}$ and strategy $S$ to compute a $2$-dimensional $2$-isogeny chain $\Phi:=\varphi_m\circ\cdots\circ\varphi_1$ of kernel $\ker(\Phi)=[4]\langle T_1, T_2\rangle$ (see \cite{Theta_dim2})\;

\tcc{Step 2.b: Gluing isogeny $f_{m+1}$ of $F_1$ in dimension 4}

Parse $X_1, \cdots, X_4,Y_1, \cdots, Y_4\longleftarrow \mathscr{B}_1$ and let $G: (P_1, \cdots, P_4)\longmapsto (\Phi(P_1, P_3), \Phi(P_2, P_4))$\;

Compute $Y'_i\longleftarrow [2^{e'-m-1}]G(Y_i)$ for all $i\in\Int{1}{4}$ and $Y'_5\longleftarrow [2^{e'-m-1}]G(Y_1+Y_2)$\;

Let $\Theta_{\mL_m}$ be the level $2$ theta structure on the codomain $(A_m,\mL_m)$ of $\Phi$ and $\Theta_{\mM_m}:=\Theta_{\mL_m}\times\Theta_{\mL_m}$\;

\caption{Computation of a 4 dimensional endomorphism derived from Kani's lemma with half available torsion.}\label{alg: Kani endo half torsion}
\end{algorithm}\begin{algorithm}
\newcommand\mycommfont[1]{\textcolor{blue}{#1}}
\SetCommentSty{mycommfont}
\LinesNumbered
\setcounter{AlgoLine}{17}

\justifying

Find a theta structure $\Theta'_{\mM_m}$ on $(A_m^2,\mM_m)$ such that $K_2(\Theta'_{\mM_m})=[4]\langle Y'_1, \cdots, Y'_4\rangle$ and compute the change of coordinates matrix $N_{24}$ from $\Theta_{\mM_m}$ to $\Theta'_{\mM_m}$ (using \cref{lemma: gluing change of basis F_1} and \cref{thm: symplectic change of basis})\;

$({\theta'}_\textbf{j}^{\mM_m}(Y'_i))_\textbf{j}\longleftarrow N_{24}\cdot (\theta_\textbf{j}^{\mM_m}(Y'_i))_\textbf{j}$ for $i\in\Int{1}{5}$\;

Using the $({\theta'}_\textbf{j}^{\mM_m}(Y'_i))_\textbf{j}$ for $i\in\Int{1}{5}$, compute $f_{m+1}$ of kernel $[4]\langle Y'_1,\cdots, Y'_4\rangle$ (Algorithm \ref{alg: dual theta null point} and \cref{rem: additional kernel points gluing})\;

\tcc{Step 2.c: Last $e_1-m-1$ isogenies of $F_1$ in dimension 4}

$Y''_i\longleftarrow [2^{e'-e_1}] f_{m+1}\circ G(Y_i)$ for all $i\in\Int{1}{4}$\;

Use \cref{alg: apply strategy dim 4} with input $Y''_1, \cdots, Y''_4$ and strategy $S_1$ to compute a $4$-dimensional $2$-isogeny chain $f_{e_1}\circ\cdots\circ f_{m+2}$ of kernel $[4]\langle Y''_1,\cdots, Y''_4\rangle$\;

\tcc{Step 3.a: First $m$ isogenies of $\widetilde{F}_2$ in dimension 2}

$\widetilde{T}_1, \widetilde{T}_2\longleftarrow [2]([a_1]P'_1+[a_2]Q'_1,-P'_2), [2](-[a_2]P'_1+[a_1]Q'_1,-R'_2)$\;

Find a theta structure $\Theta''_\mL$ on $(E_1\times E_2,\mL)$ such that $K_2(\Theta''_\mL)=[2^{m}]\langle \widetilde{T}_1, \widetilde{T}_2\rangle$ and compute the change of coordinates matrix $\widetilde{N}_{12}$ from $(x:z)$ to $\Theta''_\mL$ (using \cref{alg: change of basis dim 2} with the matrix $M_r$ of \cref{sec: change of basis dim 2 half})\;

$({\theta''}_\textbf{j}^{\mL}(\widetilde{T}_i))_\textbf{j} \longleftarrow \widetilde{N}_{12}\cdot \,^t(x_1(\widetilde{T}_i)x_2(\widetilde{T}_i), x_1(\widetilde{T}_i)z_2(\widetilde{T}_i), z_1(\widetilde{T}_i)x_2(\widetilde{T}_i), z_1(\widetilde{T}_i)z_2(\widetilde{T}_i))$ for $i\in\{1,2\}$\;

Use the coordinates $({\theta''}_\textbf{j}^{\mL}(\widetilde{T}_i))_\textbf{j}$ and strategy $S$ to compute a $2$-dimensional $2$-isogeny chain $\Psi:=\psi_m\circ\cdots\circ\psi_1$ of kernel $\ker(\Psi)=[4]\langle \widetilde{T}_1, \widetilde{T}_2\rangle$ (see \cite{Theta_dim2})\;

\tcc{Step 3.b: Gluing isogeny $g_{m+1}$ of $\widetilde{F}_2$ in dimension 4}

Parse $U_1, \cdots, U_4,V_1, \cdots, V_4\longleftarrow \mathscr{B}_1$ and let $H: (P_1, \cdots, P_4)\longmapsto (\Psi(P_1, P_3), \Psi(P_2, P_4))$\;

Compute $V'_i\longleftarrow [2^{e'-m-1}]H(V_i)$ for all $i\in\Int{1}{4}$ and $V'_5\longleftarrow [2^{e'-m-1}]H(V_1+V_2)$\;

Let $\Theta_{\mL'_m}$ be the level $2$ theta structure on the codomain $(A'_m,\mL'_m)$ of $\Psi$ and $\Theta_{\mM'_m}:=\Theta_{\mL'_m}\times\Theta_{\mL'_m}$\;

Find a theta structure $\Theta'_{\mM'_m}$ on $({A'}_m^2,\mM'_m)$ such that $K_2(\Theta'_{\mM'_m})=[4]\langle V'_1, \cdots, V'_4\rangle$ and compute the change of coordinates matrix $\widetilde{N}_{24}$ from $\Theta_{\mM'_m}$ to $\Theta'_{\mM'_m}$-coordinates (using \cref{lemma: gluing change of basis F_2} and \cref{thm: symplectic change of basis})\;

$({\theta'}_\textbf{j}^{\mM'_m}(V'_i))_\textbf{j}\longleftarrow \widetilde{N}_{24}\cdot (\theta_\textbf{j}^{\mM'_m}(V'_i))_\textbf{j}$ for $i\in\Int{1}{5}$\;

Using the $({\theta'}_\textbf{j}^{\mM'_m}(V'_i))_\textbf{j}$ for $i\in\Int{1}{5}$, compute $g_{m+1}$ of kernel $[4]\langle V'_1,\cdots, V'_4\rangle$ (Algorithm \ref{alg: dual theta null point} and \cref{rem: additional kernel points gluing})\;

\tcc{Step 3.c: Last $e_2-m-1$ isogenies of $\widetilde{F}_2$ in dimension 4}

$V''_i\longleftarrow [2^{e'-e_1}] g_{m+1}\circ H(V_i)$ for all $i\in\Int{1}{4}$\;

Use \cref{alg: apply strategy dim 4} with input $V''_1, \cdots, V''_4$ and strategy $S_2$ to compute a $4$-dimensional $2$-isogeny chain $g_{e_2}\circ\cdots\circ g_{m+2}$ of kernel $[4]\langle V''_1,\cdots, V''_4\rangle$\;

\tcc{Step 4: Computing $F_2=\widetilde{\widetilde{F}}_2$}

Compute $\widetilde{\psi}_1, \cdots, \widetilde{\psi}_m, \widetilde{g}_1, \cdots, \widetilde{g}_{e_2}$ using \cref{lemma: dual isogeny computation}\;

\KwRet $\varphi_1,\cdots, \varphi_m, f_1, \cdots, f_{e_1}, \widetilde{\psi}_1, \cdots, \widetilde{\psi}_m, \widetilde{g}_1, \cdots, \widetilde{g}_{e_2}, N_{12}, N_{24}, \widetilde{N}_{24}^{-1}, \widetilde{N}_{12}^{-1}$\;

\end{algorithm}
\begin{algorithm}
\newcommand\mycommfont[1]{\textcolor{blue}{#1}}
\SetCommentSty{mycommfont}
\SetAlgoLined
\KwData{\justifying A chain $C$ outputted by Algorithm \ref{alg: Kani endo half torsion} representing $F\in\End(E_1^2\times E_2^2)$ given by \cref{eq: definition F},  and a point $Q\in E_1^2\times E_2^2$.}
\KwResult{\justifying The Montgomery $(x:z)$-coordinates of $F(Q)$.}

\justifying
Parse $C$ as $\varphi_1,\cdots, \varphi_m, f_1, \cdots, f_{e_1}, \widetilde{\psi}_1, \cdots, \widetilde{\psi}_m, \widetilde{g}_1, \cdots, \widetilde{g}_{e_2}, N_{12}, N_{24}, \widetilde{N}_{24}^{-1},$ $\widetilde{N}_{12}^{-1}$\;
$v\longleftarrow \,^t(x(Q_1)x(Q_3), x(Q_1)z(Q_3), z(Q_1)x(Q_3), z(Q_1)z(Q_3))$\;
$({\theta'}_\textbf{i}^{\mL}(Q_1,Q_3))_\textbf{i}\longleftarrow N_{12}\cdot v$\;
$(\theta_\textbf{i}^{\mL_m}(R_1))_\textbf{i}\longleftarrow \varphi_m\circ\cdots\circ\varphi_1(({\theta'}_\textbf{i}^{\mL}(Q_1,Q_3))_\textbf{i})$\;
$w\longleftarrow \,^t(x(Q_2)x(Q_4), x(Q_2)z(Q_4), z(Q_2)x(Q_4), z(Q_2)z(Q_4))$\;
$({\theta'}_\textbf{i}^{\mL}(Q_2,Q_4))_\textbf{i}\longleftarrow N_{12}\cdot w$\;
$(\theta_\textbf{i}^{\mL_m}(R_2))_\textbf{i}\longleftarrow \varphi_m\circ\cdots\circ\varphi_1(({\theta'}_\textbf{i}^{\mL}(Q_1,Q_3))_\textbf{i})$\;
$\theta_{\textbf{i}_1,\textbf{i}_2}^{\mM_m}(R)\longleftarrow\theta_{\textbf{i}_1}^{\mL_m}(R_1)\cdot \theta_{\textbf{i}_2}^{\mL_m}(R_2)$ for $\textbf{i}_1, \textbf{i}_2\in (\Z/2\Z)^2$\;
$({\theta'}_\textbf{i}^{\mM_m}(R))_\textbf{i}\longleftarrow N_{24}\cdot (\theta_\textbf{i}^{\mM_m}(R))_\textbf{i}$\;
$({\theta'}_\textbf{i}^{\mM'_m}(S_1,S_2))_\textbf{i}\longleftarrow \widetilde{g}_{m+1}\circ\cdots\circ\widetilde{g}_{e_2}\circ f_{e_1}\circ\cdots\circ f_{m+1}(({\theta'}_\textbf{i}^{\mM_m}(R))_\textbf{i})$\;
$(\theta_{\textbf{i}}^{\mM'_m}(S_1,S_2))_\textbf{i}\longleftarrow \widetilde{N}_{24}^{-1}\cdot ({\theta'}_\textbf{i}^{\mM'_m}(S))_\textbf{i}$\;
Use \cref{alg: splitting product dim 4 to dim 2} with input $(\theta_{\textbf{i},\textbf{j}}^{\mM'_m}(S_1, S_2))_{\textbf{i},\textbf{j}\in(\Z/2\Z)^2}$ to obtain $(\theta_\textbf{i}^{\mL'_m}(S_1))_{\textbf{i}\in (\Z/2\Z)^2}$ and $(\theta_\textbf{i}^{\mL'_m}(S_2))_{\textbf{i}\in (\Z/2\Z)^2}$\;
$({\theta''}_\textbf{j}^{\mL}(T_i))_{\textbf{j}}\longleftarrow \widetilde{\psi}_1\circ\cdots\circ\widetilde{\psi}_m((\theta_\textbf{j}^{\mL'_m}(S_i))_{\textbf{j}})$ for all $i\in\{1,2\}$\;
$(x_1(F(Q))x_3(F(Q)), x_1(F(Q))z_3(F(Q)), z_1(F(Q))x_3(F(Q)),$ $z_1(F(Q))z_3(F(Q)))\longleftarrow \widetilde{N}_{12}^{-1}\cdot ({\theta''}_\textbf{j}^{\mL}(T_1))_{\textbf{j}}$\;
$(x_2(F(Q))x_4(F(Q)), x_2(F(Q))z_4(F(Q)), z_2(F(Q))x_4(F(Q)),$ $z_2(F(Q))z_4(F(Q)))\longleftarrow \widetilde{N}_{12}^{-1}\cdot ({\theta''}_\textbf{j}^{\mL}(T_2))_{\textbf{j}}$\;
Use \cref{alg: splitting dim 2 to dim 1} to recover $(x_1(F(Q)):z_1(F(Q))),\cdots, (x_4(F(Q)):z_4(F(Q)))$\;

\KwRet $(x_1(F(Q)):z_1(F(Q))),\cdots, (x_4(F(Q)):z_4(F(Q)))$\;

\caption{Evaluation of a 4 dimensional endomorphism derived from Kani's lemma with half available torsion given its representation.}\label{alg: Kani endo half torsion eval}
\end{algorithm}

\subsection{Change of coordinates in dimension 2 with half available torsion}\label{sec: change of basis dim 2 half}

By \cref{lemma: first gluing isogenies F_2}, we can compute $\widetilde{F}_2$ exactly as we would compute $F_1$ (or $F$ with the full torsion available). We start by computing a chain of $m$ $2$-isogenies $\psi_1, \cdots, \psi_m$ in dimension $2$. To compute $\psi_1$, we have to compute a theta structure $\Theta'_\mL$ satisfying $K_2(\Theta'_\mL)=\ker(\psi_1)$. $\Theta'_\mL$ is induced by the symplectic basis:
\[((0,P_2),([\mu]P_1,-[\mu a_2]P_2),([a_1]P_1+[a_2]Q_1,-P_2),(-[a_2]P_1+[a_1]Q_1,-[q]Q_2)).\]
where $(P_1, Q_1)$ is a basis of $E_1[4]$ and $(P_2, Q_2):=(\sigma(P_1),[r]\sigma(Q_1))$, with $rq \equiv 1 \mod 4$ and $\mu a_1\equiv 1 \mod 4$. Hence, the change of coordinates matrix from $(x:z)$ to $\Theta'_\mL$ can be computed with \cref{alg: change of basis dim 2} where the matrix $M_r$ in \cref{lemma: change of basis dim 2} is replaced by:
\[M_r:=\left(\begin{array}{cccc}
0 & \mu & a_1 & -a_2\\
0 & 0 & -1 & 0\\
0 & 0 & a_2 & a_1 \\
1 & -\mu a_2 & 0 & -q
\end{array}\right).\]

\newpage

\subsection{Change of coordinates in dimension 4 with half available torsion}\label{sec: change of basis dim 4 half}

\subsubsection{Finding two symplectic basis inducing dual theta structures on $\mC$}

Our goal is to find symplectic basis of $(E_1^2\times E_2^2)[2^{e'+2}]$ inducing via $F_1$ and $\widetilde{F}_2$ the same level 2 theta structure on their common codomain $\mC$ (up to a Hadamard transform). This is explained in the following lemma:

\begin{Lemma}\cite[Corollary 58]{SQISignHD_autocite}\label{lemma: matching symplectic basis}
Consider two $\zeta$-symplectic basis of $(E_1^2\times E_2^2)[2^{e'+2}]$, $(X_1,\cdots, X_4, Y_1, \cdots, Y_4)$ and $(U_1, \cdots, U_4, V_1, \cdots, V_4)$ such that:
\begin{enumerate}[label=(\roman*)]
\item $\ker(F_1)=[c_1]\langle Y_1,\cdots, Y_4\rangle$ and $\ker(\tilde{F}_2)=[c_2]\langle V_1,\cdots, V_4\rangle$;
\item $[c_2]F(X_l)=-[c_2]V_l$ and $[c_1]\widetilde{F}(U_l)=[c_1]Y_l$ for all $l\in\Int{1}{4}$;
\end{enumerate}
where $c_i:=2^{e'+2-e_i}$ for $i\in\{1,2\}$. Then the symplectic basis of $\mC[4]$:
\[\mathscr{B}_1:=([2^{e'}]F_1(X_1),\cdots, [2^{e'}]F_1(X_4), [c_1]F_1(Y_1), \cdots, [c_1]F_1(Y_4))\]
\[\mathscr{B}_2:=([2^{e'}]\widetilde{F}_2(U_1),\cdots, [2^{e'}]\widetilde{F}_2(U_4), [c_2]\widetilde{F}_2(V_1), \cdots, [c_2]\widetilde{F}_2(V_4))\]
induced by $F_1$ and $\widetilde{F}_2$ respectively are related by the standard symplectic matrix $\mathscr{B}_2=J\cdot\mathscr{B}_1$, with: 
\[J:=\left(\begin{array}{cc}
0 & -I_4\\
I_4 & 0 
\end{array}\right)\in\Sp_8(\Z/4\Z).\]
In particular, the theta coordinates induced by $F_1$ and $\widetilde{F}_2$ on $\mC$ are the dual of each other \emph{i.e.} related by a Hadamard transform.
\end{Lemma}

\sloppy
To find symplectic basis $\mathscr{B}_1:=(X_1,\cdots, X_4, Y_1, \cdots, Y_4)$ and $\mathscr{B}_2:=(U_1, \cdots, U_4, V_1, \cdots, V_4)$ of $(E_1^2\times E_2^2)[2^{e'+2}]$ as in \cref{lemma: matching symplectic basis}, the idea is to compute a basis of $\ker(F_1)$, find a symplectic complement $(X_1,\cdots, X_4)$, evaluate $F(X_1), \cdots, F(X_4)$, find a symplectic complement $(U_1,\cdots, U_4)$, and finally evaluate $\widetilde{F}(U_1), \cdots, \widetilde{F}(U_4)$. We then set $V_l:=-F(X_l)$ and $Y_l:=\widetilde{F}(U_l)$ for all $l\in\Int{1}{4}$. Evaluating $F$ and $\widetilde{F}$ can be easily done with the knowledge of $\sigma(E_1[2^{e'+2}])$, $q$, $a_1$ and $a_2$. All these operations can be done with linear algebra computations over $\Z/2^{e'+2}\Z$. We summarize them in \cref{alg: meet in the middle basis}.

\begin{algorithm}
\SetAlgoLined
\KwData{\justifying $a_1, a_2, q$ such that $a_2$ is even, $q$ is odd and $a_1^2+a_2^2+q=2^e$, two supersingular elliptic curves $E_1$ and $E_2$ defined over $\F_{p^2}$, $(P''_1,Q''_1)$ a basis of $E_1[2^{e'+2}]$, $(\sigma(P''_1),\sigma(Q''_1))$ for some $q$-isogeny $\sigma: E_1\longrightarrow E_2$.} 
\KwResult{\justifying  Two $\zeta$-symplectic basis of $(E_1^2\times E_2^2)[2^{e'+2}]$, $\mathscr{B}_1$ and $\mathscr{B}_2$ satisfying the conditions of \cref{lemma: matching symplectic basis}, with $\zeta:=e_{2^{e'+2}}(P''_1,Q''_1)$ and the symplectic change of basis matrices from $\mathscr{B}_0$ defined on \cref{line: def basis B_0} to $\mathscr{B}_1$ and $\mathscr{B}_2$.}

$r\longleftarrow 1/q \mod 2^{e'+2}$\;

$\mathscr{B}_0\longleftarrow ((P''_1,0,0,0),(0,P''_1,0,0),(0,0,\sigma(P''_1),0),(0,0,0,\sigma(P''_1)), (Q''_1,0,0,0),$ $(0,Q''_1,0,0),(0,0,[r]\sigma(Q''_1),0),(0,0,0,[r]\sigma(Q''_1))$\;\label{line: def basis B_0}

$C_1\longleftarrow\left(\begin{array}{cccc}
a_1 & 0 & -a_2 & 0\\
a_2 & 0 & a_1 & 0\\
1 & 0 & 0 & 0 \\
0 & 0 & 1 & 0
\end{array}\right)$ and $D_1\longleftarrow \left(\begin{array}{cccc}
0 & a_1 & 0 & -a_2 \\
0 & a_2 & 0 & a_1\\
0 & q & 0 & 0 \\
0 & 0 & 0 & q
\end{array}\right)$\;

Find $A_1, B_1\in M_4(\Z/2^{e'+2}\Z)$ satisfying $\,^t B_1 A_1\equiv \,^t A_1 B_1$, $\,^t D_1 C_1\equiv\,^t C_1 D_1$, $\,^t A_1 D_1-\,^t B_1 C_1\equiv I_4$ mod $2^{e'+2}$, so that $\left(\begin{array}{cc}
A _1 & C_1\\
B_1 & D_1
\end{array}\right)\in\Sp_8(\Z/2^{e'+2}\Z)$\;

$E_1\longleftarrow \left(\begin{array}{cccc}
a_1 & a_2 & q & 0 \\
-a_2 & a_1 & 0 & q\\
-1 & 0 & a_1 & -a_2 \\
0 & -1 & a_2 & a_1
\end{array}\right)$ and $E_2\longleftarrow \left(\begin{array}{cccc}
a_1 & a_2 & 1 & 0 \\
-a_2 & a_1 & 0 & 1\\
-q & 0 & a_1 & -a_2 \\
0 & -q & a_2 & a_1
\end{array}\right)$\;
$M_F\longleftarrow\Diag(E_1,E_2)$\;

$\left(\begin{array}{c}
C_2\\
D_2
\end{array}\right)\longleftarrow M_F\cdot \left(\begin{array}{c}
A_1\\
B_1
\end{array}\right)$\;

Find $C_2, D_2\in M_4(\Z/2^{e'+2}\Z)$ satisfying $\,^t B_2 A_2\equiv \,^t A_2 B_2$, $\,^t D_2 C_2\equiv\,^t C_2 D_2$, $\,^t A_2 D_2-\,^t B_2 C_2\equiv I_4$ mod $2^{e'+2}$\;

$M_2\longleftarrow \left(\begin{array}{cc}
A _1 & C_1\\
B_1 & D_1
\end{array}\right)\in\Sp_8(\Z/2^{e'+2}\Z)$\;

$E_3\longleftarrow \left(\begin{array}{cccc}
a_1 & -a_2 & -q & 0 \\
a_2 & a_1 & 0 & -q\\
1 & 0 & a_1 & a_2 \\
0 & 1 & -a_2 & a_1
\end{array}\right)$ and $E_4\longleftarrow \left(\begin{array}{cccc}
a_1 & -a_2 & -1 & 0 \\
a_2 & a_1 & 0 & -1\\
q & 0 & a_1 & a_2 \\
0 & q & -a_2 & a_1
\end{array}\right)$\;
$M_{\widetilde{F}}\longleftarrow\Diag(E_3,E_4)$\;

$\left(\begin{array}{c}
C'_1\\
D'_1
\end{array}\right)\longleftarrow M_{\tilde{F}}\cdot \left(\begin{array}{c}
A_2\\
B_2
\end{array}\right)$\;

$M'_1\longleftarrow \left(\begin{array}{cc}
A_1 & -C'_1\\
B_1 & -D'_1
\end{array}\right)\in\Sp_8(\Z/2^{e'+2}\Z)$\;

$\mathscr{B}_1\longleftarrow M'_1\cdot\mathscr{B}_0$ and $\mathscr{B}_2\longleftarrow M_2\cdot\mathscr{B}_0$\;  

\KwRet $\mathscr{B}_1, \mathscr{B}_2, M'_1, M_2$\;

\caption{Two basis of $(E_1^2\times E_2^2)[2^{e'+2}]$ inducing dual theta structures on $\mC$.}\label{alg: meet in the middle basis}
\end{algorithm}

\subsubsection{Change of coordinates before the $(m+1)$-th isogeny computation (gluing)}\label{sec: dim 4 gluing half torsion}

Let $m=v_2(a_2)$. When we compute $F_1$, we start by computing the $m$ first $2$-isogenies of the chain in dimension $2$ as we do in the full available torsion case. We also use \cref{alg: change of basis dim 2} introduced in \cref{sec: change of basis dim 2} to compute the change of theta structure prior to the computation of the first 2-dimensional isogeny $\varphi_1:E_1\times E_2\longrightarrow A_1$. Hence, after the computation of the $2$-isogeny chain $\Phi:=\varphi_m\circ\cdots\circ\varphi_1: E_1\times E_2\longrightarrow A_m$ (using the notations of \cref{lemma: first gluing isogenies} applied to $F_1$) we obtain the level $2$ product theta structure $\Theta_{mM_m}$ on $(A_m^2,\mM_m)$ induced by the symplectic basis of $A_m^2[4]$:
\begin{align*}
\mathscr{C}_1\times\mathscr{C}_1&:=(([2^m]\Phi(S_1),0),([2^m]\Phi(S_2),0),(0,[2^m]\Phi(S_1)),(0,[2^m]\Phi(S_2)),\\
&\qquad (\Phi(T_1),0),(\Phi(T_2),0),(0,\Phi(T_1)),(0,\Phi(T_2)))
\end{align*}
introduced in \cref{sec: dim 4 gluing full torsion} (with $S_1, S_2, T_1, T_2$ defined in \cref{lemma: theta structure dim 2 end chain}). 

However, this product theta structure $\Theta_m$ is not suitable to compute $f_{m+1}: A_m^2\longrightarrow B$ and subsequent isogenies of the chain, since $K_2(\Theta_{\mL})\neq \ker(f_{m+1})$. The isogeny $f_{m+1}$ is computed with the theta structure $\Theta'_{\mM_m}$ induced by the following symplectic basis of $A_m^2[4]$:
\begin{align*}\mathscr{C}:=&\ ([2^{e'}]f_m\circ\cdots\circ f_1(X_1),\cdots, [2^{e'}]f_m\circ\cdots\circ f_1(X_4),[2^{e'-m}]f_m\circ\cdots\circ f_1(Y_1),\\
&\cdots, [2^{e'-m}]f_m\circ\cdots\circ f_1(Y_4))
\end{align*}
where $\mathscr{B}_1:=(X_1, \cdots, X_4, Y_1,\cdots, Y_4)$ is an output of \cref{alg: meet in the middle basis}. We then have to compute the change of coordinates between $\Theta_{\mM_m}$ and $\Theta'_{\mM_m}$. This can be done with the following lemma:

\begin{Lemma}\label{lemma: gluing change of basis F_1}
Let $\left(\begin{array}{cc} A & C\\ B & D\end{array}\right)\in\Sp_8(\Z/2^{e'+2}\Z)$ be the basis change matrix from $\mathscr{B}_0$ to $\mathscr{B}_1$, where $\mathscr{B}_0$ has been defined on \cref{line: def basis B_0} of \cref{alg: meet in the middle basis}. Then the change of basis matrix from $\mathscr{C}_1\times\mathscr{C}_1$ to $\mathscr{C}$ is:
\[\left(\begin{array}{c|c}
-a_1\cdot B_1-a_2\cdot A_1-B_3 & (-a_1\cdot D_1-a_2\cdot C_1-D_3)/2^m\\
a_1\cdot A_1-a_2\cdot B_1+q\cdot A_3 & (a_1\cdot C_1-a_2\cdot D_1+q\cdot C_3)/2^m\\
-a_1\cdot B_2-a_2\cdot A_2-B_4 & (-a_1\cdot D_2-a_2\cdot C_2-D_4)/2^m\\
a_1\cdot A_2-a_2\cdot B_2+q\cdot A_4 & (a_1\cdot C_2-a_2\cdot D_2+q\cdot C_4)/2^m\\
\hline
2^m\cdot A_3 & C_3\\
2^m(\mu\cdot B_1+\mu a_2\cdot A_3) & \mu\cdot D_1+\mu a_2\cdot C_3\\
2^m\cdot A_4 & C_4\\
2^m(\mu\cdot B_2+\mu a_2\cdot A_4) & \mu\cdot D_2+\mu a_2\cdot C_4 
\end{array}\right),\]
where $\mu \equiv 1/a_1 \mod 4$ and the $A_i, B_i, C_i, D_i$ are the $i$-th lines of $A, B, C, D$ respectively. 
\end{Lemma}

\begin{proof}
The change of basis matrix can be computed via 4-th Weil pairings as in the proof of \cref{lemma: matrix V_i}.
\end{proof}

The same method applies for the computation of $\widetilde{F}_2$. Using notations introduced in \cref{sec: change of basis dim 2 half}, the product theta structure we obtain on ${A'}_m^2$ is induced by the symplectic basis of ${A'}_m^2[4]$:
\begin{align*}
\mathscr{C}_2\times\mathscr{C}_2&:=(([2^m]\Psi(S_1),0),([2^m]\Psi(S_2),0),(0,[2^m]\Psi(S_1)),(0,[2^m]\Psi(S_2)),\\
&\qquad (\Psi(T_1),0),(\Psi(T_2),0),(0,\Psi(T_1)),(0,\Psi(T_2)))
\end{align*}
with $\Psi:=\psi_m\circ\cdots\circ\psi_1$,
\[S_1:=([2^{m+1}]Q'_1,[a]\sigma(P'_1)+[b]\sigma(Q'_1)), \quad S_2:=([\mu]P'_1,[c]\sigma(P'_1)+[d]\sigma(Q'_1)),\]
\[T_1:=([a_1]P'_1+[a_2]Q'_1,-\sigma(P'_1)) \quad \mbox{and} \quad T_2:=(-[a_2]P'_1+[a_1]Q'_1,-\sigma(Q'_1)),\]
$P'_1=[2^{e'-m}]P''_1$, $Q'_1=[2^{e'-m}]Q''_1$ (where $(P''_1, Q''_1)$ is the input basis of \cref{alg: meet in the middle basis}), $a\equiv 2^{m+1}a_2/q$, $b\equiv -(1+2^{m+1}a_1)/q$, $\mu\equiv (1-2^{m+1}q)/a_1$, $d\equiv -\mu a_2/q \mod 2^{m+2}$ and $c=2^{m+1}$.

\begin{Lemma}\label{lemma: gluing change of basis F_2}
Let $\left(\begin{array}{cc} A & C\\ B & D\end{array}\right)\in\Sp_8(\Z/2^{e'+2}\Z)$ be the change of basis matrix from $\mathscr{B}_0$ to $\mathscr{B}_2$, where $\mathscr{B}_0$ has been defined on \cref{line: def basis B_0} of \cref{alg: meet in the middle basis}. Then change of basis matrix from $\mathscr{C}_2\times\mathscr{C}_2$ to $\widetilde{\mathscr{C}}$ is:
\[\left(\begin{array}{c|c}
-a_1\cdot B_1+a_2\cdot A_1+B_3 & (-a_1\cdot D_1+a_2\cdot C_1+D_3)/2^m\\
a_1\cdot A_1+a_2\cdot B_1-q\cdot A_3 & (a_1\cdot C_1+a_2\cdot D_1-q\cdot C_3)/2^m\\
-a_1\cdot B_2+a_2\cdot A_2+B_4 & (-a_1\cdot D_2+a_2\cdot C_2+D_4)/2^m\\
a_1\cdot A_2+a_2\cdot B_2-q\cdot A_4 & (a_1\cdot C_2+a_2\cdot D_2-q\cdot C_4)/2^m\\
\hline
-2^m\cdot A_3 & -C_3\\
2^m(\mu\cdot B_1+\mu a_2\cdot A_3) & \mu\cdot D_1+\mu a_2\cdot C_3\\
-2^m\cdot A_4 & -C_4\\
2^m(\mu\cdot B_2+\mu a_2\cdot A_4) & \mu\cdot D_2+\mu a_2\cdot C_4 
\end{array}\right),\]
where $\mu \equiv 1/a_1 \mod 4$ and the $A_i, B_i, C_i, D_i$ are the $i$-th lines of $A, B, C, D$ respectively. 
\end{Lemma}

\subsection{Recovering the product theta structure on an intermediate product of abelian surfaces}\label{sec: intermediate product theta structure}

When computing the "second part" $\widetilde{F}_2$ of $F$, we also have to compute a chain of $2$-isogenies $\psi_i$ in dimension $2$ of length $m=v_2(a_2)$ and a gluing isogeny $g_{m+1}: {A'}_m^2\longrightarrow B'$ (\cref{lemma: first gluing isogenies}). To recover $F_2=\widetilde{\widetilde{F}}_2$, we compute the dual of every isogeny in the chain. In particular, we compute $\widetilde{g}_{m+1}: B'\longrightarrow {A'}_m^2$ and $\widetilde{\varphi}_i$ for all $i\in\Int{1}{m}$. To be able to evaluate the composition $\widetilde{g}_{m}\circ \widetilde{g}_{m+1}=(\widetilde{\psi}_m\times\widetilde{\psi}_m)\circ\widetilde{g}_{m+1}$, we need to convert the theta coordinates of images of $\widetilde{g}_{m+1}$ for a non-product  theta structure $\Theta'_{\mM'_m}$ on $({A'}_m^2,\mM'_m)$ into product theta coordinates (for $\Theta_{\mM'_m}=\Theta_{\mL'_m}\times\Theta_{\mL'_m}$). To translate $\Theta'_{\mM'_m}$-coordinates into $\Theta_{\mM'_m}$-coordinates, we can simply act by the inverse of the change of coordinates matrix from $\Theta_{\mM'_m}$ to $\Theta'_{\mM'_m}$-coordinates that has been computed to obtain the gluing isogeny $g_{m+1}$ (using \cref{lemma: gluing change of basis F_2,thm: symplectic change of basis}). 

We then need to split the $\Theta_{\mM'_m}$-coordinates on $({A'}_m^2,\mM'_m)$ into couples of $\Theta_{\mL'_m}$-coordinates on $(A'_m,\mL'_m)$. Our approach is similar to \cref{sec: splitting change of basis}. Let $x, y\in A'_m$. Given $(\theta_{\textbf{i},\textbf{j}}^{\mM'_m}(x,y))_{(\textbf{i},\textbf{j})\in(\Z/2\Z)^4}$, we want to compute $(\theta_\textbf{i}^{\mL'_m}(x))_{\textbf{i}\in (\Z/2\Z)^2}$ and $(\theta_\textbf{i}^{\mL'_m}(y))_{\textbf{i}\in (\Z/2\Z)^2}$ up to a projective constant. By \cref{lemma: product theta structure}, we have for all $\textbf{i}, \textbf{j}\in (\Z/2\Z)^2$, $\theta_{\textbf{i},\textbf{j}}^{\mM'_m}(x,y)=\theta_{\textbf{i}}^{\mL'_m}(x)\cdot\theta_{\textbf{j}}^{\mL'_m}(y)$. Hence we can start by finding $\textbf{i}_0, \textbf{j}_0\in (\Z/2\Z)^2$ such that $\theta_{\textbf{i}_0,\textbf{j}_0}^{\mM'_m}(x,y)\neq 0$ and then compute $\theta_{\textbf{i},\textbf{j}_0}^{\mM'_m}(x,y)/\theta_{\textbf{i}_0,\textbf{j}_0}^{\mM'_m}(x,y)=\theta_{\textbf{i}}^{\mL'_m}(x)/\theta_{\textbf{i}_0}^{\mL'_m}(x)$ and $\theta_{\textbf{i}_0,\textbf{j}}^{\mM'_m}(x,y)/\theta_{\textbf{i}_0,\textbf{j}_0}^{\mM'_m}(x,y)=\theta_{\textbf{j}}^{\mL'_m}(y)/\theta_{\textbf{j}_0}^{\mL'_m}(y)$ for all $\textbf{i},\textbf{j}\in (\Z/2\Z)^2$. This is explained in \cref{alg: splitting product dim 4 to dim 2}. 

\begin{algorithm}
\SetAlgoLined
\KwData{\justifying Product $\Theta_{\mM'_m}$-coordinates $(\theta^{\mM'_m}_{\textbf{i},\textbf{j}}(x,y))_{\textbf{i},\textbf{j}\in(\Z/2\Z)^2}$ of $(x,y)\in {A'}_m^2$.}
\KwResult{\justifying The $\Theta_{\mL'_m}$-coordinates $(\theta^{\mL'_m}_\textbf{i}(x))_{\textbf{i}\in(\Z/2\Z)^2}$ and $(\theta^{\mL'_m}_\textbf{i}(x))_{\textbf{i}\in(\Z/2\Z)^2}$.}
Find $\textbf{i}_0, \textbf{j}_0\in (\Z/2\Z)^2$ such that $\theta_{\textbf{i}_0,\textbf{j}_0}^{\mM'_m}(x,y)\neq 0$\;
$x_{\textbf{i}_0}\longleftarrow 1$, $y_{\textbf{j}_0}\longleftarrow 1$\;
Compute $x_\textbf{i}\longleftarrow \theta^{\mM'_m}_{\textbf{i},\textbf{j}_0}(x,y)/\theta^{\mM'_m}_{\textbf{i}_0,\textbf{j}_0}(x,y)$ for all $\textbf{i}\in(\Z/2\Z)^2\setminus\{\textbf{i}_0\}$\;
Compute $y_\textbf{j}\longleftarrow \theta^{\mM'_m}_{\textbf{i}_0,\textbf{j}}(x,y)/\theta^{\mM'_m}_{\textbf{i}_0,\textbf{j}_0}(x,y)$ for all $\textbf{j}\in(\Z/2\Z)^2\setminus\{\textbf{j}_0\}$\;

\KwRet $(x_\textbf{i})_{\textbf{i}\in(\Z/2\Z)^2}, (y_\textbf{j})_{\textbf{j}\in(\Z/2\Z)^2}$\;

\caption{Splitting product theta coordinates (from dimension 4 to dimension 2).}\label{alg: splitting product dim 4 to dim 2}
\end{algorithm}

\subsection{Recovering the product theta structure in dimension 2}

When we evaluate $F=F_2\circ F_1$, the last isogeny of the chain to be evaluated is the $2$-dimensional splitting isogeny $\widetilde{\psi}_1: A'_1\longrightarrow E_1\times E_2$. The resulting image points are expressed in non-product $\Theta''_{\mL}$-coordinates. We have to translate these points into  $(x:z)$-Montgomery coordinates. Given $\Theta''_{\mL}$-coordinates $({\theta''}_i^{\mL}(R_1,R_2))_i$ of a point $(R_1, R_2)\in E_1\times E_2$, we can apply the inverse of the change of coordinates matrix from $(x:z)$ to $\Theta''_{\mL}$ computed in \cref{sec: change of basis dim 2 half}. We then obtain $(x(R_1)x(R_2):x(R_1)z(R_2):z(R_1)x(R_2):z(R_1)z(R_2))$. If $z(R_1)z(R_2)\neq 0$, we can then compute $(x(R_1)/z(R_1):1)$ and $(x(R_2)/z(R_2):1)$ as follows: $x(R_1)/z(R_1)=x(R_1)z(R_2)/z(R_1)z(R_2)$ and $x(R_2)/z(R_2)=z(R_1)x(R_2)/z(R_1)z(R_2)$. Otherwise, we have $R_1=0$ or $R_2=0$ and we can also recover the Montgomery $(x:z)$-coordinates (with the convention $(x(0):z(0))=(1:0)$). We refer to \cref{alg: splitting dim 2 to dim 1} for the complete conversion procedure of $\Theta''_{\mL}$-coordinates into Montgomery $(x:z)$-coordinates.

\begin{algorithm}
\SetAlgoLined
\KwData{\justifying $\Theta''_{\mL}$-coordinates $({\theta''}^{\mL}_\textbf{i}(R))_{\textbf{i}\in (\Z/2\Z)^2}$ of a point $R:=(R_1,R_2)\in E_1\times E_2$ and the inverse of the change of coordinates matrix $N^{-1}$ from $(x:z)$ to $\Theta''_{\mL}$.}
\KwResult{\justifying Montgomery $(x:z)$-coordinates $(x(R_1):z(R_1)), (x(R_2):z(R_2))$.}

$(x(R_1)x(R_2):x(R_1)z(R_2):z(R_1)x(R_2):z(R_1)z(R_2))\longleftarrow N^{-1}\cdot ({\theta''}^{\mL}_\textbf{i}(R))_\textbf{i}$\;

\uIf{$z(R_1)z(R_2)\neq 0$}{
$t\longleftarrow 1/(z(R_1)z(R_2))$\; 
$x_1\longleftarrow x(R_1)z(R_2)\cdot t$\; 
$x_2\longleftarrow z(R_1)x(R_2)\cdot t$\;
\KwRet $(x_1:1), (x_2:1)$\;
}

\uElseIf{$z(R_1)x(R_2)=0$ and $x(R_1)z(R_2)\neq 0$}{
$t\longleftarrow 1/(x(R_1)z(R_2))$\; 
$x_2\longleftarrow x(R_1)x(R_2)\cdot t$\;
\KwRet $(1:0), (x_2:1)$\;
}

\uElseIf{$z(R_1)x(R_2)\neq 0$ and $x(R_1)z(R_2)=0$}{
$t\longleftarrow 1/(z(R_1)x(R_2))$\;
$x_1\longleftarrow x(R_1)x(R_2)\cdot t$\;
\KwRet $(x_1:1), (1:0)$\;
}

\uElse{
\KwRet $(1:0), (1:0)$\;
}

\caption{Splitting non product theta coordinates (from dimension 2 to dimension 1).}\label{alg: splitting dim 2 to dim 1}
\end{algorithm}

\newpage

\section{Basic arithmetic optimisations}

\subsection{Batch inversion}\label{sec: batch inversion}

Let $k$ be a base field in which we want to invert several elements. In general inversions are much more costly than multiplications over $k$ (e.g. when $k$ is a finite field). In \cref{alg: batch inversion}, we present how to invert $n$ elements with only one inversion at the expense of $3(n-1)$ multiplications.

\begin{algorithm}
\SetAlgoLined
\KwData{\justifying $a_1, \cdots, a_n\in k^*$.}
\KwResult{\justifying $1/a_1, \cdots, 1/a_n$.}

$b_1\longleftarrow a_0$ 
\For{$i=2$ \KwTo $n$}{
$b_i\longleftarrow b_{i-1}\cdot a_i$ \tcp*{$b_i=a_1\cdots a_i$}
}
$c_1\longleftarrow 1/b_n$\;\label{line: invert}
\For{$i=2$ \KwTo $n$}{
$c_i\longleftarrow c_{i-1}\cdot a_{n-i+2}$ \tcp*{$c_i=1/(a_1\cdots a_{n-i+1})$}
}
$d_1\longleftarrow c_n$\;
\For{$i=2$ \KwTo $n$}{
$d_i\longleftarrow c_{n-i+1}\cdot b_{i-1}$\tcp*{$d_i=1/(a_1\cdots a_i)\cdot (a_1\cdots a_{i-1})=1/a_i$}
}
\KwRet $d_1, \cdots, d_n$\;

\caption{Batch inversion.}\label{alg: batch inversion}
\end{algorithm}

If we are willing to work projectively and obtain $\lambda/a_1, \cdots, \lambda/a_n$ with $\lambda\in k^*$ a projective constant when $a_1, \cdots, a_n\in k^*$ are given, we can simply remove the inversion on \cref{line: invert} ($c_1\longleftarrow 1$) and compute $3(n-1)$ multiplications only. Though considered and recommended, this optimization is not implemented in our dimension 4 code.

\subsection{Recursive Hadamard transform}\label{sec: recursive Hadamard}

For all $g\in\N^*$, let $H_g$ be the Hadamard transform over $k^{(\Z/2\Z)^g}$:
\[H_g: (x_\textbf{i})_{\textbf{i}\in(\Z/2\Z)^g}\longmapsto \left(\sum_{\textbf{i}\in(\Z/2\Z)^g}(-1)^{\langle \textbf{i}|\textbf{j}\rangle}x_\textbf{i}\right)_{\textbf{j}\in(\Z/2\Z)^g}.\]
A naive evaluation of $H_g$ would cost $2^{2g}$ additions/subtractions which can become costly. Instead, we propose a recursive method to compute $H_g$. We notice that for all $g\geq 2$ and $(x,y)\in (k^{(\Z/2\Z)^{g-1}})^2$:
\[H_g(x,y)=\left(H_1(H_{g-1}(x)_{(j_1,\cdots, j_{g-1})},H_{g-1}(y)_{(j_1,\cdots, j_{g-1})}))_{j_{g}}\right)_{\textbf{j}\in(\Z/2\Z)^g},\]
where:
\[\forall x\in k^{\Z/2\Z}, \quad H_1(x_0,x_1)=(x_0+x_1,x_0-x_1).\]
We can then apply \cref{alg: recursive Hadamard} to evaluate $H_g$ with only $g\cdot 2^g$ additions/subtractions. For $g=4$, this decreases the complexity from $256$ to $64$ additions/subtractions.

\begin{algorithm}
\SetAlgoLined
\KwData{\justifying $(x_\textbf{i})_{\textbf{i}\in(\Z/2\Z)^g}$.}
\KwResult{\justifying $H((x_\textbf{i})_{\textbf{i}\in(\Z/2\Z)^g})$.}

\eIf{$g=1$}{
\KwRet $(x_0+x_1,x_0-x_1)$\;
}{
$x, y\longleftarrow (x_{\textbf{i},0})_{\textbf{i}\in(\Z/2\Z)^{g-1}}, (x_{\textbf{i},1})_{\textbf{i}\in(\Z/2\Z)^{g-1}}$\;
$z, t\longleftarrow H_{g-1}(x), H_{g-1}(y)$ \tcp*{Recursive calls}
\For{$\textbf{j}\in(\Z/2\Z)^g$}{
$u_\textbf{j}\longleftarrow z_{(j_1,\cdots, j_{g-1})}+(-1)^{j_g}t_{(j_1,\cdots, j_{g-1})}$\;
}
\KwRet $(u_\textbf{j})_{\textbf{j}\in (\Z/2\Z)^g}$\;
}

\caption{Recursive Hadamard transform.}\label{alg: recursive Hadamard}
\end{algorithm}

\newpage

\section{Optimal strategies for isogenies derived from Kani's lemma}\label{sec: optimal strategies}

\subsection{Definition of optimal strategies}\label{sec: def optimal strategies}

Let us assume we want to compute a $2^e$-isogeny $F:\mathcal{A}\longrightarrow \mathcal{B}$ (in dimension $g$), decomposed as a chain of $2$-isogenies:
\[A_1=\mathcal{A}\overset{f_1}{\relbar\joinrel\relbar\joinrel\longrightarrow} A_2 \quad \cdots \quad A_{e} \overset{f_e}{\relbar\joinrel\relbar\joinrel\longrightarrow} A_{e+1}=\mathcal{B},\]
and that we are given a basis $\mathscr{B}_{K''}$ of a maximal isotropic subgroup $K''\subseteq \mathcal{A}[2^{e+2}]$ such that $\ker(F)=[4]K''$. For all $i\in\Int{1}{e}$, we need to know $\mathscr{B}_{i-1,e-i}:=[2^{e-i}]f_{i-1}\circ\cdots\circ f_1(\mathscr{B}_{K''})$ in order to compute $f_i$ (using Algorithm \ref{alg: dual theta null point}). 

Hence, computing $F$ reduces to computing the leaves $\mathscr{B}_{i-1,e-i}$ of the binary computation tree whose:
\begin{itemize}
\item vertices are the basis $\mathscr{B}_{i,j}:=[2^j]f_i\circ\cdots\circ f_1(\mathscr{B}_{K''})$ for all $i, j \in \N$ such that $i+j\leq e-1$;
\item left edges are doublings $\mathscr{B}_{i,j-1}\overset{[2]}{\longrightarrow}\mathscr{B}_{i,j}$;
\item right edges are $2$-isogeny evaluations $\mathscr{B}_{i-1,j}\overset{f_i}{\longrightarrow}\mathscr{B}_{i,j}$.
\end{itemize}

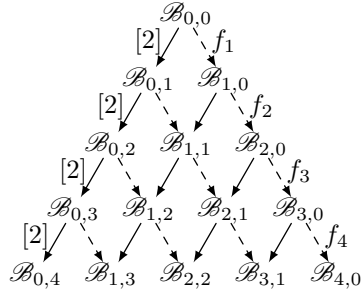
\begin{figure}[tbh]

\begin{tikzpicture}[line cap=round,line join=round,>=triangle 45,x=1cm,y=0.866cm,scale=1]
\clip(-6,-4.5) rectangle (6,0.5);

\draw (0,0) node {$\mathscr{B}_{0,0}$};
\draw [-latex,line width=0.5pt] (-0.1,-0.2)--(-0.4,-0.8);
\draw (-0.5,-0.4) node {$[2]$};
\draw (-0.5,-1) node {$\mathscr{B}_{0,1}$};
\draw [-latex,line width=0.5pt] (-0.6,-1.2)--(-0.9,-1.8);
\draw (-1,-1.4) node {$[2]$};
\draw (-1,-2) node {$\mathscr{B}_{0,2}$};
\draw [-latex,line width=0.5pt] (-1.1,-2.2)--(-1.4,-2.8);
\draw (-1.5,-2.4) node {$[2]$};
\draw (-1.5,-3) node {$\mathscr{B}_{0,3}$};
\draw [-latex,line width=0.5pt] (-1.6,-3.2)--(-1.9,-3.8);
\draw (-2,-3.4) node {$[2]$};
\draw (-2,-4) node {$\mathscr{B}_{0,4}$};

\draw [-latex,dashed,line width=0.5pt] (0.1,-0.2)--(0.4,-0.8);
\draw (0.5,-0.4) node {$f_1$};
\draw [-latex,dashed,line width=0.5pt] (-0.4,-1.2)--(-0.1,-1.8);
\draw [-latex,dashed,line width=0.5pt] (-0.9,-2.2)--(-0.6,-2.8);
\draw [-latex,dashed,line width=0.5pt] (-1.4,-3.2)--(-1.1,-3.8);

\draw (0.5,-1) node {$\mathscr{B}_{1,0}$};
\draw [-latex,line width=0.5pt] (0.4,-1.2)--(0.1,-1.8);
\draw (0,-2) node {$\mathscr{B}_{1,1}$};
\draw [-latex,line width=0.5pt] (-0.1,-2.2)--(-0.4,-2.8);
\draw (-0.5,-3) node {$\mathscr{B}_{1,2}$};
\draw [-latex,line width=0.5pt] (-0.6,-3.2)--(-0.9,-3.8);
\draw (-1,-4) node {$\mathscr{B}_{1,3}$};

\draw [-latex,dashed,line width=0.5pt] (0.6,-1.2)--(0.9,-1.8);
\draw (1,-1.4) node {$f_2$};
\draw [-latex,dashed,line width=0.5pt] (0.1,-2.2)--(0.4,-2.8);
\draw [-latex,dashed,line width=0.5pt] (-0.4,-3.2)--(-0.1,-3.8);

\draw (1,-2) node {$\mathscr{B}_{2,0}$};
\draw [-latex,line width=0.5pt] (0.9,-2.2)--(0.6,-2.8);
\draw (0.5,-3) node {$\mathscr{B}_{2,1}$};
\draw [-latex,line width=0.5pt] (0.4,-3.2)--(0.1,-3.8);
\draw (0,-4) node {$\mathscr{B}_{2,2}$};

\draw [-latex,dashed,line width=0.5pt] (1.1,-2.2)--(1.4,-2.8);
\draw (1.5,-2.4) node {$f_3$};
\draw [-latex,dashed,line width=0.5pt] (0.6,-3.2)--(0.9,-3.8);

\draw (1.5,-3) node {$\mathscr{B}_{3,0}$};
\draw [-latex,line width=0.5pt] (1.4,-3.2)--(1.1,-3.8);
\draw (1,-4) node {$\mathscr{B}_{3,1}$};

\draw [-latex,dashed,line width=0.5pt] (1.6,-3.2)--(1.9,-3.8);
\draw (2,-3.4) node {$f_4$};

\draw (2,-4) node {$\mathscr{B}_{4,0}$};

\end{tikzpicture}
\captionsetup{width=\textwidth}
\caption{Computational structure of the $2^e$ isogeny $F$ with $e=5$.}\label{fig: computation tree}

\end{figure}

Such a tree is displayed in \cref{fig: computation tree} for $e=5$\footnote{This tree is inspired from \cite[Figure 2]{SQISignHD_autocite}.}. The computation tree can only be evaluated depth first and left first since the leaf $\mathscr{B}_{i-1,e-i}$ has to be computed prior to any evaluation by $f_i$. However, evaluating all the vertices $\mathscr{B}_{i,j}$ would be a waste of computational resources leading to a quadratic complexity $O(e^2)$. Optimal strategies consist in navigating the computation tree depth first and left first with a minimal number of doublings and evaluations to evaluate the leaves $\mathscr{B}_{i-1,e-i}$.
 
As in \cite{SIDH}, we can represent the computation tree as a \emph{discrete equilateral triangle} $T_e$ formed by points of the unit triangular equilateral lattice delimited by the $x$ axis and the straight lines $y=\sqrt{3}x$ and $y=\sqrt{3}(e-1-x)$:
\[T_e:=\left\{\left(r+\frac{s}{2},\frac{s\sqrt{3}}{2}\right)\middle| \ r,s \in\N, \quad r+s\leq e-1\right\}\]
In $T_e$, \emph{edges} are unit segments connecting two points of $T_e$. A \emph{left edge} is a segment of positive slope and a \emph{right edge} is a segment of negative slope. Edges are oriented in the direction of decreasing $y$ coordinates. This defines an oriented graph structure on $T_e$. Vertices on $x,y\in T_e$ are ordered $x\rightarrow y$ if there exists a path from $x$ to $y$. On a subgraph of $T_e$, the \emph{root} is the initial points and \emph{leaves} are final points.

\begin{Definition}
A \emph{strategy} $S$ of $T_e$ is a subgraph of $T_e$ having a unique root. In the following, we only consider strategies that are:
\begin{enumerate}
\item \emph{full}, meaning that $S$ contains all leaves of $T_e$.
\item \emph{well-formed}, meaning that there is only one path going through interior point of $S$ and no leaf in $S$ distinct from the leaves of $T_e$.
\end{enumerate} 
Such a (full and well formed) strategy of $T_e$ is also called a \emph{strategy of depth $e-1$}. We denote $|S|=e$ its number of leaves.
\end{Definition} 

To compare strategies, we fix a measure $(\alpha,\beta)\in\R_+^2$ on them, where $\alpha$ is the cost of a left edge (accounting for doubling cost) and $\beta$ is the cost of a right edge (accounting for evaluation cost). Given such a measure, an \emph{optimal strategy} of depth $e-1$ is a strategy of $T_e$ with minimal cost. 

We define the \emph{tree topology} of a strategy $S$ of depth $e-1$ as the binary tree with $e$ leaves obtained by forgetting internal vertices of out degree less than two and keeping the same connectivity structure. Conversely, to any binary tree $T$ with $e$ leaves we associate a \emph{canonical strategy} $S_T$ of depth $e-1$ recursively as follows. If $e=1$, we take $S_T:=T_1$. If $e\geq 2$, we consider the left and right branches $T'$ and $T''$ of $T$ respectively and consider the canonical strategies $S':=S_{T'}$ and $S_{T''}$ associated to them. Let $S''$ be the translate of $S_{T''}$ by $|S'|$ to the right. Let $r'$ and $r''$ be the roots of $S'$ and $S''$ in $T_e$ respectively and $r$ be the root of $T_e$. Then the shortest paths $rr'$ and $rr''$ from $r$ to $r'$ and $r''$ respectively are respectively made of left edges only and right edges only. We can then consider the strategy $S_{T}:=rr'\cup rr''\cup S'\cup S''$. 

\begin{figure}

\begin{minipage}{7.5cm}

\centering
\begin{tikzpicture}[line cap=round,line join=round,>=triangle 45,x=1cm,y=0.866cm,scale=0.5]
\clip(-0.5,0.9) rectangle (4.5,4.1);

\draw[fill=black] (2,4) circle (0.05);

\draw[fill=black] (1.5,3) circle (0.05);
\draw[fill=black] (2.5,3) circle (0.05);

\draw[fill=black] (1,2) circle (0.05);
\draw[fill=black] (2,2) circle (0.05);
\draw[fill=black] (3,2) circle (0.05);

\draw[fill=black] (0.5,1) circle (0.05);
\draw[fill=black] (1.5,1) circle (0.05);
\draw[fill=black] (2.5,1) circle (0.05);
\draw[fill=black] (3.5,1) circle (0.05);

\draw [line width=0.5pt] (2,4)--(1.5,3);
\draw [line width=0.5pt] (2,4)--(2.5,3);

\draw [line width=0.5pt] (1.5,3)--(1,2);
\draw [line width=0.5pt] (2.5,3)--(2,2);
\draw [line width=0.5pt] (2.5,3)--(3,2);

\draw [line width=0.5pt] (1,2)--(0.5,1);
\draw [line width=0.5pt] (1,2)--(1.5,1);
\draw [line width=0.5pt] (2,2)--(2.5,1);
\draw [line width=0.5pt] (3,2)--(3.5,1);

\end{tikzpicture}\begin{tikzpicture}[line cap=round,line join=round,>=triangle 45,x=1cm,y=0.866cm,scale=0.5]
\clip(-0.5,0.9) rectangle (4.5,4.1);

\draw[fill=black] (2,4) circle (0.05);

\draw[fill=black] (1.5,3) circle (0.05);
\draw[fill=black] (2.5,3) circle (0.05);

\draw[fill=black] (1,2) circle (0.05);
\draw[fill=black] (2,2) circle (0.05);
\draw[fill=black] (3,2) circle (0.05);

\draw[fill=black] (0.5,1) circle (0.05);
\draw[fill=black] (1.5,1) circle (0.05);
\draw[fill=black] (2.5,1) circle (0.05);
\draw[fill=black] (3.5,1) circle (0.05);

\draw [line width=0.5pt] (2,4)--(1.5,3);
\draw [line width=0.5pt] (2,4)--(2.5,3);

\draw [line width=0.5pt] (1.5,3)--(1,2);
\draw [line width=0.5pt] (2.5,3)--(3,2);

\draw [line width=0.5pt] (1,2)--(0.5,1);
\draw [line width=0.5pt] (1,2)--(1.5,1);
\draw [line width=0.5pt] (3,2)--(2.5,1);
\draw [line width=0.5pt] (3,2)--(3.5,1);

\end{tikzpicture}\begin{tikzpicture}[line cap=round,line join=round,>=triangle 45,x=1cm,y=0.866cm,scale=0.5]
\clip(-0.5,0.9) rectangle (4.5,4.1);

\draw[fill=black] (2,4) circle (0.05);

\draw[fill=black] (1.5,3) circle (0.05);
\draw[fill=black] (2.5,3) circle (0.05);

\draw[fill=black] (1,2) circle (0.05);
\draw[fill=black] (2,2) circle (0.05);
\draw[fill=black] (3,2) circle (0.05);

\draw[fill=black] (0.5,1) circle (0.05);
\draw[fill=black] (1.5,1) circle (0.05);
\draw[fill=black] (2.5,1) circle (0.05);
\draw[fill=black] (3.5,1) circle (0.05);

\draw [line width=0.5pt] (2,4)--(1.5,3);
\draw [line width=0.5pt] (2,4)--(2.5,3);

\draw [line width=0.5pt] (1.5,3)--(1,2);
\draw [line width=0.5pt] (1.5,3)--(2,2);
\draw [line width=0.5pt] (2.5,3)--(3,2);

\draw [line width=0.5pt] (1,2)--(0.5,1);
\draw [line width=0.5pt] (2,2)--(1.5,1);
\draw [line width=0.5pt] (3,2)--(2.5,1);
\draw [line width=0.5pt] (3,2)--(3.5,1);

\end{tikzpicture}
\captionsetup{width=\textwidth}
\caption{Three strategies of depth $3$ sharing the same tree topology. The middle one is canonical.}
\end{minipage}\hspace{0.5cm}\begin{minipage}{4cm}
\definecolor{ao}{rgb}{0,0.5,0}

\centering
\begin{tikzpicture}[line cap=round,line join=round,>=triangle 45,x=1cm,y=0.866cm,scale=0.5]
\clip(-0.5,0.9) rectangle (4.5,4.1);

\draw[ao,fill=ao] (2,3) circle (0.05);

\draw[ao,fill=ao] (1,2) circle (0.05);
\draw[ao,fill=ao] (3,2) circle (0.05);

\draw[ao,fill=ao] (0.5,1) circle (0.05);
\draw[ao,fill=ao] (1.5,1) circle (0.05);
\draw[ao,fill=ao] (2.5,1) circle (0.05);
\draw[ao,fill=ao] (3.5,1) circle (0.05);

\draw [color=ao,line width=0.5pt] (2,3)--(1,2);
\draw [color=ao,line width=0.5pt] (2,3)--(3,2);

\draw [color=ao,line width=0.5pt] (1,2)--(0.5,1);
\draw [color=ao,line width=0.5pt] (1,2)--(1.5,1);
\draw [color=ao,line width=0.5pt] (3,2)--(2.5,1);
\draw [color=ao,line width=0.5pt] (3,2)--(3.5,1);

\end{tikzpicture}
\captionsetup{width=\textwidth}
\caption{Tree topology of the strategies on the left.}

\end{minipage}
\end{figure}

The following result has been proved in \cite{SIDH}:

\begin{Lemma}\cite[Lemma 4.3]{SIDH}
The canonical strategy is minimal, with respect to any measure, among all the strategies sharing the same tree topology.
\end{Lemma}

It follows that we can restrict to canonical strategies to find optimal strategies in the following. If $S$ is a canonical strategy, we can consider its left and right branches $S'$ and $S''$ as follows. If $S$ has $i$ leaves to the left of its root, we define $S':=S\cap T_i$ and $S'':=S\cap ((i,0)+T_{|S|-i})$.

\begin{Lemma}\cite[Lemma 4.5]{SIDH}
Let $S$ be an optimal (canonical) strategy and let $S'$ and $S''$ be its left and right branches respectively. Then, $S'$ and $S''$ translated by $-|S'|$ are optimal strategies of $T_{|S'|}$ and $T_{|S''|}$ respectively.
\end{Lemma}  

\begin{proof}
The proof is very natural. By \cite[Lemma 4.3]{SIDH}, we know that $S$ is a canonical strategy, so $S'$ and $S''$ are well defined. If $S'$ were not optimal, then by substituting an optimal strategy for $S'$ inside $S$, we obtain a strategy with measure lower than $\mu(S)$. Contradiction. The same argument holds for $S''$.
\end{proof}

As pointed out in \cite{SIDH}, this suggests a dynamic programming approach to compute optimal strategies. For $e=1$, the only optimal strategy is trivially $S=T_1$. Now, if we assume that we have computed optimal strategies $S_1, \cdots, S_{e-1}$ of $T_1, \cdots, T_{e-1}$ of respective measures $\mu(S_1), \cdots, \mu(S_{e-1})$, then the optimal strategy $S_e$ will have left branch $S_i$ and right branch $S_{e-i}$ where:
\[i:=\underset{1\leq j\leq e-1}{\argmin} (\mu(S_j)+\mu(S_{e-j})+(e-j)\alpha+j\beta).\]

\subsection{Optimal strategies for isogenies derived from Kani's lemma}\label{sec: constrained optimal strategies}

Suppose we want to compute a $2^e$-isogeny $F\in\End(E_1^2\times E_2^2)$ derived from Kani's lemma as in \cref{eq: definition F} with full available $2^{e+2}$-torsion. Then quasi-linear divide and conquer strategies introduced in \cref{sec: def optimal strategies} also apply but they have to account for the already computed gluing isogeny chain $f_{m+1}\circ\cdots\circ f_1$ from \cref{lemma: first gluing isogenies}, so the strategy:
\begin{itemize}
\item Must be of length $e-m$ instead of $e$ (to start at $f_{m+1}\circ\cdots\circ f_1$, considered as \emph{one} isogeny).
\item Must take into account the higher cost of evaluation by the "first" isogeny $f_{m+1}\circ\cdots\circ f_1$. 
\end{itemize} 

We now consider a different measure for strategies than the one introduced in \cref{sec: def optimal strategies} taking into account the additional cost of the "first" isogeny evaluation. We denote by $\mu'$ this measure parametrized by $(\alpha,\beta,\gamma)\in\R_+^3$ where $\alpha$ is the cost of a left edge (accounting for doubling cost), $\beta$ is the cost of a right edge not starting from the root (accounting for a generic evaluation cost) and $\gamma$ is the cost of a right edge starting from the root (accounting for the evaluation by the "first" isogeny). Such an optimal strategy (for the new measure $\mu'$) is called an optimal strategy \emph{with more weight at the beginning}.

\cite[Lemma 4.3]{SIDH} generalizes to strategies with more weight at the beginning for the measure $\mu'$, so we can restrict to canonical strategies. Indeed, the two key ingredients used in the proof of \cite[Lemma 4.3]{SIDH} still hold. First, if there is a path from $x$ to $y$ in $T_{e}$ (we denote $x\rightarrow y$), then all paths going from $x$ to $y$ have the same measure $\mu'(xy)$. Second, if $x\rightarrow y$, $y\rightarrow y'$ and $y\rightarrow y''$, then $\mu'(xy)+\mu'(yy')+\mu'(yy'')\leq\mu'(xy')+\mu'(xy'')$. Hence, we can also generalize \cite[Lemma 4.5]{SIDH}.

\begin{Lemma}
Let $S$ be a (canonical) optimal strategy with more weight at the beginning such that $|S|\geq 2$ and let $S'$ and $S''$ be its left and right branches respectively. Then, $|S'|\geq 2$, $S'$ is an optimal strategy of $T_{|S'|}$ for $\mu'$ with more weight at the beginning and $S''$ translated by $-|S'|$ is an optimal strategy of $T_{|S''|}$ for $\mu$ (without more weight at the beginning).
\end{Lemma}

Hence, a dynamic programming approach is still valid here. Assuming we have computed optimal strategies of $T_1, \cdots, T_{n-1}$ with more weight at the beginning $S'_1, \cdots, S'_{n-1}$ and without more weight at the beginning $S_1, \cdots, S_{n-1}$ respectively, we can compute an optimal strategy with more weight at the beginning $S'_{n}$ of $T_n$ with left branch $S'_i$ and right branch $S_{n-i}$, where $i\geq 1$ is given by:
\[i:=\underset{1\leq j\leq n-1}{\argmin} (\mu'(S'_j)+\mu(S_{n-j})+(n-j)\alpha+(j-1)\beta+\gamma),\]
$\mu$ is the measure introduced in \cref{sec: def optimal strategies} parametrized by $(\alpha,\beta)$ and $\mu'$ is the new measure parametrized by $(\alpha,\beta,\gamma)$.

\begin{Remark}\label{rem: strat half torsion case}
Note that this approach also applies to the parts $F_1$ and $\widetilde{F}_2$ of $F$ in the half available torsion case.
\end{Remark}

\subsection{Using optimal strategies in isogeny computations}\label{sec: using optimal strategies}

As suggested in \cite[§ 1.3.8]{SIKE_spec}, we can represent any strategy $S$ in a unique way as a sequence of integers $(s_1,\cdots, s_{t-1})$ by considering the tree topology $T_S$ of $S$ (as defined in \cref{sec: def optimal strategies}). To establish this sequence $(s_1,\cdots, s_{t-1})$, we write down for every internal node of the tree $T_S$ the number of leaves to its right and walk on it depth-first left-first.

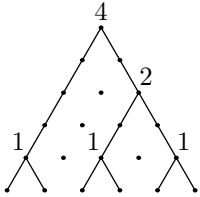
\begin{figure}[!h]
\centering
\begin{tikzpicture}[line cap=round,line join=round,>=triangle 45,x=1cm,y=0.866cm,scale=0.5]
\clip(-0.5,-0.1) rectangle (6.5,6);

\draw[fill=black] (3,5) circle (0.05);

\draw[fill=black] (2.5,4) circle (0.05);
\draw[fill=black] (3.5,4) circle (0.05);

\draw[fill=black] (2,3) circle (0.05);
\draw[fill=black] (3,3) circle (0.05);
\draw[fill=black] (4,3) circle (0.05);

\draw[fill=black] (1.5,2) circle (0.05);
\draw[fill=black] (2.5,2) circle (0.05);
\draw[fill=black] (3.5,2) circle (0.05);
\draw[fill=black] (4.5,2) circle (0.05);

\draw[fill=black] (1,1) circle (0.05);
\draw[fill=black] (2,1) circle (0.05);
\draw[fill=black] (3,1) circle (0.05);
\draw[fill=black] (4,1) circle (0.05);
\draw[fill=black] (5,1) circle (0.05);

\draw[fill=black] (0.5,0) circle (0.05);
\draw[fill=black] (1.5,0) circle (0.05);
\draw[fill=black] (2.5,0) circle (0.05);
\draw[fill=black] (3.5,0) circle (0.05);
\draw[fill=black] (4.5,0) circle (0.05);
\draw[fill=black] (5.5,0) circle (0.05);

\draw [line width=0.5pt] (3,5)--(2.5,4);
\draw [line width=0.5pt] (3,5)--(3.5,4);

\draw [line width=0.5pt] (2.5,4)--(2,3);
\draw [line width=0.5pt] (3.5,4)--(4,3);

\draw [line width=0.5pt] (2,3)--(1.5,2);
\draw [line width=0.5pt] (4,3)--(3.5,2);
\draw [line width=0.5pt] (4,3)--(4.5,2);

\draw [line width=0.5pt] (1.5,2)--(1,1);
\draw [line width=0.5pt] (3.5,2)--(3,1);
\draw [line width=0.5pt] (4.5,2)--(5,1);

\draw [line width=0.5pt] (1,1)--(0.5,0);
\draw [line width=0.5pt] (1,1)--(1.5,0);
\draw [line width=0.5pt] (3,1)--(2.5,0);
\draw [line width=0.5pt] (3,1)--(3.5,0);
\draw [line width=0.5pt] (5,1)--(4.5,0);
\draw [line width=0.5pt] (5,1)--(5.5,0);

\draw (3,5.5) node {$4$};

\draw (0.8,1.5) node {$1$};

\draw (4.2,3.5) node {$2$};

\draw (2.8,1.5) node {$1$};

\draw (5.2,1.5) node {$1$};

\end{tikzpicture}
\captionsetup{width=\textwidth}
\caption{Strategy of depth $5$ represented by $(4,1,2,1,1)$.}
\end{figure}

Given a strategy and a basis of the kernel, it is natural to compute the isogeny chain recursively, as proposed in \cite[§ 1.3.8]{SIKE_spec}. An iterative version of the same algorithm derived from \cite[Algorithm 2]{Jesus2023} and \cite{Theta_dim2} has been implemented in this work. We present it in Algorithm \ref{alg: apply strategy}.

Algorithm~\ref{alg: apply strategy dim 4} applies specifically to dimension $4$ isogenies derived from Kani's lemma when the first isogenies of the chain involving gluings $f_1, \cdots, f_{m+1}$ are already computed. The changes between Algorithms \ref{alg: apply strategy} and \ref{alg: apply strategy dim 4} are modest and are due to the fact that $f_{m+1}\circ\cdots\circ f_1$ is given on entry and that we want to avoid an unnecessary doubling corresponding to the extreme left edge of the strategy (going to the leaf at the origin, which is not used). Note that Algorithm \ref{alg: apply strategy dim 4} not only applies to full isogeny computation $F=f_e\circ\cdots\circ f_1$ but also to partial chain computation when we cannot access the $2^{e+2}$-torsion (as for the strategy computation, following \cref{rem: strat half torsion case}).

\begin{algorithm}[H]
\SetAlgoLined
\KwData{\justifying A level $2$ theta structure $(\mA,\mL,\Theta_\mL)$, a basis $\mathscr{B}_{K''}:=(T''_1,\cdots, T''_g)$ of a maximal isotropic subgroup $K''\subseteq\mA[2^{e+2}]$ such that $[2^{e+1}]K''=K_2(\Theta_\mL)$ and a strategy $S=(s_1,\cdots, s_{t-1})$ of depth $e-1$.}
\KwResult{\justifying A $2^e$-isogeny $F: \mA\longrightarrow\mB$ of kernel $[4]K''$ expressed as a chain of $2$-isogenies $f_i: A_i\longrightarrow A_{i+1}$ ($1\leq i\leq e$).}

$l\longleftarrow 1$\;
$L_{levels}\longleftarrow [0]$\;
$L_{basis}\longleftarrow [\mathscr{B}_{K''}]$\;
\For{$i=1$ \KwTo $e$}{
$\mathscr{B}\longleftarrow$ last element of $L_{basis}$\;
\While{$\sum_{x\in L_{levels}} x\neq e-l$}{
Append $s_l$ to $L_{levels}$\;
$\mathscr{B}\longleftarrow [2^{s_l}]\mathscr{B}$\;
Append $\mathscr{B}$ to $L_{basis}$\;
$l\longleftarrow l+1$\;
}
Use Algorithm \ref{alg: dual theta null point} with input $\mathscr{B}$ to compute the isogeny $f_i$ of kernel $[4]\langle\mathscr{B}\rangle$\;
Remove the last elements of $L_{levels}$ and $L_{basis}$\;
$L_{basis}\longleftarrow [f_i(\mathscr{C})\mid \mathscr{C}\in L_{basis}]$ (Algorithm \ref{alg: isogeny eval})\;
}
\KwRet $f_1, \cdots, f_e$\;

\caption{Computing an isogeny chain with a strategy.}\label{alg: apply strategy}
\end{algorithm}

\begin{algorithm}[H]
\SetAlgoLined
\KwData{\justifying A basis $\mathscr{B}_{K''}:=(T''_1,\cdots, T''_4)$ of a maximal isotropic subgroup $K''\subseteq E_1^2\times E_2^2[2^{e+2}]$ such that $[2^{e+1}]K''=K_2(\Theta_\mL)$ and $[4]K''=\ker(F)$ where $F\in\End(E_1^2\times E_2^2)$ is given by  \cref{eq: definition F}, the $2^{m+1}$-isogeny chain $f_{m+1}\circ\cdots\circ f_1$ of \cref{lemma: first gluing isogenies} and a strategy $S=(s_1,\cdots, s_{t-1})$ of depth $e-m-1$ with more weight at the beginning.}
\KwResult{\justifying A chain of $2$-isogenies $f_1, \cdots, f_e$ such that $F=f_e\circ \cdots\circ f_1$.}

$l\longleftarrow 1$\;
$L_{levels}\longleftarrow [0]$\;
$L_{basis}\longleftarrow [\mathscr{B}_{K''}]$\;
\For{$i=m+1$ \KwTo $e$}{
$\mathscr{B}\longleftarrow$ last element of $L_{basis}$\;
\While{$\sum_{x\in L_{levels}} x\neq e-l$}{
Append $s_l$ to $L_{levels}$\;
\If{$i>m+1$ or $\sum_{x\in L_{levels}} x\neq e-l$}{
\tcc{We avoid a useless doubling when $i=m+1$}
$\mathscr{B}\longleftarrow [2^{s_l}]\mathscr{B}$\;
Append $\mathscr{B}$ to $L_{basis}$\;
}
$l\longleftarrow l+1$\;
}
Remove the last element of $L_{levels}$\;
\eIf{$i>m+1$}{
Use Algorithm \ref{alg: dual theta null point} with input $\mathscr{B}$ to compute the isogeny $f_i$ of kernel $[4]\langle\mathscr{B}\rangle$\;
Remove the last element of $L_{basis}$\;
$L_{basis}\longleftarrow [f_i(\mathscr{C})\mid \mathscr{C}\in L_{basis}]$\;
}{
$L_{basis}\longleftarrow [f_{m+1}\circ\cdots\circ f_1(\mathscr{C})\mid \mathscr{C}\in L_{basis}]$\;
}
}
\KwRet $f_1, \cdots, f_e$\;

\caption{Computing an isogeny chain derived from Kani's lemma in dimension $4$ with a strategy.}\label{alg: apply strategy dim 4}
\end{algorithm}

\bibliographystyle{amsplain}

\end{document}